\documentclass[acmsmall]{acmart}

\usepackage{multirow}

\usepackage{subfigure}
\usepackage{pifont}
\usepackage[normalem]{ulem}
\usepackage{bm}
\usepackage{url}
\usepackage{hyperref}
\hypersetup{
    colorlinks=true,
    linkcolor=blue,
    filecolor=magenta,      
    urlcolor=cyan,
}

\graphicspath{{pictures/}}

\usepackage{algorithm}
\usepackage{algorithmic}
\usepackage{setspace}

\usepackage{tcolorbox}

\newcommand{\summary}[1]{
\begin{center}
\begin{tcolorbox}[colback=gray!15, colframe=black, boxsep=-0.15cm, middle=-0.15cm]
\textbf{\ding{46} Summary}
$\blacktriangleright$
{#1}
$\blacktriangleleft$
\end{tcolorbox}
\end{center}
}

\AtBeginDocument{%
  \providecommand\BibTeX{{%
    \normalfont B\kern-0.5em{\scshape i\kern-0.25em b}\kern-0.8em\TeX}}}

\setcopyright{acmcopyright}
\copyrightyear{2023}
\acmYear{2023}

\acmJournal{TOSEM}
\acmVolume{0}
\acmNumber{0}
\acmArticle{1}
\acmMonth{0}

\title{Abstract Syntax Tree for Programming Language Understanding and Representation: How Far Are We?}

\begin{document}

\author{Weisong Sun} \email{weisongsun@smail.nju.edu.cn}
\orcid{0000-0001-9236-8264}
\affiliation{
  \institution{State Key Laboratory for Novel Software Technology, Nanjing University}
  \city{Nanjing}
  \state{Jiangsu}
  \country{China}
  \postcode{210093}
}
\affiliation{
  \institution{School of Computer Science and Engineering, Nanyang Technological University}
  \state{50 Nanyang Avenue}
  \country{Singapore}
  \postcode{639798}
}

\author{Chunrong Fang} \email{fangchunrong@nju.edu.cn}
\orcid{0000-0002-9930-7111}
\authornote{\textbf{Chunrong Fang is the corresponding author.}}
\affiliation{
  \institution{State Key Laboratory for Novel Software Technology, Nanjing University}
  \city{Nanjing}
  \state{Jiangsu}
  \country{China}
  \postcode{210093}
}

\author{Yun Miao}     
\email{miaoyun001my@gmail.com}
\orcid{0000-0003-0347-6939}
\affiliation{
  \institution{State Key Laboratory for Novel Software Technology, Nanjing University}
  \city{Nanjing}
  \state{Jiangsu}
  \country{China}
  \postcode{210093}
}

\author{Yudu You}
\email{nju_yyd@163.com}
\orcid{0009-0006-2193-3696}
\affiliation{
  \institution{State Key Laboratory for Novel Software Technology, Nanjing University}
  \city{Nanjing}
  \state{Jiangsu}
  \country{China}
  \postcode{210093}
}

\author{Mengzhe Yuan}
\email{shiroha123321@gmail.com}
\orcid{0009-0003-9281-2662}
\affiliation{
  \institution{State Key Laboratory for Novel Software Technology, Nanjing University}
  \city{Nanjing}
  \state{Jiangsu}
  \country{China}
  \postcode{210093}
}

\author{Yuchen Chen}     \email{yuc.chen@outlook.com}
\orcid{0000-0002-3380-5564}
\affiliation{
  \institution{State Key Laboratory for Novel Software Technology, Nanjing University}
  \city{Nanjing}
  \state{Jiangsu}
  \country{China}
  \postcode{210093}
}

\author{Quanjun Zhang}    \email{quanjun.zhang@smail.nju.edu.cn}
\orcid{0000-0002-2495-3805}
\affiliation{
  \institution{State Key Laboratory for Novel Software Technology, Nanjing University}
  \city{Nanjing}
  \state{Jiangsu}
  \country{China}
  \postcode{210093}
}

\author{An Guo}     
\email{guoan218@smail.nju.edu.cn}
\orcid{0009-0005-8661-6133}
\affiliation{
  \institution{State Key Laboratory for Novel Software Technology, Nanjing University}
  \city{Nanjing}
  \state{Jiangsu}
  \country{China}
  \postcode{210093}
}

\author{Xiang Chen}
\email{xchencs@ntu.edu.cn}
\orcid{0000-0002-1180-3891}
\affiliation{
  \institution{Nantong University}
  \city{Nantong}
  \state{Jiangsu}
  \country{China}
  \postcode{226019}
}
\author{Yang Liu}    
\email{yangliu@ntu.edu.sg}
\orcid{0000-0001-7300-9215}
\affiliation{
  \institution{School of Computer Science and Engineering, Nanyang Technological University}
  \state{50 Nanyang Avenue}
  \country{Singapore}
  \postcode{639798}
}

\author{Zhenyu Chen}
\email{zychen@nju.edu.cn}
\orcid{0000-0002-9592-7022}
\affiliation{
  \institution{State Key Laboratory for Novel Software Technology, Nanjing University}
  \city{Nanjing}
  \state{Jiangsu}
  \country{China}
  \postcode{210093}
}

\renewcommand{\shortauthors}{W. Sun and C. Fang, Y. Miao, Y. You, M. Yuan, Y. Chen, Q. Zhang, A. Guo, X. Chen, Y. Liu, X. Chen.}

\begin{abstract}
Programming language understanding and representation (a.k.a code representation learning) has always been a hot and challenging task in the field of software engineering. 
It aims to apply deep learning techniques to produce numerical representations of the source code features while preserving its semantics. 
These representations can be used for facilitating subsequent code-related tasks, e.g., code summarization. 
The abstract syntax tree (AST), a fundamental code feature, illustrates the syntactic information of the source code and has been widely used in code representation learning. 
It is commonly acknowledged that AST-based code representation is critical to solving code-related tasks. However, there is still a lack of systematic and quantitative evaluation of how well AST-based code representation facilitates subsequent code-related tasks. Additionally, learning an AST-based code representation is an extremely complex endeavor involving three intertwining stages, including, AST parsing, AST preprocessing, and AST encoding. The solutions available in each stage are diverse. There is currently a lack of guidance on how to select solutions at each stage to get the most out of AST.

In this paper, we first conduct a comprehensive empirical study to explore the effectiveness of the AST-based code representation in facilitating follow-up code-related tasks. 
To do so, we compare the performance of models trained with code token sequence (Token for short) based code representation and AST-based code representation on three popular types of code-related tasks, including code clone detection, code search, and code summarization. 
Surprisingly, the overall quantitative statistical results demonstrate that models trained with AST-based code representation consistently perform worse across all three tasks compared to models trained with Token-based code representation. 
Our further quantitative analysis reveals that models trained with AST-based code representation outperform models trained with Token-based code representation in certain subsets of samples across all three tasks. For instance, in the code summarization task, such samples constitute as much as 39\% of the total, while in the code search task, they account for 28\%. 
This implies that a more in-depth investigation is needed to understand how to effectively utilize AST to enhance a model's ability to solve different code-related tasks. 
Therefore, we also conduct comprehensive experiments to evaluate and reveal the impact of the choice of AST parsing/preprocessing/encoding methods on AST-based code representation and subsequent code-related tasks. 
The experiments involve four AST parsing methods, six AST preprocessing methods, and four AST encoding methods, all of which are widely utilized in existing AST-based code representation research. 
The experimental results showcase that the impact of different methods at different stages varies for different code-related tasks. 
As ASTs are now being used in practice under various contexts (a.k.a., code-related tasks), the results in this paper called for more research on context-specific AST-based code representation learning in the future. 
Our study provides future researchers with detailed guidance on how to select solutions at each stage to fully exploit AST.
\end{abstract}



\maketitle

\section{Introduction}
\label{sec:introduction}
The ascendancy of deep learning (DL) in software engineering has engendered a growing interest in code representation learning for program understanding and representation. Code representation learning aims to use DL techniques to learn the distributed vector representation of (program) code features, which supersedes the feature engineering and selection step in classic machine learning methods. The distributed vector representation of code features (code representation/embedding, for short) preserves code semantics and facilitates subsequent code intelligent tasks, such as code clone detection~\cite{2016-DL-Code-Clone-Detection, 2020-Functional-Code-Clone-Detection}, neural code search~\cite{2018-Neural-Code-Search, 2021-Multimodal-Representation-NCS, 2022-TranCS}, neural code summarization~\cite{2018-Deep-Code-Comment-Generation, 2021-Project-Level-Encoding-Code-Summarization, 2023-EACS}, etc.

The procedure of code representation learning can be decomposed into two sequential steps: extracting code features from program source code and encoding code features using DL techniques, where the code embeddings produced by DL techniques aim to capture the semantics of the code features. The choice of code features has a great impact on code representation learning.
In existing code representation learning efforts, typical code features include code token sequences (Token), abstract syntax trees (AST), control flow graphs (CFG), data flow graphs (DFG), etc. Researchers utilize different code features depending on the kind of information that needs to be extracted, such as Token for lexical information, AST for syntactic information, and CFG/DFG for semantic information~\cite{2022-code-representation-survey}. 
Syntactic differences are considered to be the key differences between programming languages and natural languages. Therefore, AST, which expresses syntactic information, has received widespread attention and is widely used in code representation learning research~\cite{2016-DL-Code-Clone-Detection, 2019-Code2Vec, 2023-Fold2Vec, 2019-ASTNN, 2020-Rencos, 2020-Retrieve-and-Refine-Comment-Generation, 2022-Automatic-Source-Code-Summarization-With-GNN, 2018-Learning-Represent-Programs-with-Graphs, 2019-Ast-attendgru, 2020-Hybrid-DeepCom, 2018-Deep-Code-Comment-Generation, 2019-Code-Summarization-with-Extended-Tree-LSTM, 2021-Authorship-Attribution-Language-agnostic-Approach, 2021-CAST, 2016-TBCNN, 2021-API2Com, 2017-Supervised-Deep-Features-for-Clone-Detection, 2019-Multi-modal-Attention-for-Code-Retrieval, 2021-BASTS, 2021-TabCS, 2022-AST-trans}. 

It is generally believed that Token contains lexical information, while AST contains lexical information and the syntactic structure of source code~\cite{2019-ASTNN}. 
Previous research~\cite{2019-ASTNN, 2020-Hybrid-DeepCom, 2022-FcarCS} indicates that the source code comprises rich syntactic information and cannot be directly considered as plain text. 
It is considered crucial to model the code's syntactic knowledge via AST which is often done in practice. AST has been widely used to model the source code (i.e, learning AST-based code representation) for many follow-up code-related tasks, such as code classification~\cite{2019-ASTNN, 2022-UAST}, code clone detection~\cite{2019-ASTNN, 2019-Recursive-Aggregation-of-AST-for-Clone-Detection, 1998-Clone-Detection-Using-ASTs}, code search~\cite{2019-Multi-modal-Attention-for-Code-Rerieval, 2021-CRaDLe, 2020-NJACS, 2020-Multi-Perspective-Architecture-for-CS, 2021-TabCS, 2021-Multimodal-Representation-NCS}, code summarization~\cite{2021-BASTS, 2021-CAST}, etc. Nevertheless, some other studies have demonstrated that Token outperforms AST in facilitating DL models to learn code semantics, or the incorporation of AST leads to reduced DL models' performance~\cite{2021-FCCA, 2021-API2Com, 2019-Multi-modal-Attention-for-Code-Retrieval}. 
Hence, it necessitates conducting a systematic and quantitative evaluation of the extent to which AST-based code representation facilitates subsequent code-related tasks.

\begin{figure*}[!t]
  \centering
  \includegraphics[width=0.9\textwidth]{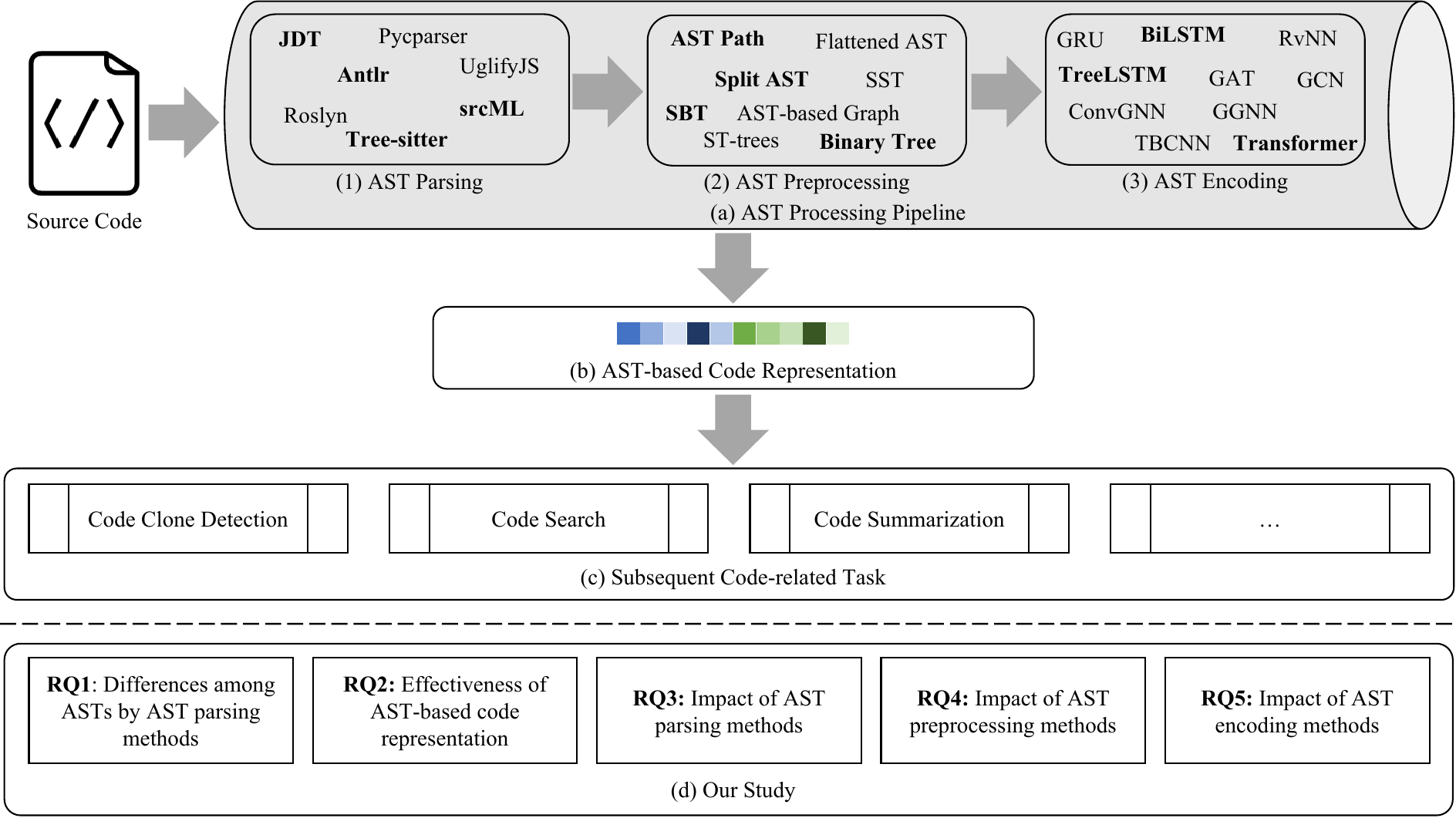}
  \caption{Overview of our empirical study}
  \label{fig:overview_of_AST_processing_and_usage}
\end{figure*}

Furthermore, learning AST-based code representation and using it to solve code-related tasks is extremely challenging. Fig.~\ref{fig:overview_of_AST_processing_and_usage}(a)--(c) show the procedure of using AST for code representation learning to solve code-related tasks. 
It is observed that given a piece of source code (usually a method/function), it goes through the AST processing pipeline to obtain an AST-based code representation. Then, the AST-based code representation that preserves code semantics is fed into distinct task models to solve different code-related tasks.  
As shown in Fig.~\ref{fig:overview_of_AST_processing_and_usage}(a), the internal design of the AST processing pipeline is complex and varied, which consists of three core and intertwining stages: AST parsing, AST preprocessing, and AST encoding. 

\textbf{For AST Parsing}: This stage is responsible for parsing the source code into AST. Different AST parsing methods adopt different lexical rules and grammar rules to parse the source code. 
Currently, there are many AST parsing methods (also called AST parsers) available for users, such as JDT~\cite{JDT}, ANTLR~\cite{antlr}, Tree-sitter~\cite{tree-sitter}, javalang~\cite{javalang}, srcML~\cite{2013-srcML}, JavaParser~\cite{JavaParser}, etc. 
It is noteworthy that these parsers would generate distinct ASTs for the same source code due to different lexical rules and grammar rules. 
As DL models are highly sensitive to input data, users must consider whether the differences in ASTs generated by different parsers might affect the model's learning of AST-based code representation and subsequent code-related tasks.  
Utkin et al.~\cite{2022-Impact-of-Code-Parsers-on-Models} conduct a preliminary empirical study to explore the impact of different AST parsers on the performance of Code2Seq~\cite{2019-Code2seq} and TreeLSTM~\cite{2015-Improved-Semantic-Representations} on the task of method name prediction (MNP, for short). 
Their experimental results demonstrate that ASTs generated by different AST parsers vary in size and abstraction level. Such variation affects the models' MNP performance. 
As mentioned in their paper~\cite{2022-Impact-of-Code-Parsers-on-Models}, extending the findings to other code-related tasks (e.g., code clone detection, code search, and code summarization) is still an important topic that needs to be investigated. 

\textbf{For AST Preprocessing}: This stage is responsible for preprocessing the AST to simplify the complexity of the AST~\cite{2017-Supervised-Deep-Features-for-Clone-Detection, 2019-Multi-modal-Attention-for-Code-Retrieval}, or to adapt to the input requirements of different DL models~\cite{2022-AST-trans}. 
As shown in Fig.~\ref{fig:overview_of_AST_processing_and_usage}(a), existing research has successively proposed a variety of AST preprocessing methods. 
These methods can be divided into two categories according to the form of the AST data they output: sequential AST preprocessing methods and structural AST preprocessing methods. 
As its name implies, the sequential AST preprocessing method processes the given AST into sequential data (e.g., SBT~\cite{2018-Deep-Code-Comment-Generation} or AST Path~\cite{2018-Path-based-Representation-Predicting-Program-Properties}), while the structural AST preprocessing method processes the given AST into structural data (e.g., Binary Tree~\cite{2017-Supervised-Deep-Features-for-Clone-Detection} and Split AST~\cite{2021-CAST}). 
The effectiveness of these preprocessing methods has been more or less verified in corresponding papers. 
However, horizontal comparisons of these methods are still lacking, even though such comparisons can help guide subsequent users to choose a method suitable for their specific scenarios.  
It is important to note that there are significant differences in node and structure information between sequential AST data and structural AST data. Additionally, there are noticeable differences among different sequential/structural AST data as well. 
However, the impact of these differences on the model's ability to learn AST-based code representation and its performance on subsequent code-related tasks is still lacking systematic investigation. 

\textbf{For AST Encoding}: This stage is responsible for transforming the sequential/structural AST data into numerical vector representations (i.e., AST-based code representation). 
Such representations can be used for solving various code-related tasks. 
To generate accurate and semantic-preserving AST-based code representation, as shown in Fig.~\ref{fig:overview_of_AST_processing_and_usage}(a), software engineering researchers have proposed AST encoding methods and tried various neural network architectures. 
According to the form of the AST data they target, existing AST encoding methods can be categorized into two types: sequence models (e.g., BiLSTM~\cite{schuster1997bidirectional} and Transformer~\cite{2017-Transformer}) and tree-structured models (e.g., TreeLSTM~\cite{2015-Improved-Semantic-Representations} and AST-Trans~\cite{2022-AST-trans}). 
The sequence models and tree-structure models are designed for encoding sequential AST data and structural AST data, respectively. 
Just as with AST preprocessing methods, there exist notable differences between sequence models and tree-structured models. Moreover, there are significant differences among different sequence/tree-structure models.  
However, there is still a lack of systematic evaluation of the impact of these differences on the produced AST-based code representation and subsequent code-related tasks. 

Based on the above brief summary of the three stages of the AST processing pipeline, it is evident that: (1) there is a wide array of choices for AST parsing/preprocessing/encoding methods; (2) there is still a lack of systematic evaluation of the impact of the choice of these methods on AST-based code representation and subsequent code-related tasks.

To address the above gaps, in this paper, we conduct a comprehensive empirical study to (1) explore the effectiveness of the AST-based code representation in improving subsequent code-related tasks; (2) reveal the impact of the choice of AST parsing/preprocessing/encoding methods on AST-based code representation and subsequent code-related tasks. 
All experiments are performed on three popular types of code-related tasks, including a code-to-code matching task, i.e., code clone detection~\cite{2016-DL-Code-Clone-Detection}, a text-to-code matching task, i.e., code search~\cite{2018-Neural-Code-Search}, and a code-to-text generating task, i.e., code summarization~\cite{2020-Transformer-based-Approach-for-Code-Summarization}. We quantitatively evaluate the effectiveness of the involved models on commonly used metrics for different tasks. A total of 10 metrics are used in the three tasks. BigCloneBench~\cite{2014-Benchmark-Inter-project-Code-Clones} and CodeSearchNet~\cite{2019-CodeSearchNet-Challenge} are selected as experimental datasets, which are widely used in existing code-related studies. 
The purpose of this study is to provide a systematic and generalized understanding of AST processing and applications, which could facilitate learning AST-based code representation and solving many code-related tasks. 
Specifically, as illustrated in Fig.~\ref{fig:overview_of_AST_processing_and_usage}(d), we conduct comprehensive empirical studies to answer the following five research questions (RQs):
\begin{description}
    \item[\textbf{RQ1:}] What are the differences among the ASTs parsed by different AST parsing methods? To answer this question, we follow~\cite{2022-Impact-of-Code-Parsers-on-Models} and use five metrics (including tree size, tree depth, branching factor, unique types, and unique tokens) to characterize and compare the ASTs generated by four commonly used AST parsing methods, including JDT, srcML, ANTLR, and Tree-sitter. The detailed experimental design and results are reported in Section~\ref{subsec:Answer_to_RQ1}.
    
    \item[\textbf{RQ2:}] Does AST improve the expressiveness of code representation and facilitate subsequent code-related tasks? To answer this question, we conduct a comparative analysis of the performance of eight models trained with four distinct types of inputs including Token, SBT, SBT w/o Token, and Token + SBT w/o Token on all three code-related tasks. Among the eight models, four are built on the BiLSTM architecture and four are built on the Transformer architecture. The detailed experimental design and results are reported in Section~\ref{subsec:Answer_to_RQ2}.
    
    \item[\textbf{RQ3:}] How do AST parsing methods affect the performance of AST-based code representation on subsequent code-related tasks? To answer this question, we compare the performance of eight models trained with the ASTs generated by the same four AST parsing methods as \textbf{RQ2} on all three code-related tasks. Similar to \textbf{RQ2}, among the eight models, four are built on the BiLSTM architecture and four are built on the Transformer architecture. The detailed experimental design and results are reported in Section~\ref{subsec:Answer_to_RQ3}.
    
    \item[\textbf{RQ4:}] How do AST preprocessing methods affect the performance of AST-based code representation on subsequent code-related tasks? To answer this question, we compare the performance of six models trained with the sequential/structural AST data produced by six AST preprocessing methods, including three sequential AST preprocessing methods (including BFS, SBT, and AST Path) and three structural AST preprocessing methods (including Raw AST, Binary Tree, and Split AST) on all three code-related tasks. Among the six models, three are built on the BiLSTM architecture and three are built on the TreeLSTM architecture. The detailed experimental design and results are reported in Section~\ref{subsec:Answer_to_RQ4}.
    
    \item[\textbf{RQ5:}] How do AST encoding methods affect the performance of AST-based code representation on subsequent code-related tasks? To answer this question, we compare the performance of four models built on four classic types of AST encoding methods, including BiLSTM, Transformer, TreeLSTM, and AST-Trans, on all three code-related tasks. The detailed experimental design and results are reported in Section~\ref{subsec:Answer_to_RQ5}.
\end{description}

Based on the extensive experimental results, we conclude the following findings:
 \begin{itemize}
    \item The ASTs generated by different AST parsing methods differ in size and abstraction level. The size (in terms of tree size and tree depth) and abstraction level (in terms of unique types and unique tokens) of the ASTs generated by JDT are the smallest and highest, respectively. On the contrary, ASTs generated by ANTLR exhibit the largest size and the lowest abstraction level. Tree-sitter and srcML are both intermediate in structure size and abstraction level between JDT and ANTLR.
    
    \item Overall, the contribution of AST to the expressiveness of code representation is weaker compared to Token. Similarly, its promotion impact on subsequent code-related tasks is also weaker than Token. 
    Nonetheless, it is worth noting that the models trained with AST outperform the models trained with Token on certain samples of all three code-related tasks, such as pairs of code snippets have low token similarity in code clone detection, code snippets require more high-level abstract summaries in code summarization, and code snippets semantically match but contain fewer query words in code search.
    In these samples, the syntactic information contained in the AST plays a vital role in improving the expressiveness of code representation as well as code-related tasks. 
    
    \item Among all four AST parsing methods, JDT generates ASTs with the smallest tree size, shallowest tree depth, and highest abstraction level, and yields the most favorable outcomes for all three code-related tasks. 
    Although the ASTs generated by srcML, ANTLR, and Tree-sitter are generally richer than those of JDT, this richness may cause redundancy and potentially bring a higher learning burden to the model at the same time, which may not be conducive to improving the model’s performance on code-related tasks.
    
    \item The AST data produced by the six AST preprocessing methods has noticeable differences in terms of AST node and structure information. 
    Both AST node and structure information contained in the AST data can enhance AST-based code representation and subsequent code-related tasks. 
    Different code-related tasks have varying requirements for AST node and structure information, leading to differences in performance among different AST preprocessing methods across distinct tasks. 
    For example, in code clone detection and code search tasks, the AST preprocessing methods that preserve complete AST node (i.e., token) and structure information (e.g., SBT and Raw AST) are optimal. In the code summarization task, when using tree-structured models as AST encoding methods (e.g., TreeLSTM), removing the redundant node and structure information (e.g., Binary Tree) yields the most significant improvement.
    
    \item The four AST encoding methods (models) we investigate have different contributions to AST-based code representation and subsequent code-related tasks. 
    Among the two sequence models, Transformer performs better than BiLSTM on all three code-related tasks. 
    Among the two tree-structured models, TreeLSTM outperforms AST-Trans on the code clone task, but the reverse is true on the code summarization and code search tasks. 
    Among all four models, Transformer performs best overall. There is still room for improvement in designing models suitable for learning AST-based code representation from structural AST data (e.g., Raw AST).
\end{itemize}

In summary, we make the following contributions in this paper.
\begin{itemize}
        \item We conduct a systematic and quantitative evaluation of the effectiveness of AST-based code representations on three popular types of code-related tasks. 
        To better grasp under what circumstances AST-based code representation can facilitate code-related tasks, we conduct a more detailed qualitative analysis of code characteristics in samples where models trained using AST performed better. 
        The results of quantitative evaluation and qualitative analysis demonstrate that the current application of AST in code-related tasks is still insufficient, and the value of AST is not maximized. As ASTs are now being used in practice under various contexts (many code-related tasks), the results in this paper called for more research on context-specific AST-based code representation learning in the future.
        
        \item We conduct a comprehensive evaluation of the impact of the choice of AST parsing, preprocessing, and encoding methods on AST-based code representation and subsequent code-related tasks. Our study can provide future researchers with detailed guidance on how to select solutions at each stage to fully exploit AST.
        
        \item To facilitate replication and further research in this area, we release the source code and the dataset of our experiments to help other researchers replicate and extend our study~\cite{2023-AST4PLU}.
\end{itemize}

The remainder of this paper is organized as follows. Section~\ref{sec:background} provides the research background. Section~\ref{sec:study_design} describes our study design. Section~\ref{sec:results_and_findings} presents experimental results and concludes our findings. Section~\ref{sec:threats_to_validity} discusses some threats to validity. Section~\ref{sec:related_work} introduces the related work. 
We conclude the paper in Section~\ref{sec:conclusion}.

\section{Background}
\label{sec:background}
\subsection{Abstract Syntax Tree}
\label{subsec:abstract_syntax_tree}
An Abstract Syntax Tree (AST) is one of the most commonly used code features in code representation learning~\cite{2022-code-representation-survey}. 
It uniquely represents a source code snippet in a given language and grammar. 
As the program is highly structured data compared to plain text, many code-related works attempt to extract the structure information behind the source code to capture syntactic information of the source code~\cite{2022-Learning-Program-Semantics-Empirical-Study}. 
Currently, the specific definitions of AST vary slightly across different sources. The representative one is the definition given by Alon et al.~\cite{2019-Code2seq}. They formalize the AST of a method/function code snippet as follows:
\begin{definition}[Abstract Syntax Tree]
    \label{def:abstract_syntax_tree}
    An Abstract Syntax Tree (AST) for a method is a tuple $\langle N, T, X, s, \delta, \phi \rangle$ where $N$ is a set of non-terminal nodes, $T$ is a set of terminal nodes, $X$ is a set of values, $s \in N$ is the root node, $\delta: N \to (N \cup T)^*$ is a function that maps a non-terminal node to a list of its children, $^*$ represents closure operation, and $\phi: T \to X$ is a function that maps a terminal node to an associated value. Every node except the root appears exactly once in all the lists of children. 
\end{definition}

\begin{figure}[htbp]
    \centering
	\includegraphics[width=0.9\linewidth]{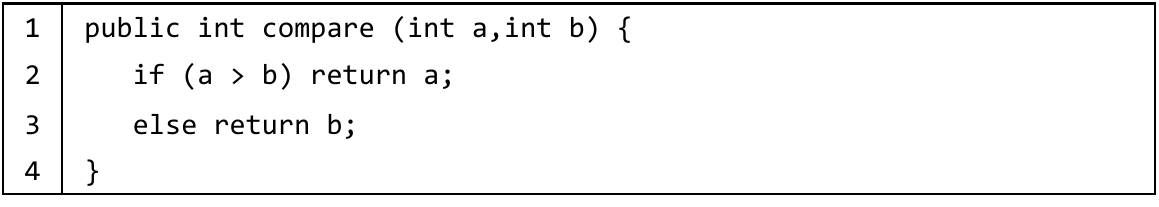}
    \caption{A Java code snippet $s_1$}
    \label{fig:example_of_AST_code}
\end{figure}

\begin{figure*}[htbp]
    \centering
	\includegraphics[width=\linewidth]{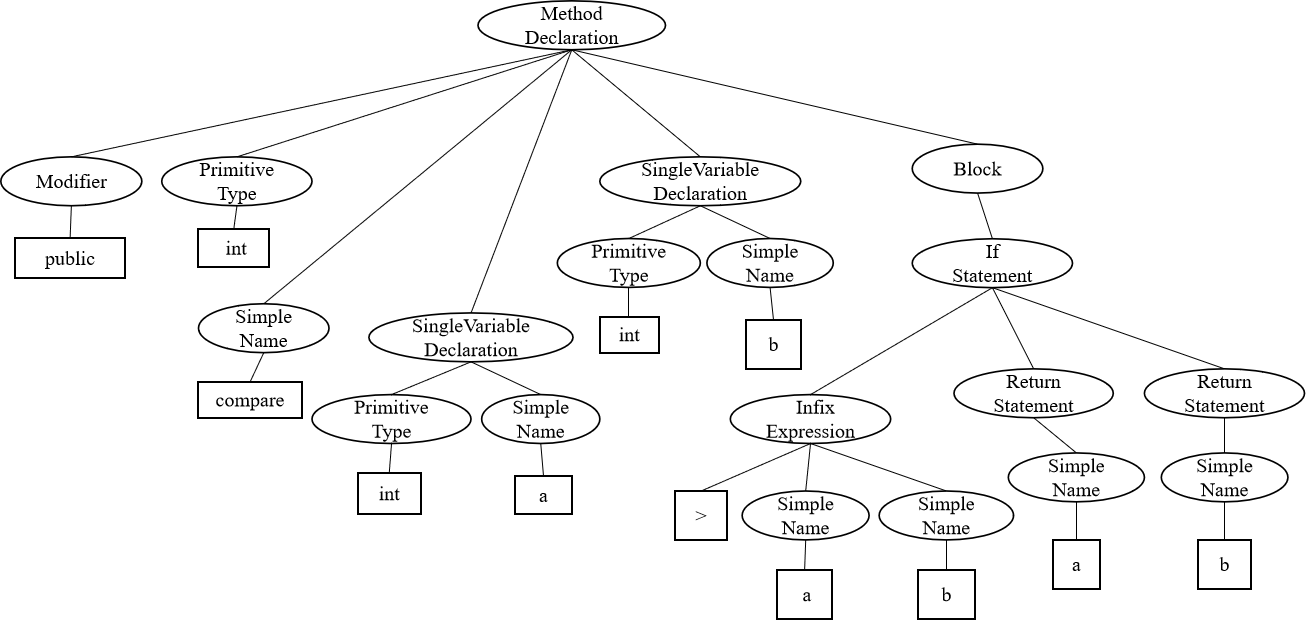}
    \caption{The AST generated by JDT for code snippet $s_1$}
    \label{fig:example_of_AST}
\end{figure*}

In the above definition, terminals are considered as leaf nodes of AST. In fact, many researchers~\cite{2018-Tree-to-Tree-Neural-Network, 2020-Improved-Code-Summarization-Via-GNN, 2021-Code-Summarization-for-Smart-Contracts, 2023-Fold2Vec, 2022-Impact-of-Code-Parsers-on-Models} also regard the values of the terminals as the leaf nodes. In this paper, we also follow the latter.

\noindent\textbf{Example of AST.} Fig.~\ref{fig:example_of_AST} shows an example of AST, which is generated by JDT (an AST parsing method detailed in Section~\ref{subsec:AST_Parsing_method}) for the Java code snippet $s_1$ shown in Fig.~\ref{fig:example_of_AST_code}. 
In the AST shown in Fig.~\ref{fig:example_of_AST}, nodes like $\mathtt{MethodDeclaration}$, $\mathtt{SingleVariableDeclaration}$ and $\mathtt{IfStatement}$ are non-terminals; nodes like $\mathtt{Modifier}$,  $\mathtt{PrimitiveType}$ and $\mathtt{SimpleName}$ are terminals; and leaf nodes like $\mathtt{public}$, $\mathtt{int}$ and $\mathtt{compare}$ are the corresponding values of the terminals. 
It is evident that, compared to the source code of $s_1$, its AST is significantly more complex. 
In addition, as mentioned in Section~\ref{sec:introduction}, different AST parsing methods (e.g., ANTLR and Tree-sitter) would generate distinct ASTs for the same source code due to different lexical rules and grammar rules. 
Therefore, this paper focuses on investigating the current progress of feature engineering and application of AST, aiming to provide guidance for subsequent researchers on how to effectively leverage complex ASTs to enhance code representation and subsequent code-related tasks.

\subsection{Code Representation Learning}
\label{subsec:code_representation_learning}
Code representation learning aims to apply DL techniques to convert source code features (e.g., Token and AST) into distributed, real-valued vector representation (also known as code representations/embeddings).  
Code representations condense the semantics of code features, and two pieces of code with similar semantics will be located in close proximity to one another in the vector space. Code representation enables the application of DL techniques in various code-related tasks.

Code representation learning can be divided into two sequential processes: code feature extraction and code feature representation. 
Code feature extraction is responsible for extracting code features, such as method name~\cite{2018-DeepCodeSearch, 2018-Neural-Code-Search, 2021-TabCS}, Token~\cite{2018-DeepCodeSearch, 2018-Neural-Framework-Code-Summarization, 2021-FCCA, 2021-TabCS, 2021-Multimodal-Representation-NCS, fang2021self}, API sequences~\cite{2018-TL-CodeSum, 2018-DeepCodeSearch, 2021-MuCoS, 2021-Search-for-Compatible-Code}, CFG~\cite{2018-DeepSim, 2019-Multi-modal-Attention-for-Code-Retrieval, 2021-CRaDLe}, etc. 
AST that we investigate in this paper is also a fundamental code feature, illustrating the syntactic structure of the source code. 
More details on other code features are discussed in Section~\ref{subsec:code_features}. 
Code feature representation is responsible for applying DL techniques to convert source code features into code representations. 
Typically, the code representation generated for Token is called Token-based code representation. 
In this paper, we focus on the code representation generated for AST, namely AST-based code representation. 
More details on code representations for other code features are discussed in Section~\ref{subsec:code_representation_learning_related_work}. 
To generate high-quality and semantic-preserving AST-based code representation, existing researchers have successively proposed a large number of AST preprocessing methods, and introduced a lot of neural network models from the fields of CV and NLP as AST encoding methods. 
AST preprocessing methods aim to transform the raw AST generated by AST parsing methods into a data form that is streamlined and easy for AST encoding models to learn. 
As mentioned in Section~\ref{sec:introduction}, common data forms encompass sequential AST data (e.g., SBT~\cite{2018-Deep-Code-Comment-Generation}) and structural AST data (e.g., Binary Tree~\cite{2017-Supervised-Deep-Features-for-Clone-Detection}). 
Correspondingly, in order to process different forms of AST data, existing research has introduced various sequence models (e.g., BiLSTM~\cite{schuster1997bidirectional}) and tree-structured models (e.g., TreeLSTM~\cite{2015-Improved-Semantic-Representations}). 
In this paper, we focus on evaluating the impact of various choices of AST preprocessing and encoding methods, aiming to guide future researchers on how to effectively exploit these methods to enhance AST-based code representation as well as their own code-related tasks. 

\section{Study Design}
\label{sec:study_design}

\subsection{Research Questions}
\label{subsec:research_questions}
In this paper, the evaluation presented in this work aims to answer the following research questions:

\noindent\textbf{RQ1: What are the differences among the ASTs parsed by different AST parsing methods?}

This research question systematically compares ASTs generated by different AST parsing methods. Their differences are revealed through multiple quantitative measurements, including tree size, tree depth, unique types, and unique tokens, applicable programming languages (detailed in Section~\ref{subsec:Answer_to_RQ1}).

\noindent\textbf{RQ2: Does AST improve the expressiveness of code representation and facilitate subsequent code-related tasks?}

This research question seeks to figure out whether code representations fused AST perform better on subsequent code-related tasks than unfused ones. Some existing works~\cite{2019-ASTNN, 2020-Hybrid-DeepCom, 2022-FcarCS} claim that AST can facilitate code representation, while other works hold the opposite attitude~\cite{2021-FCCA, 2021-API2Com, 2019-Multi-modal-Attention-for-Code-Retrieval}. In this paper, we validate the usefulness of AST in facilitating three subsequent code-related tasks, including code clone detection, code search, and code summarization. More details of the three code-related tasks are discussed in Section~\ref{subsubsec:code_related_tasks}.

\noindent\textbf{RQ3: How do AST parsing methods affect the performance of AST-based code representation on subsequent code-related tasks?}

As mentioned in \textbf{RQ1}, the ASTs generated by different AST parsing methods may be different. These differences may have implications for code representation and the models used for subsequent code-related tasks. This research question aims to expose these implications. In this paper, we investigate the four commonly used AST parsing tools, including JDT~\cite{JDT}, srcML~\cite{2013-srcML}, ANTLR~\cite{antlr}, and Tree-sitter~\cite{tree-sitter} (detailed in Section~\ref{subsec:AST_Parsing_method}).

\noindent\textbf{RQ4: How do AST preprocessing methods affect the performance of AST-based code representation on subsequent code-related tasks?}

Considering that the original AST generated by the AST parsing method may be too large and complex to be learned by the code representation model, some researchers have proposed some AST preprocessing methods to alleviate this. These preprocessing methods change (or even break) the structure and content of the original AST. These changes may have implications for AST-based code representation and subsequent code-related task models. This research question aims to expose these implications. In this paper, we investigate six commonly used AST preprocessing methods, including Breadth-first Search (BFS), SBT~\cite{2018-Deep-Code-Comment-Generation}, AST Path~\cite{2018-Path-based-Representation-Predicting-Program-Properties}, Binary Tree~\cite{2017-Supervised-Deep-Features-for-Clone-Detection, 2019-Multi-modal-Attention-for-Code-Retrieval, 2021-FCCA} and Split AST~\cite{2021-BASTS} (detailed in Section~\ref{subsec:AST_preprocessing_method}). 

\noindent\textbf{RQ5: How do AST encoding methods affect the performance of AST-based code representation on subsequent code-related tasks?}

Different neural network architectures/models (e.g., LSTM and Transformer) differ in capturing data features. For the data characteristics of different AST preprocessing results (e.g., sequential data or structured data), existing research has also tried a variety of neural network models to capture the features of these AST data. This research question seeks to expose the impact of AST encoding methods on AST-based code representation and subsequent code-related task models. In this paper, we investigated four AST encoding methods, including BiLSTM~\cite{schuster1997bidirectional} and Transformer~\cite{2017-Transformer} for sequential AST data, TreeLSTM~\cite{2015-Improved-Semantic-Representations} and AST-Trans~\cite{2022-AST-trans} for structured AST data (detailed in Section~\ref{subsec:AST_Encoding_Method}).

\subsection{AST Parsing Method}
\label{subsec:AST_Parsing_method}
AST parsing methods/tools (parsers, for short) are used to convert source code into corresponding abstract syntax trees. There are many AST parsers available for different scenarios. In this paper, we conduct experiments to investigate four commonly used AST parsers as follows.

\textbf{Eclipse Java Development tools (JDT)}~\cite{JDT}: JDT contains a set of plugins to support Java in Eclipse IDE, including code highlighting, code refactoring, and AST parsing. It has been widely used by researchers in solving various software engineering tasks, such as code comment generation~\cite{2018-Deep-Code-Comment-Generation}, code clone detection~\cite{2019-Recursive-Aggregation-of-AST-for-Clone-Detection}, and code understanding~\cite{2021-MulCode}.

\textbf{srcML}~\cite{2013-srcML}: srcML is a lightweight and highly scalable AST parser, which converts source code into an XML representation, where the markup tags identify elements of the AST. Currently, it supports C/C++, C\#, and Java. It has been used in multiple works.
For example, LeClair et al.~\cite{2019-Ast-attendgru}, Wei et al.~\cite{2020-Retrieve-and-Refine-Comment-Generation}, and Shahbazi et al.~\cite{2021-API2Com} leverage srcML to parse AST in their code summarization techniques. 

\textbf{ANTLR}~\cite{antlr}: ANTLR takes the grammar of the target programming language as input and generates a corresponding AST parser. Currently, ANTLR supports many programming languages including C++, Java, and Python. In this paper, we use the open-source ANTLR Java grammar~\footnote{\url{https://github.com/antlr/grammars-v4/tree/master/java/java}} to generate the Java parser. Shi et al.\cite{2021-CAST} and Hua et al.\cite{2021-FCCA} employ ANTLR as the AST parser in their code summarizer CAST and clone detector FCCA, respectively. 
 
\textbf{Tree-sitter}~\cite{tree-sitter}: Like ANTLR, Tree-sitter can generate a parser for a target programming language using its grammar. Tree-sitter focuses on real-time usage in text editors or IDEs, and thus, it can parse code incrementally and is robust enough to parse code with syntax errors. In this paper, we use the most popular Tree-sitter Java grammar~\footnote{https://github.com/tree-sitter/tree-sitter-java} to generate the Java parser. Gu et al.~\cite{2021-Multimodal-Representation-NCS} and Guo et al.~\cite{2022-UniXcoder} adopt Tree-sitter to parse source code into AST in the code search tool and the pre-training task, respectively.

\subsection{AST Preprocessing Method}
\label{subsec:AST_preprocessing_method}

In this section, we introduce several typical AST preprocessing methods that are investigated, including structure-based traversal (SBT)~\cite{2018-Deep-Code-Comment-Generation}, AST path~\cite{2018-Path-based-Representation-Predicting-Program-Properties}, binary tree~\cite{2017-Supervised-Deep-Features-for-Clone-Detection, 2019-Multi-modal-Attention-for-Code-Retrieval, 2021-FCCA}, and split AST~\cite{2021-BASTS}.

\textbf{Breadth-first Search (BFS).} BFS is a traditional AST traversal method, which starts at the root node and explores all nodes at the present depth prior to moving on to the nodes at the next depth level. Through BFS traversal, users can get the node sequence of an AST. 

\textbf{Structure-based Traversal (SBT)}. SBT proposed by Hu et al.~\cite{2018-Deep-Code-Comment-Generation} is a popular AST traversal method and has been widely used in the field of code representation~\cite{2019-Ast-attendgru, 2020-Retrieve-and-Refine-Comment-Generation}. SBT converts a tree into a sequence consisting of tree nodes and brackets, from which we can restore the original tree. SBT representation is produced via the tree preorder traversal algorithm. The detailed procedure is as follows: 1) from the root node, SBT first uses a pair of brackets to represent the tree structure and puts the root node itself behind the right bracket. 2) Then, it traverses the subtrees of the root node and puts all root nodes of subtrees into the brackets. 3) Finally, it traverses each subtree recursively until all nodes are traversed and the final sequence is produced. 

\textbf{AST Path.} AST path proposed by Alon et al.~\cite{2018-Path-based-Representation-Predicting-Program-Properties} is a common AST traversal method, which represents the AST using a set of paths extracted from the AST.  An AST path is a path between two terminals in the AST, which is formally defined as follows.

\begin{definition}[AST Path]
    Given an $AST:= \langle N, T, X, s, \delta, \phi \rangle$ (see Definition~\ref{def:abstract_syntax_tree}), an AST-path of length $k$ is a sequence of the form: $n_1d_1 \cdots n_kd_kn_{k+1}$, where $n_1, n_{k+1} \in T$ are terminals, for $i \in [2 \cdots k]: n_i \in N$ are non-terminals, and for $i \in [1 \cdots k]: d_i \in \{\uparrow, \downarrow\}$ are movement directions (either up or down in the tree). If $d_i = \uparrow$, then: $n_i \in \delta(n_{i+1})$; if $d_i = \downarrow$, then: $n_{i+1} \in \delta(n_i)$. For an AST-path $p$, we use $start(p)$ to denote $n_1$ - the starting terminal of $p$; and $end(p)$ to denote $n_{k+1}$ - its final terminal.
\end{definition}

We use path-context as the final input of the encoding models. A path-context is a tuple of an AST path and the values of two terminals, which are formally defined as follows.

\begin{definition}[Path-context]
    Given an AST Path $p$, its path-context is a triplet $\langle x_{start}, p, x_{end}\rangle$ where $x_{start} = \phi(start(p))$ and $x_{end} = \phi(end(p))$ are the values associated with the start and end terminals of $p$. 
\end{definition}

Following previous work~\cite{2019-Code2Vec}, we use maximum $length$ -- the maximal value of $k$ -- to limit the length of an AST Path, and the maximum $width$ -- the maximal difference in child index between two child nodes of the same intermediate node -- to limit the horizontal distance between nodes in an AST Path, thereby limiting the size of the training data and reduce sparsity. These values are determined empirically as hyperparameters. 

\textbf{Binary Tree.} Considering the computational cost, some existing code representation works~\cite{2017-Supervised-Deep-Features-for-Clone-Detection, 2019-Multi-modal-Attention-for-Code-Retrieval, 2021-FCCA} transform an otherwise complex AST into simple-structured a binary tree. In a binary tree, each non-leaf node has at most two children. In an AST, different kinds of nodes may have different numbers of children, which can cause problems in parameter-sharing~\cite{2017-Supervised-Deep-Features-for-Clone-Detection}. Usually, to avoid this problem, ASTs are transformed into binary trees whose nodes only have two or zero children. The process contains the following two steps: 1) split nodes with more than two children, and generate a new right child together with the old left child as its children, and then put all children except the leftmost as the children of this new node; repeat this operation in a top-down way until only nodes with zero, one, and two children are left; 2) combine nodes with one child with its child. Now only nodes with zero or two children remain and the AST is transformed into a binary tree. 

\textbf{Split AST.} Due to the complexity of programs, the ASTs of the program source code are usually large and deep, leading to long training time and gradient vanishing problems for DL models~\cite{2021-CAST}. To overcome this problem, some researchers propose to split the large and deep AST into a set of small and shallow subtrees. For example, Zhang et al.~\cite{2019-ASTNN} split an AST into small statement trees, each of which represents a statement in the source code. Shi et al.~\cite{2021-CAST} split an AST into a set of subtrees that are reconstructible. After learning the representation of subtrees, they reconstruct the split ASTs and learn the AST's representation from all subtrees’ representation by a tree-based neural model. Lin et al.~\cite{2021-BASTS} first split the code of a method based on the blocks in the dominator tree of the control flow graph (CFG), and then generate a split AST for each split code.

In this paper, we mainly investigate the state-of-the-art AST split solution proposed by Lin et al.~\cite{2021-BASTS}. The reason why we investigate the solution by Lin et al. is detailed in Section~\ref{subsec:threats_to_internal_validity}. Split ASTs in~\cite{2021-BASTS} are constructed as follows: 

(1) Construct a CFG to capture control flow relationships among statements. In the following, we will omit the sample index and use $G = (\mathcal{U}(G), \mathcal{E}(G))$ to denote the CFG for the current sample. A CFG is a directed graph where each node $u \in \mathcal{U}(G)$ represents a statement and each edge $e \in \mathcal{E}(G) = (u \to u')$ represents the control flow from the statement $u$ to the statement $u' \in \mathcal{U}(G)$.

(2) Construct the dominator tree on the CFG. A dominator tree is a directed graph $DT = (\mathcal{U}(DT), \mathcal{E}(DT))$, where $\mathcal{U}(DT) = \mathcal{U}(G)$, and $\mathcal{E}(DT) \subseteq \mathcal{E}(G)$. Each edge is connected, i.e., $e \in \mathcal{E}(DT) = (u\to u')$ if and only if $u$ dominates $u'$. $u$ dominates $u'$ if and only if every path in CFG from the start node to $u'$ goes through $u$, where $u, u'\in \mathcal{U}(DT)$.

(3) Partition the dominator tree into the blocks as split code. Each block in the dominator tree is a set of consecutive nodes containing no branches except at the very end. Thus, every edge $e \in \mathcal{E}(DT) = (u \to u')$ is removed if $u'$ has more than one in-coming edges, or $u$ has more than one out-going edges. In the end, we will have multiple disconnected subgraphs, each of which corresponds to a set of statements in the code, i.e., a piece of split code.

(4) Add the method declaration statement in the original complete code to the beginning of each split code, and extract the AST for each split code, i.e., the split AST.

\begin{figure}[htbp]
    \centering
    \includegraphics[width=0.8\linewidth]{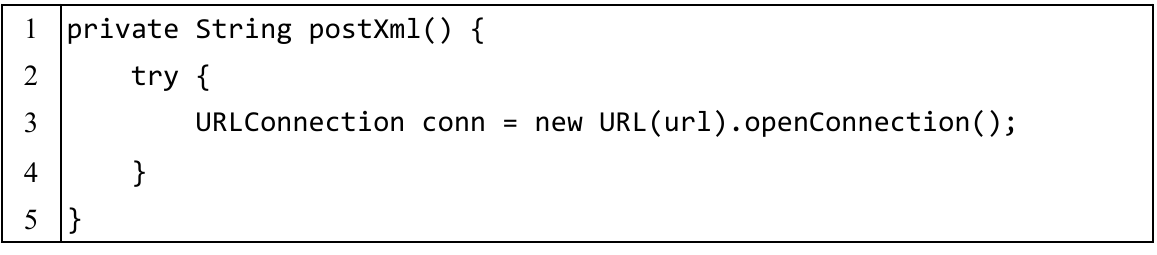}
    \caption{A Java code snippet $s_2$}
    \label{fig:AST-preprocessing-case-code}
\end{figure}

\begin{figure*}[!t]
    \centering 
    \includegraphics[width=\linewidth]{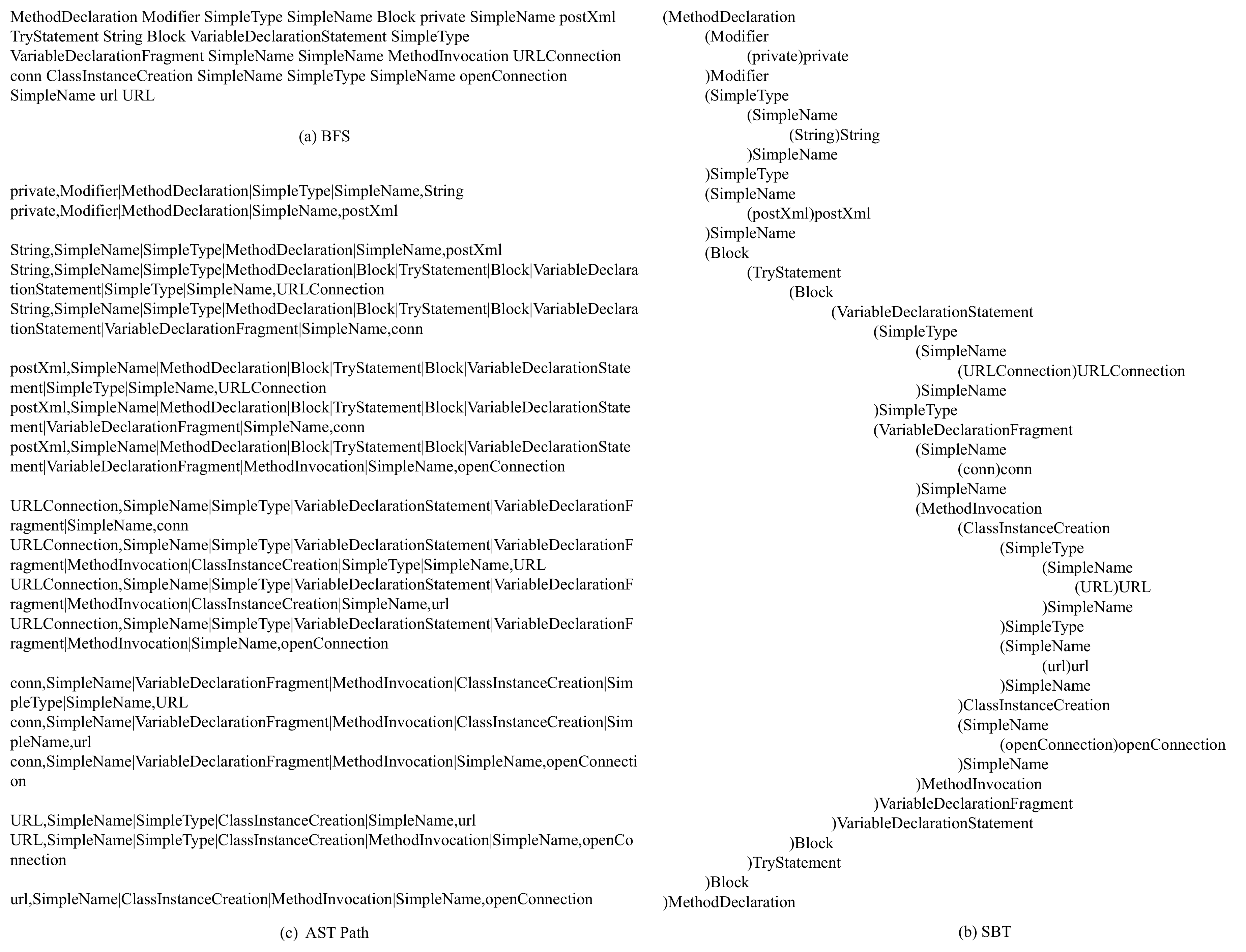}
    \caption{BFS, SBT, AST Path of the code snippet $s_2$. Using JDT as the AST parser.}
    \label{fig:AST-preprocessing-case-sequntial}
\end{figure*}

\begin{figure*}[!t]
    \centering
    \includegraphics[width=\linewidth]{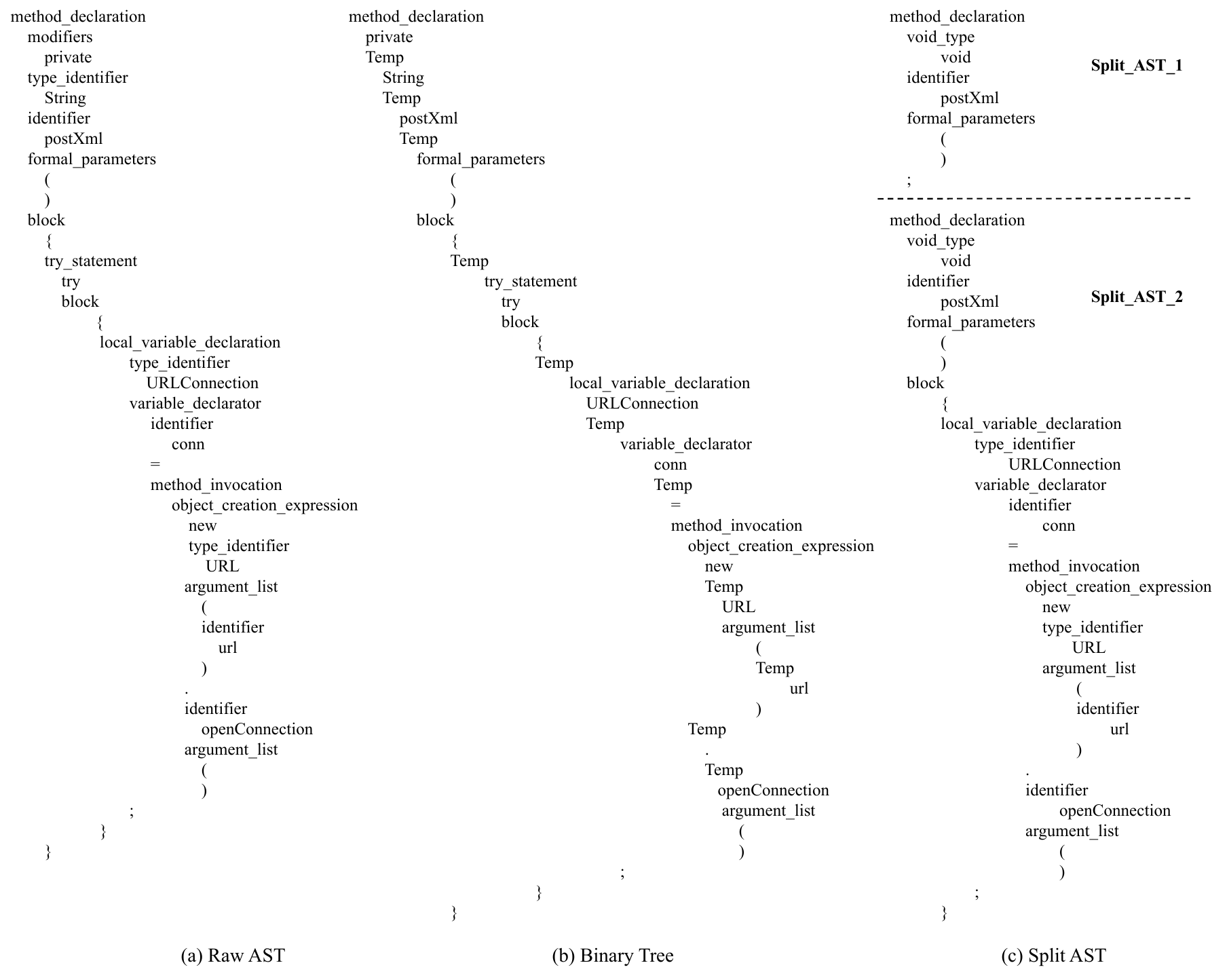}
    \caption{Raw AST, Binary Tree, and Split AST of the code snippet $s_2$. Using Tree-sitter as the AST parser.}
    \label{fig:AST-preprocessing-case-structural}
\end{figure*}

\textbf{Example of data produced by AST preprocessing methods. } Fig.~\ref{fig:AST-preprocessing-case-code} shows a piece of Java code snippet $s_2$. Fig.~\ref{fig:AST-preprocessing-case-sequntial} shows BFS, SBT, and AST Path of code snippet $s_2$ and Fig.~\ref{fig:AST-preprocessing-case-structural} shows Raw AST, Binary Tree, and Split AST of $s_2$. 
The complete information contained in an AST consists of its nodes and structures. Different AST preprocessing methods differ in retaining the node and structure information of the AST. As illustrated in Fig.~\ref{fig:AST-preprocessing-case-sequntial} and Fig.~\ref{fig:AST-preprocessing-case-structural}, SBT and Raw AST retain all nodes and complete structure information. 
BFS also preserves every node of the AST while disregarding much of the structure information. 
Binary tree, converting the raw AST into a binary tree and merging nodes with only one child node with their child nodes, removes redundant intermediate nodes, and retains complete structure information. 
AST Path and Split AST divide the complete AST into smaller components, facilitating model learning, but it comes at the cost of losing partial node or structure information of the AST.
It is worth noting that AST Path, due to constraints on the input size of the encoding models, is often limited in width, length, and quantity. Consequently, a significant portion of node and structure information tends to be discarded. Similarly, during the process of converting the source code into the Split AST~\cite{2021-CAST}, some statements in the source code (e.g., $\mathtt{catch}$ clause) are not added to the split code set, leading to the loss of node information.

\subsection{AST Encoding Method}
\label{subsec:AST_Encoding_Method}
In this section, we will introduce two types of AST encoding methods: sequence models and tree-structured models.

\subsubsection{Sequence models}
\label{subsubsec:sequence_models}
\
\newline
\noindent\textbf{(i) Bidirectional Long Short Term Memory (BiLSTM).}

The LSTM architecture~\cite{1997-LSTM} addresses the problem of learning long-term dependencies of RNN by introducing a memory cell that is able to preserve state over long periods of time~\cite{2015-Improved-Semantic-Representations}. In the work~\cite{2015-Improved-Semantic-Representations}, the LSTM unit at each time step $t$ is defined to be a collection of vectors in $\mathbb{R}^{d}$ ($\mathbb{R}$ is the set of real numbers, and $d$ is the memory dimension of the LSTM): an input gate $i_{t}$, a forget gate $f_{t}$, an output gate $o_{t}$, a memory cell $c_{t}$ and a hidden state $h_{t}$. The entries of the gating vectors $i_{t}$, $f_{t}$ and $o_{t}$ are in $[0,1]$. The LSTM transition equations are the following: 
\begin{equation}
    \small
    \begin{aligned}
    & i_{t}=\sigma\left(W^{(i)}x_{t}+U^{(i)}h_{t-1}+b^{(i)}\right),\\
    & f_{t}=\sigma\left(W^{(f)}x_{t}+U^{(f)}h_{t-1}+b^{(f)}\right),\\
    & o_{t}=\sigma\left(W^{(o)}x_{t}+U^{(o)}h_{t-1}+b^{(o)}\right),\\
    & u_{t}=tanh\left(W^{(u)}x_{t}+U^{(u)}h_{t-1}+b^{(u)}\right),\\
    & c_{t}=i_{t}\odot u_{t}+f_{t}\odot c_{t-1},\\
    & h_{t}=o_{t}\odot tanh(c_{t})
    \end{aligned}
\end{equation}
where $x_{t}$ is the input at the current time step; $W, U$ are the weighted metrics; $b$ is the bias vector; $\sigma$ denotes the logistic sigmoid function; $tanh$ denotes the hyperbolic tangent function; $\odot$ denotes element-wise multiplication.

Bidirectional LSTM (BiLSTM)~\cite{schuster1997bidirectional} uses two LSTMs at each layer. One LSTM takes the original sequence as input and the other takes the reversed sequence as input. The $h_t$ at the final timestamp can be used as the code representation denoted $\textbf{z}=h_{t_{final}}$.

\textbf{(ii) Transformer.} Transformer~\cite{2017-Transformer} follows an encoder-decoder structure, which utilizes a multi-headed self-attention mechanism and positional encoding. For the code clone detection and the code search task, the decoder part of Transformer is not used since a decoder is not needed. For the code summarization task, both encoder and decoder are used.

\textbf{Encoder~\cite{2021-API2Com}:} Encoder combines multiple identical layers where each layer consists of two sub-layers. The first sub-layer forms a multi-headed self-attention structure while the other one is a fully connected layer. Both sub-layers are followed by another layer which normalizes the output of each sub-layer. The encoder maps an input sequence of symbol representations $\textbf{x}=(x_1, \dots, x_n)$ to a sequence of continuous representations $\textbf{z}=(z_1, \dots, z_n)$. 

\textbf{Decoder~\cite{2021-API2Com}:} Decoder has a similar structure as encoder except that it includes one additional sub-layer. This extra sub-layer conducts multi-head attention on the encoder’s output. Moreover, the self-attention sub-layer is reformed to avoid attending to subsequent positions. Given $\textbf{z}$, the decoder generates an output sequence $\textbf{y}=(y_1, ..., y_m)$ of symbols one element at a time. 

\textbf{Multi-head attention mechanism~\cite{2021-API2Com}:} Transformer’s attention aims to map a query (Q) and a collection of key (K) - value (V) pairs to vectors which are calculated as a weighted sum of the values. The attention is computed by multiplying queries by keys, divided by $\sqrt{d_{k}}$, where $d_{k}$ is the dimension of the key vector. Then, a softmax function is applied to attain the weights for values.
\begin{equation}
    \small
    Attention(Q,K,V)=softmax(QK^T/\sqrt{d_{k}})V
\end{equation} 
In order to improve the performance, a new mechanism called multi-head attention is added to the self-attention technique. The intuition is to conduct self-attention process multiple times separately, with different weight matrices. However, the feed-forwarding layer only accepts a single input. Therefore, all the results are concatenated and multiplied by an additional weight matrix $W$, as shown in the following formula. Here, $d_{model}$ is the input dimension, $d_k$ is the dimension of the key vector, $d_v$ is the dimension of the value vector, $h=8$ is the number of parallel attention layers, or heads.
\begin{equation}
\small
    \begin{split}
    MultiHead(Q,K,V)=Concat(head_{1},...,head_{h})W^O\\
    {\rm where}\quad head_{i}=Attention(QW_{i}^Q,KW_{i}^K,VW_{i}^V)
    \end{split}
\end{equation}
Where $W_{i}^Q\in\mathbb{R}^{d_{model}\times d_{k}}$, $W_{i}^K\in\mathbb{R}^{d_{model}\times d_{k}}$, $W_{i}^v\in\mathbb{R}^{d_{model}\times d_{v}}$ and $W^O\in\mathbb{R}^{hd_{v}\times d_{model}}$. 

\textbf{Positional Encoding~\cite{2021-API2Com}:} Given that Transformer does not include recurrence, some information regarding relative and absolute tokens' positions needs to be taken into account. To achieve this goal, a positional encoding layer is added at the lowest
part of the encoder and decoder stacks. The layers should have equal dimensions as the output results of embedding should be added to positional encodings. There are many choices of positional encodings. In this work, we use sine and cosine functions of different frequencies:
\begin{equation}
    \small
    \begin{aligned}
      &  PE_{(pos,2i)}=sin(pos/10000^{2i/d_{model}}),\\
      &  PE_{(pos,2i+1)}=cos(pos/10000^{2i/d_{model}})
    \end{aligned}   
\end{equation}
where $pos$ is the position and $i$ is the dimension.

Transformer has demonstrated strong feature learning ability on NLP tasks~\cite{2021-TreeBERT} and also proved to be effective in software engineering studies~\cite{2021-API2Com, 2020-Transformer-based-Approach-for-Code-Summarization, 2022-UniXcoder}.

\subsubsection{Tree-structured models}
\label{subsubsec:tree-structured_models}
\
\newline

\textbf{(i) Tree-Structured Long Short-Term Memory Networks (TreeLSTM)}
TreeLSTM is first proposed by Tai~\cite{2015-Improved-Semantic-Representations} to capture the syntactic properties of natural language. TreeLSTM is a generalization of LSTMs to model tree-structured topologies. Given a tree, let $C(j)$ denote the set of children of node $j$. The TreeLSTM unit at each node $j$ is defined to be a collection of vectors in $\mathbb{R}^{d}$, where $\mathbb{R}$ is the set of real numbers, $d$ is the memory dimension of the TreeLSTM: an input gate $i_{j}$, forget gates $f_{jk}$ where $k\in C(j)$, an output gate $o_{j}$, a memory cell $c_{j}$ and a hidden state $h_{j}$. The entries of the gating vectors $i_{j}$, $f_{j}$ and $o_{j}$ are in $[0,1]$.

\textbf{Child-Sum TreeLSTMs~\cite{2015-Improved-Semantic-Representations}:} The Child-Sum TreeLSTM can be used on tree structures where the branching factor is arbitrary and children's orders are not considered. The Child-Sum TreeLSTM transition equations are the following:
\begin{equation}
    \small
    \begin{aligned}
    & \tilde{h}_{j}=\sum_{k\in C(j)}{h_k},\\
    & i_{j}=\sigma\left(W^{(i)}x_{j}+U^{(i)}\tilde{h}_{j}+b^{(i)}\right),\\
    & f_{jk}=\sigma\left(W^{(f)}x_{j}+U^{(f)}h_{k}+b^{(f)}\right),\\
    & o_{j}=\sigma\left(W^{(o)}x_{j}+U^{(o)}\tilde{h}_{j}+b^{(o)}\right),\\
    & u_{j}=tanh\left(W^{(u)}x_{j}+U^{(u)}\tilde{h}_{j}+b^{(u)}\right),\\
    & c_{j}=i_{j}\odot u_{j}+\sum_{k\in C(j)}{f_{jk}\odot c_{k}},\\
    & h_{j}=o_{j}\odot tanh(c_{j})
    \end{aligned}
\end{equation}
where $x_{j}$ is the input at node $j$, $W, U$ are the weighted metrics, $b$ is the bias vector, $\sigma$ denotes the logistic sigmoid function, $tanh$ denotes the hyperbolic tangent function, $\odot$ denotes element-wise multiplication. The hidden state $h_j$ at the root node is considered the representation of the source code, that is code representation $\textbf{z}=h_{root}$.

\textbf{N-ary TreeLSTMs~\cite{2015-Improved-Semantic-Representations}:} The $N$-ary TreeLSTM can be used on tree structures where the branching factor is at most $N$ and where children are ordered, i.e., they can be indexed from 1 to $N$. For any node $j$, write the hidden state and memory cell of its $k$th child as $h_{jk}$ and $c_{jk}$ respectively. The $N$-ary TreeLSTM transition equations are the following:
\begin{equation}
\small
    \begin{aligned}
    & i_{j}=\sigma\left(W^{(i)}x_{j}+\sum_{\ell =1}^{N}{U_{\ell}^{(i)}h_{j\ell}}+b^{(i)}\right),\\
    & f_{jk}=\sigma\left(W^{(f)}x_{j}+\sum_{\ell =1}^{N}{U_{k\ell}^{(f)}h_{j\ell}}+b^{(f)}\right),\\
    & o_{j}=\sigma\left(W^{(o)}x_{j}+\sum_{\ell =1}^{N}{U_{\ell}^{(o)}h_{j\ell}}+b^{(o)}\right),\\
    & u_{j}=tanh\left(W^{(u)}x_{j}+\sum_{\ell =1}^{N}{U_{\ell}^{(u)}h_{j\ell}}+b^{(u)}\right),\\
    & c_{j}=i_{j}\odot u_{j}+\sum_{\ell =1}^{N}{f_{j\ell}\odot c_{j\ell}},\\
    & h_{j}=o_{j}\odot tanh(c_{j})
    \end{aligned}
\end{equation}
where $k=1, 2, \dots, N$, $x_{j}$ is the input at node $j$, $W, U$ are the weighted metrics, $b$ is the bias vector, $\sigma$ denotes the logistic sigmoid function, $tanh$ denotes the hyperbolic tangent function, $\odot$ denotes element-wise multiplication. Similar to Child-Sum TreeLSTM, code representation $\textbf{z}=h_{root}$.

TreeLSTM has been widely adopted to capture the syntactic features in ASTs. CDLH~\cite{2017-Supervised-Deep-Features-for-Clone-Detection} uses TreeLSTM to learn representations of code fragments for clone detection where code fragments are parsed to ASTs. Wang et al.~\cite{2020-Reinforcement-Learning-Guided-Code-Summarization} apply TreeLSTM to extract the important information from ASTs for the task of code summarization.

\textbf{(ii) AST-Trans.}
AST-Trans~\cite{2022-AST-trans} is a simple variant of the Transformer model to efficiently handle the tree-structured AST. AST-Trans exploits ancestor-descendant and sibling relationship matrices to represent the tree structure, and uses these matrices to dynamically exclude irrelevant nodes. 

AST-Trans has the same encoder and decoder structure as the Transformer, while replacing the singe-head self-attention with tree-structured attention. The absolute position embedding from the original Transformer is replaced with relative position embeddings defined by the two relationship matrices to better model the dependency.

For an AST, it will be firstly linearized into a sequence, which means being transformed into SBT~\cite{2018-Deep-Code-Comment-Generation} in our experiment. Then the ancestor-descendent and sibling relationships among its nodes will be denoted through two specific matrices. Based on the matrices, tree-structured attention is adopted to better model these two relationships. In the following part we will introduce the construction of relationship matrices and tree-structured attention. 

\textbf{Construction of relationship matrices.} We define two kinds of relationships between AST nodes: ancestor-descendant ($A$) and sibling ($S$) relationships. And we use two position matrices $A_{N\times N}$ and $S_{N\times N}$ to represent the ancestor-descendent and sibling relationships respectively, where $N$ is the total number of nodes in AST. The $i$-th node in the linearized AST is denoted as $n_i$. $A_{ij}$ is the distance of the shortest path between $n_i$ and $n_j$ in the AST. $S_{ij}$ is horizontal sibling distance between $n_i$ and $n_j$ if they satisfy the sibling relationship. If one relationship is not satisfied, its value in the matrix will be infinity. Note that we consider the relative relationship between two nodes, which means $A_{ij} = -A_{ji}$ and $S_{ij} = -S_{ji}$ if a relationship exists between $n_i$ and $n_j$.

Formally, we use $SPD(i,j)$ and $SID(i,j)$ to denote the Shorted Path Distance and horizontal Sibling Distance between $n_i$ and $n_j$. The values in the relationship matrices are defined as:
\begin{equation}
\label{eq:Aij}
\small
    A_{ij}=
        \begin{cases}
            SPD(i,j) & \text{if $|SPD(i,j)|\leq P$ }\\
            \infty& \text{otherwise }
        \end{cases}
\end{equation}

\begin{equation}
\label{eq:Sij}
\small
    S_{ij}=
        \begin{cases}
            SID(i,j) & \text{if $|SID(i,j)|\leq P$ }\\
            \infty& \text{otherwise }
        \end{cases}
\end{equation}

$P$ is a pre-defined threshold and nodes with relative distance beyond $P$ will be ignored. We set $P=7$ according to
the code provided by the original paper~\cite{2022-AST-trans}.

\textbf{Tree-structured Attention.} Tree-structured attention is built on standard self-attention with relative position embeddings and disentangled attention. Tree-structured attention transforms an input sequence $\textbf{x}=(x_1, ...,x_n)$ ($x_i\in\mathbb{R}^d$ which stands for the embedding of $n_i$) into a sequence of output vectors $\textbf{o}=(o_1, ...,o_n)$ ($o_i\in\mathbb{R}^d$).

The relative distance defined under the linear relationship is replaced with $\delta_R(i,j)$ where $R$ stands for either the ancestor-descendent relationship $A$ or the sibling relationship $B$ in the tree structure. $\delta_R(i,j)$ reflects the pairwise distance between $n_i$ and $n_j$ in relationship $R$. Denote $P$ as the max relative distance, $\delta_R(i,j)$ is defined as:

\begin{equation}
\small
    \delta_R(i,j) =
        \begin{cases}
        R_{ij}+P+1 & \text{if $R_{ij}\in[-P,P]$ }\\
        0 & \text{if $R_{ij}=\infty$}
        \end{cases}
\end{equation}

$R_{ij}$ refers to either $A_{ij}$ defined in Eq \ref{eq:Aij} or $S_{ij}$ defined in Eq \ref{eq:Sij}.

As there are two kinds of relationships, each head only considers one relationship so that it will not add any additional parameter on top of the standard Transformer. $h_A$ heads will use $\delta_A(i,j)$ and the rest $h_S$ heads will use $\delta_S(i,j)$. Information from the two relationships will be merged together through multi-head attention. The output vector $\textbf{o}=(o_1, ...,o_n)$  is computed as below:

\begin{gather}
    \small
    \alpha_{i,j}=Q(x_i)K(x_j)^T+Q(x_i){K^P_{\delta_R(i,j)}}^T+Q^P_{\delta_R(j,i)}K(x_j)^T \\
    o_i=\sum^{j\in\left\{j|\delta_R(i,j)>0\right\}}_j \sigma(\frac{ \alpha_{i,j}}{\sqrt{3d}})(V(x_j)+V^P_{R_{ij}})
\end{gather}
where $Q, K:\mathbb{R}^d \rightarrow \mathbb{R}^m$ are query and key functions respectively, $V:\mathbb{R}^d\rightarrow\mathbb{R}^d$ is a value function, $\sigma$ is a scoring function (e.g. softmax or hardmax), $Q^P, K^P \in \mathbb{R}^{(2P+1)\times m}$ represent the query and key projection matrices of relative positions,$V^P$ represents the value project matrix of relative distances, $K^P_{\delta_R(i,j)}$is the $\delta_R(i,j)$-th row of $K^P$ and $Q^P_{\delta_R(i,j)}$is the $\delta_R(i,j)$-th row of $Q^P$, $V^P_{R_{ij}}$is the $R_{ij}$-th row of $V^P$. Note that only the attention weights for node pairs where $\delta_{R(i,j)}>0$ are computed.

\subsection{Experimental Setup}
\label{subsec:experimental_setup}

\subsubsection{Code-related Tasks}
\label{subsubsec:code_related_tasks} 
\

\noindent\textbf{(1) Code Clone Detection.} 
This task is to measure the similarity between two code fragments to predict whether they are code clone pairs. Generally, there are four clone types. \textbf{Type-1:} Identical fragments except for variations in comments and layout~\cite{roy2007survey}. \textbf{Type-2:} Identical fragments except for variations in identifier names and literal values in addition to Type-1 differences~\cite{roy2007survey}. \textbf{Type-3:} Syntactically similar fragments that differ at the statement level. The fragments have statements added, modified, or removed with respect to each other, in addition to Type-2 differences~\cite{roy2007survey}. \textbf{Type-4:} Syntactically dissimilar fragments that implement the same functionality~\cite{roy2007survey}. Type-1, Type-2, and Type-3 clones indicate textual similarity whereas Type-4 clones indicate functional similarity.

Code Clone Detection is a binary classification problem to predict whether a given pair of codes has the same semantics, with the $F_1$-score used as the evaluation metric. Given $n$ code fragments $\{C_{1},..., C_{n}\}$ where $C_{i}$ is the $i$-th raw code fragment, and pairwise labels to indicate whether two code fragments belong to a clone pair or not: $y_{ij}$=1 if $(C_{i}, C_{j})$ is a clone pair, $y_{ij}$=0 if $(C_{i}, C_{j})$ is not a clone pair, then the training set is represented by a set of triplets $\mathcal{D}=\{(C_{i}, C_{j},y_{ij})|i,j\in[n], i<j\}$, where $[n]={1,2,...,n}$~\cite{2017-Supervised-Deep-Features-for-Clone-Detection}. Given a pair of code fragments, we first convert them into ASTs using AST parsers, and preprocess the ASTs if necessary. Then we use different encoding methods to encode the input as vectors $z_{i}$ and $z_{j}$. Given $z_{i}$ and $z_{j}$, their distance is measured by $z_{ij}=|z_{i}-z_{j}|$ for semantic relatedness~\cite{2015-Improved-Semantic-Representations}. We use a linear layer and a softmax function to map $z_{ij}$ to $r_{ij}\in R^{2}$, where $r_{ij}[0]$ and $r_{ij}[1]$ represents the probability of $y_{ij}=0$ or $y_{ij}=1$ separately. We use cross-entropy as the loss function, which is defined as
\begin{equation}
J(\Theta,\hat{y},y)=\sum{(-(y\cdot log(\hat{y})+(1-y)\cdot log(1-\hat{y})))}
\end{equation}
Our goal is to minimize the loss and we use Adam as the optimizer. The training will stop if the model's $F_1$-score on the evaluation dataset doesn't exceed previous best $F_1$-score for more than 3 continuous epochs. After training, the model with the best $F_1$-score is stored. For new code pairs $(C_{i}, C_{j})$, they will be fed into the reloaded model and get $r_{ij}$ and $\hat{y}=r_{ij}[1]$. We get the prediction of whether $(C_{i}, C_{j})$ is a clone pair by
\begin{equation}
\small
    Prediction=
    \begin{cases}
    True& \text{$\hat{y}>\delta$ }\\
    False& \text{$\hat{y}\le\delta$ }
    \end{cases}
\end{equation}
where $\delta$ is the threshold. We enumerate $\delta$ from 0.01 to 0.99 with step 0.01 and evaluate the model on the evaluation dataset. The $\delta$ with the best model $F_1$-score is chosen as the final $\delta$.

\noindent\textbf{(2) Code Search.}
The task is to find specific code snippets according to a natural language query. Previous studies mainly relied on the textual similarity between source code and natural language query. Since the rise of DL technology, today it is more often to jointly embed code snippets and natural language descriptions into a high-dimensional vector space, in such a way that a code snippet and its corresponding description have similar vectors. Using the unified vector representation, code snippets related to a natural language query can be retrieved according to their vectors. Semantically related words can also be recognized and irrelevant/noisy keywords in queries can be handled. According to related works, the loss function tends to be a triplet loss function, which is defined as:
\begin{equation}
    \small
    L(a,p,n)=max(d(a,p) - d(a,n)+\alpha,0)
\end{equation}
where $d(x, y)$ measures the distance between $x$ and $y$, and $\alpha$ is the margin that ensures the positive code snippet is closer to the anchor than the negative code snippet. We can consider different distance metrics for $d$, such as the Euclidean and Cosine distances. To construct a triplet, first a description is chosen randomly from the dataset as an anchor and then its corresponding code snippet is selected as positive and a random code snippet is selected as negative for the anchor accordingly.

\noindent\textbf{(3) Code Summarization}
Code summarization is the task of generating natural language descriptions for the given code snippets. It can help developers quickly understand and maintain source code~\cite{2023-EACS}. 
With the availability of large-scale data, deep learning methods are widely used to improve the performance of code summarization tools. The pioneers find that some Seq2Seq models, such as RNN~\cite{2014-RNN-Encoder-Decoder, 2015-Attention-Neural-Machine-Translation} and LSTM~\cite{2016-CODE-NN}, are capable of modeling the semantic relations between code snippets and summary. In order to capture the structural information, some researchers use TreeLSTM to process structural representations of code~\cite{2015-Improved-Semantic-Representations, 2019-Code-Summarization-with-Extended-Tree-LSTM}. 
However, RNN-based methods have poor long-term dependence. To break through the bottleneck that may be encountered when modeling long sequences, some Transformer-based methods~\cite{2021-SiT, 2020-Transformer-based-Approach-for-Code-Summarization} are proposed, which show superior performance for the task. 

Generally, the code summary model consists of two parts: an encoder and a decoder. The encoder uses the required input data to generate the context vectors $\boldsymbol{e^{En}}$. Then, the decoder takes in $\boldsymbol{e^{En}}$ and unfolds it into the target sequence. The words in the target sequence are predicted one by one, which means the word with the highest probability is selected from the vocabulary each time. 

Specifically, when both encoder and decoder are based on LSTM, the initial hidden state and cell state of the decoder are initialized by $\boldsymbol{e^{En}}$. The dynamic model is as follows:
\begin{equation}
    \begin{aligned}
    & \boldsymbol{h_{t}, c_{t}}=f(y_{t-1}, \boldsymbol{h_{t-1}, c_{t-1}}) \\
    & p(y_{t}|Y_{<t}, X)=g(y_{t-1}, \boldsymbol{h_{t-1}}) &
    \end{aligned}
\end{equation}
where $f(\cdot)$ and $g(\cdot)$ are activation functions. $\boldsymbol{h_{t}}$ and $\boldsymbol{c_{t}}$ are the hidden state and the cell state at time $t$. $y_{t}$
is the predicted target word at $t$. $Y_{<t}$ denotes the history ${y_{1}, y_{2}, \dots ,y_{t-1}}$.

When both encoder and decoder are based on Transformer, the memory of the decoder is initialized by $\boldsymbol{e^{En}}$. The dynamic model is as follows:
\begin{equation}
    \begin{aligned}
    & p(y_{t}|Y_{<t}, X)=k(Y_{<t}, \boldsymbol{e^{En}}) &
    \end{aligned}
\end{equation}
where $k(\cdot)$ is activation function.

In our experiment, when the encoder is BiLSTM, Transformer, or AST-Trans, the decoder is the same neural network architecture, which means BiLSTM, Transformer, or AST-Trans. When the encoder is TreeLSTM, the decoder is BiLSTM.

\subsubsection{Datasets}
\label{subsubsec:datasets} \ 

\noindent\textbf{(1) Dataset for Code Clone Detection.}

\textbf{BigCloneBench.} BigCloneBench is a well-known dataset of method-level Java code clones provided by Svajlenko et al.~\cite{2014-Benchmark-Inter-project-Code-Clones}, which is a widely used large code clone benchmark~\cite{2021-CodeXGLUE}. It consists of known true and false positive clones from a big data inter-project Java repository. Lu et al.~\cite{2021-CodeXGLUE} filter the BigCloneBench dataset by discarding code snippets without any tagged true or false clone pairs. In this paper, we directly use the filtered BigCloneBench dataset provided by Lu et al. in CodeXGLUE~\footnote{\url{https://github.com/microsoft/CodeXGLUE}}~\cite{2021-CodeXGLUE} and remove the data if there exists as least one AST parser that cannot parse the code. 
Row 2 of Table~\ref{tab:datasets_statistics} presents the statistics of the BigCloneBench dataset, including programming language, training set size, validation set size, and test set size.

\begin{table}[htbp]
    \centering
    \caption{Dataset statistics}
    \label{tab:datasets_statistics}
    
    \begin{tabular}{|c|c|ccc|}
        \hline
        Dataset & Language & Training & Validation & Test \\
        \hline
        BigCloneBench & Java & 900,713 & 415,416 & 415,416 \\
        \hline
        CodeSearchNet & Java & 164,923 & 5,183 & 10,955 \\
        \hline
    \end{tabular}
\end{table}

\noindent\textbf{(2) Dataset for Code Search and Code Summarization.}

\textbf{CodeSearchNet.} The CodeSearchNet dataset~\cite{2019-CodeSearchNet-Challenge} is a vast collection of methods accompanied by their respective comments, written in Go, Java, JavaScript, PHP, Python, and Ruby. These methods are sourced from open-source projects hosted on GitHub. This dataset is widely used in studying code search~\cite{2023-Graphsearchnet, 2023-Fold2Vec, 2021-Multimodal-Representation-NCS, 2022-UniXcoder, 2021-TabCS} and code summarization~\cite{2021-API2Com, 2022-UniXcoder, 2021-BASTS, 2022-Automatic-Source-Code-Summarization-With-GNN}. Analogously, Lu et al.~\cite{2021-CodeXGLUE} show that some comments contain content unrelated to the code snippets and perform data cleaning on the CodeSearchNet dataset. Therefore, in this paper, we directly use the clean version of the CodeSearchNet dataset provided by them in CodeXGLUE. 
In our specific case, we consider Java as the only programming language in our experiments, so we concentrated on utilizing the Java-related data from the CodeSearchNet dataset while disregarding data in other languages. Row 2 of Table~\ref{tab:datasets_statistics} presents the statistics of the Java corpus of the CodeSearchNet dataset.

\subsubsection{Evaluation Metrics}
\label{subsubsec:evaluation_metrics}\ 

\noindent\textbf{(1) Evaluation Metrics for Code Clone Detection}

\textbf{Precision ($P$)} Precision measures among all of the clone pairs detected by a clone detection approach, how many of them are true clone pairs:
\begin{equation}
    \small
    P = \frac{\text{\# of true clone pairs}}{\text{Total \# of detected clone pairs}}
    \label{equ:precision}
\end{equation}

\noindent\textbf{Recall ($R$)} Recall measures among all known true clone pairs, how many of them are detected by a clone detection approach:
\begin{equation}
    \small
    R = \frac{\text{\# of true clone pairs detected}}{\text{Total \# of known true clone pairs}}
    \label{equ:recall}
\end{equation}

\noindent\textbf{$F_1$-score ($F_1$)} $F_1$-score is a harmonic mean of precision and recall.
\begin{equation}
    \small
    F_1 = 2\times\frac{P\times R}{P+R}
    \label{equ:f1}
\end{equation}

Following previous work~\cite{2017-Supervised-Deep-Features-for-Clone-Detection, 2019-ASTNN, 2022-UniXcoder}, we use Recall, Precision, and $F_1$-score as the evaluation metric in the code clone detection task.

\noindent\textbf{(2) Evaluation Metrics for Code Search}

\textbf{SuccessRate@k ($SR@k$)}~\cite{2019-Multi-modal-Attention-for-Code-Retrieval} measures the percentage of queries for which more than one correct result could exist in the top $k$ ranked results, which
is calculated as follows:
\begin{equation}
\small
    SR@k = \dfrac{1}{\left| Q \right|}\sum_{i=1}^{\left|Q\right|}\delta(FRank_{Q_i}\leqslant k)
    \label{equ:SR@k}
\end{equation}
where $Q$ is a set of queries, $\delta(\cdot)$ is a function that returns 1 if the input is true and returns 0 otherwise and $FRank_{Q_i}$ refers to the rank position of the correct result for the $i$-th query in $Q$. 
SuccessRate@k is important because a better code search engine should allow developers to discover the needed code by inspecting fewer returned results. The higher the SuccessRate value is, the better the code search performance is.

\textbf{Mean Reciprocal Rank($MRR$)}~\cite{2019-Multi-modal-Attention-for-Code-Retrieval} is the average of the reciprocal ranks of results of a set of queries $Q$. The reciprocal rank of a query is the inverse of the rank of the first hit result. The MRR is defined as:
\begin{equation}
\small
    MRR = \dfrac{1}{\left| Q \right|}\sum_{i=1}^{\left|Q\right|}\dfrac{1}{FRank_{Q_i}}
    \label{equ:MRR}
\end{equation}
where $\lvert Q \rvert$ is the size of the query set. The higher the MRR value is, the better the code search performance is.

\noindent\textbf{(3) Evaluation Metrics for Code Summarization}

We use three automatic metrics BLEU~\cite{2002-BLEU}, METEOR ~\cite{2005-METEOR}, and ROUGE ~\cite{2004-ROUGE},  to evaluate the quality of the generated comments in the code summarization task. These three metrics are widely used in code summarization~\cite{2016-CODE-NN, 2017-Transformer, 2018-TL-CodeSum, 2018-Improving-Code-Summarization-via-DRL, 2021-SiT, 2022-SCRIPT}.

\textbf{BLEU-4}, the abbreviation for BiLingual Evaluation Understudy ~\cite{2002-BLEU}, is widely used for evaluating the quality of generated summaries~\cite{2016-CODE-NN, 2018-TL-CodeSum, 2018-Improving-Code-Summarization-via-DRL}. It compares n-grams in the predicted and target summaries. Typical implementations of BLEU scores set the range of $n$ from 1 to 4. An average BLEU score is then computed by combining these individual n-gram scores using predetermined weights. It
is computed as:

\begin{equation}
    \small
    BLEU = BP * exp(\sum_{n=1}^N w_{n}\log p_{n})
\end{equation}

\begin{equation}
    \small
    BP= \left \{
        \begin{array}{ll}
        1,                    & if |g|>|r|\\
        e^{1-\frac{|r|}{|g|}},     & if |g|\le|r|\\
        \end{array}
        \right.
\end{equation}
where $N = 1, 2, 3, 4$ and $w_n = \frac{1}{N}$. $p_n$ is the n-gram precision~\cite{2022-Evaluation-Neural-Code-Summarization}. $BP$ represents the brevity penalty. $g$ and $r$ denote a generated (predicted) summary and a reference summary, respectively. $|g|$ and $|r|$ denote the lengths of $g$ and $r$, respectively. In this paper, we follow~\cite{2021-SiT, 2022-SCRIPT} and show the standard BLEU score which provides a cumulative score of 1-, 2-, 3-, and 4-grams~\cite{2021-Why-My-Code-Summarization-Not-Work}. 

\textbf{METEOR}~\cite{2005-METEOR} is introduced to address the concerns of using BLEU ~\cite{2002-BLEU}. It is also widely used to evaluate the quality of generated summaries~\cite{2020-Rencos, 2020-Reinforcement-Learning-Guided-Code-Summarization, 2021-Code-Summarization-for-Smart-Contracts}. It combines n-gram precision and n-gram recall by taking their harmonic mean to compute a measure of similarity. Suppose $m$ is the number of mapped unigrams between the reference summary $r$ and the generated summary $g$, respectively. Then, precision ($P_{unigram}$), recall ($R_{unigram}$), and $METEOR$ are computed as follows:

\begin{equation}
     \small
    P_{unigram} = \frac{m}{|g|},R_{unigram} = \frac{m}{|r|}
\end{equation}

\begin{equation}
    \small
    METEOR = (1-\gamma*frag^\beta)*
    \frac{P_{unigram}*R_{unigram}}{\alpha*P_{unigram}+(1-\alpha)*R_{unigram}}
\end{equation}
where $frag$ is the fragmentation fraction; $\alpha$, $\beta$, and $\gamma$ are three penalty parameters. In this paper, we set $\alpha$, $\beta$, $\gamma$ to 0.9, 3.0, and 0.5 respectively according to~\cite{2020-Rencos}. 

\textbf{ROUGE-L}~\cite{2004-ROUGE} evaluates how much reference text appears in the generated text. It is also widely used to evaluate the quality of generated code summaries~\cite{2021-Project-Level-Encoding-Code-Summarization, 2021-BASTS, 2021-API2Com}. Based on the longest common subsequence (LCS), it uses the F-score, which is the harmonic mean of precision and recall. Specifically, the LCS-based F-measure ($F_{lcs}$) is called ROUGE-L, and $F_{lcs}$ is computed as follows:

\begin{equation}
     \small
    P_{lcs} = \frac{LCS(r,g)}{|g|},
    R_{lcs} = \frac{LCS(r,g)}{|r|}
\end{equation}

\begin{equation}
     \small
    F_{lcs} = \frac{(1+\beta^{2})P_{lcs}R_{lcs}}
    {R_{lcs}+\beta^{2}P_{lcs}}
\end{equation}
where $r$ and $g$ also denote the reference summary and the generated summary, respectively; $\beta$ is set to 1.2 as in~\cite{2018-Improving-Code-Summarization-via-DRL, 2020-Rencos, 2021-CoCoSum}.

\section{Results and Findings}
\label{sec:results_and_findings}

\subsection{Answer to RQ1: What are the differences among the ASTs parsed by different AST parsing methods?}
\label{subsec:Answer_to_RQ1}

As mentioned in Section~\ref{sec:introduction}, different AST parsing methods use different lexical and grammatical rules, resulting in ASTs with different structures and node labels. 
Therefore, before applying AST to subsequent code-related tasks, it is necessary to figure out whether and where the Raw ASTs generated by different AST parsing methods are different. 
This will help reveal which differences have an impact on AST-based code representation as well as subsequent code-related tasks. 
We follow Utkin, et al.~\cite{2022-Impact-of-Code-Parsers-on-Models} and use five metrics to characterize the ASTs generated by AST parsing methods. 

\textbf{Tree size}. It is defined as the total number of nodes in an AST.

\textbf{Tree depth}. It is defined as the number of nodes in the path from the root of an AST to its deepest node.

\textbf{Branching factor}: It is defined as the average number of children of non-leaf nodes in an AST.

\textbf{Unique types}: It is defined as the number of unique non-leaf nodes in an AST. Lower values of unique types represent a higher level of abstraction used in an AST parsing method as it can represent the same code snippet in a more compact way~\cite{2022-Impact-of-Code-Parsers-on-Models}.

\textbf{Unique tokens}: It is defined as the number of unique sub-tokens in AST leaf nodes, which often represent code tokens. Lower values of unique tokens also reflect a higher level of abstraction (e.g., whether an AST parsing method keeps binary operators as code tokens or as node types~\cite{2022-Impact-of-Code-Parsers-on-Models}.

Tree size, tree depth, and branching factor reflect the difference in the structure of ASTs and can be used to estimate how different the trees are size-wise. Unique types and unique tokens reflect the difference in node labels of ASTs. Lower values in unique types and unique tokens mean a higher level of abstraction.

Specifically, we compare the ASTs generated by the four commonly used AST parsing methods, i.e., JDT, srcML, ANTLR, and Tree-sitter, which are discussed in detail in Section~\ref{subsec:AST_Parsing_method}. We utilize them to generate ASTs for Java code snippets in the BigCloneBench and CodeSearchNet datasets, then compute the five metrics for each AST parsing method. 

\begin{table*}[!t]
    \centering  
    \footnotesize
    \tabcolsep=1.4pt
    \caption{Comparison of ASTs generated by different AST parsing methods in the five metrics. Column Parsing lists AST parsing methods for comparison.}
    \label{tab:comparison_of_ASTs_on_five_metrics}
    \begin{tabular}{ccccccccccccc}
        \toprule
        
        \multirow{2}{*}{Dataset} & \multirow{2}{*}{Parsering} & \multicolumn{2}{c}{Tree Size} & \multicolumn{2}{c}{Tree Depth} & \multicolumn{2}{c}{Branch Factor} & \multicolumn{2}{c}{Unique Types} & \multicolumn{2}{c}{Unique Tokens} & \multirow{2}{*}{Language Support} \\
        
        \cmidrule(lr){3-4} \cmidrule(lr){5-6} \cmidrule(lr){7-8} \cmidrule(lr){9-10} \cmidrule(lr){11-12}

        & & Mean & Median & Mean & Median & Mean & Median & Mean & Median & Mean & Median & \\

        \midrule
        
        \multirow{4}{*}{BigCloneBench}& JDT & 312 & 213 & 13 & 12 & \textbf{2} & \textbf{2} & 22 & 22 & 58 & 48 & Java \\
        
        & srcML & 529 & 356 & 22 & 21 & 1 & 1 & 26 & 26 & 67 & 55 & Java/C/C++/C\# \\
        
        & ANTLR & \textbf{709} & \textbf{480} & \textbf{27} & \textbf{26} & \textbf{2} & \textbf{2} & \textbf{37} & \textbf{37} & \textbf{76} & \textbf{65} & Any \\
        
        & Tree-sitter & 485 & 329 & 14 & 13 & \textbf{2} & \textbf{2} & 28 & 27 & \textbf{76} & \textbf{65} & Any \\
        
        \midrule
        
        \multirow{4}{*}{CodeSearchNet}& JDT & 117 & 83 & 10 & 10 & \textbf{2} & \textbf{2} & 16 & 15 & 29 & 24 & Java \\
        
        & srcML & 193 & 135 & 18 & 17 & 1 & 1 & 23 & 23 & 37 & 29 & Java/C/C++/C\# \\
        
        & ANTLR & \textbf{262} & \textbf{184} & \textbf{22} & \textbf{21} & \textbf{2} & \textbf{2} & \textbf{31} & \textbf{31} & \textbf{42} & \textbf{36} & Any \\
        
        & Tree-sitter & 181 & 126 & 11 & 11 & \textbf{2} & \textbf{2} & 20 & 19 & \textbf{42} & \textbf{36} & Any \\

        \midrule
        
        \multirow{4}{*}{Average}& JDT & 214.5 & 148 & 11.5 & 11 & \textbf{2} & \textbf{2} & 19 & 18.5 & 43.5 & 36 & -- \\
        
        & srcML & 361 & 245.5 & 20 & 19 & 1 & 1 & 24.5 & 24.5 & 52 & 42 & -- \\
        
        & ANTLR & \textbf{485.5} & \textbf{332} & \textbf{24.5} & \textbf{23.5} & \textbf{2} & \textbf{2} & \textbf{34} & \textbf{34} & \textbf{59} & \textbf{50.5} & -- \\
        
        & Tree-sitter & 333 & 227.5 & 12.5 & 12 & \textbf{2} & \textbf{2} & 24 & 23 & \textbf{59} & \textbf{50.5} & -- \\
        
        \bottomrule
    \end{tabular}
\end{table*}

Table~\ref{tab:comparison_of_ASTs_on_five_metrics} presents the mean and median values of the computed metrics for each AST parsing method. 
From rows 3--6 of Table~\ref{tab:comparison_of_ASTs_on_five_metrics}, it is observed that in terms of structure (i.e., tree size, tree depth, and branch factor), the ASTs generated by ANTLR achieve the maximum mean and median values. This means that ANTLR tends to generate larger, deeper ASTs than the other three. Compared with ANTLR, JDT is the other extreme. The ASTs generated by JDT achieve the minimum mean and median values in both tree size and tree depth. 
The same phenomenon that ANTLR and JDT tend to generate larger and smaller ASTs, respectively, can also be observed in the CodeSearchNet dataset. 
In terms of abstraction level, ANTLR generates ASTs with more unique types and unique tokens, followed by Tree-sitter and srcML, and then JDT. 

\summary{The ASTs generated by different AST parsing tools differ in structure size (especially in tree size and tree depth), and abstraction level. The AST generated by JDT is the smallest, shallowest, and has the highest level of abstraction, whereas ANTLR is the opposite. Tree-sitter and srcML are both intermediate in structure size and abstraction level between JDT and ANTLR.}

\subsection{Answer to RQ2: Does AST improve the expressiveness of code representation and facilitate subsequent code-related tasks?}
\label{subsec:Answer_to_RQ2}
Previous works~\cite{2018-DL-Similarities-from-Code-Representations, 2019-ASTNN, 2019-Code-Summarization-with-Extended-Tree-LSTM, 2022-M2TS} hold on to the point that AST contains syntactic knowledge, which contributes more to modeling source code than token information, but there is no solid evidence. To figure out whether AST can improve the effectiveness of code representation, we conduct a comparative analysis of models trained with four distinct types of inputs including Token, SBT, SBT w/o Token, and Token + SBT w/o Token, on three code-related tasks, i.e., code clone detection, code search, and code summarization. The specifics of the four inputs are outlined below:

\textbf{Token}: Use the complete token sequence of the code snippet as input.

\textbf{SBT}: Use SBT sequences as input. Here code snippets are parsed into AST using JDT. SBT sequences are generated using the AST preprocessing method proposed by Hu et al.~\cite{2018-Deep-Code-Comment-Generation} detailed in Section~\ref{subsec:AST_preprocessing_method}.

\begin{figure}[htbp]
    \centering
    \includegraphics[width=\linewidth]{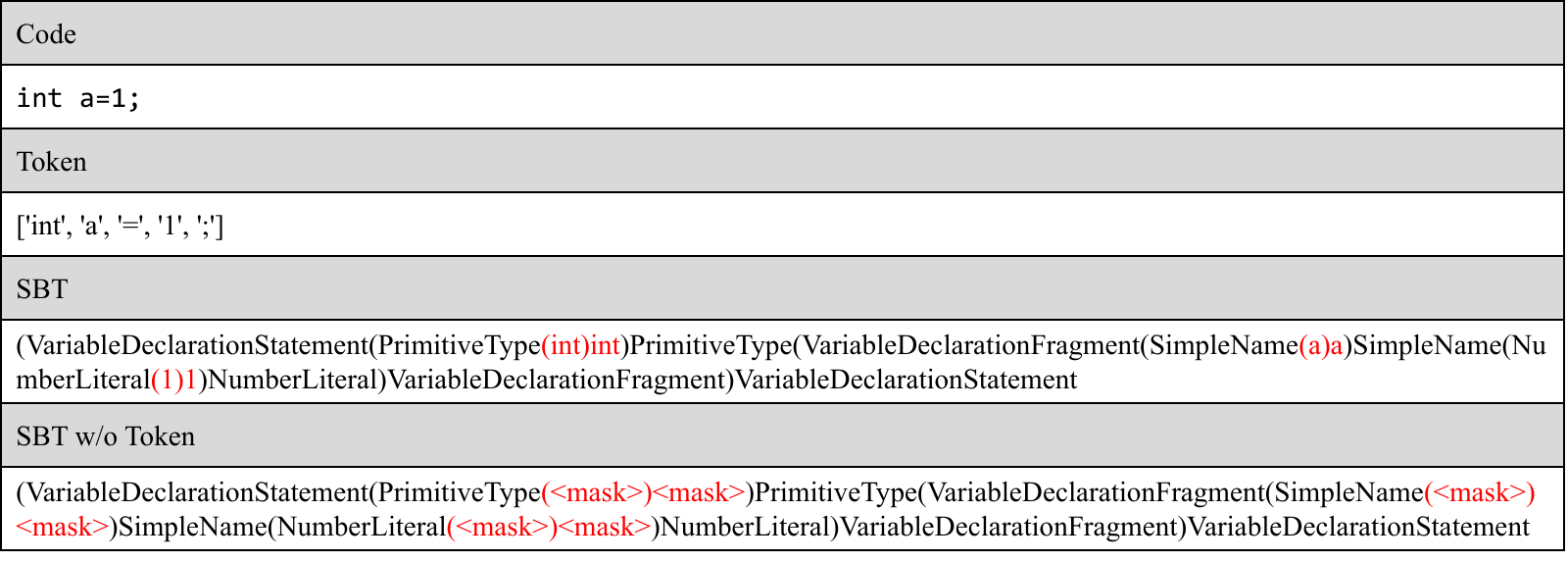}
    \caption{An example of Token, SBT and SBT w/o Token}
    \label{fig:RQ2-token-sbt-sbtwotoken-example}
\end{figure}

\textbf{SBT w/o Token}: The difference between SBT and SBT w/o Token is that we change every leaf node label to \textless mask\textgreater\ in SBT w/o Token. Intuitively, we can regard SBT w/o Token as only retaining the structural information of the AST. Fig.~\ref{fig:RQ2-token-sbt-sbtwotoken-example} shows an example of Token, SBT, and SBT w/o Token.

\textbf{Token + SBT w/o Token}: 
We feed the model with Token and SBT w/o Token, and let it learn the embedding vectors for these two inputs, respectively. Then, we concatenate these two vectors on the second dimension and use a linear layer to compress the concatenated vector to get the final embedding vector, which is the same size as the Token embedding vector and SBT w/o Token embedding vector. 
In fact, SBT and Token + SBT w/o Token can be regarded as two combinations of token information and syntactic information.

In this RQ, we uniformly use JDT as the AST parsing method. This is because ASTs generated by JDT are the smallest and least deep (detailed in Section~\ref{subsec:Answer_to_RQ1}), making them easier for the model to learn (detailed in Section~\ref{subsec:Answer_to_RQ3}) and helping to reveal whether AST inherently aids in enhancing code representation and code-related tasks. To eliminate randomness, we conduct experiments on two neural network architectures, BiLSTM and Transformer, which are widely used in existing code-related task models.

\begin{table*}[!t]
    \centering
    \footnotesize
    \tabcolsep=1.4pt
    \caption{Overall performance of models trained with four types of inputs on three code-related tasks}
    \label{tab:RQ2-code-clone-summarization-search}
    \begin{tabular}{cccccccccccc}
        \toprule
        
        \multirow{2}{*}{Model} & \multirow{2}{*}{Input} & \multicolumn{3}{c}{Code Clone Detection} & \multicolumn{3}{c}{Code Summarization} & \multicolumn{4}{c}{Code Search} \\

        \cmidrule(lr){3-5} \cmidrule(lr){6-8} \cmidrule(lr){9-12}
        
        & & Recall & Precision & $F_1$ & BLEU & METEOR & ROUGE-L & SR@1 & SR@5 & SR@10 & MRR \\
        
        \midrule 
        
        \multirow{4}{*}{BiLSTM}& Token & \textbf{88.92} & \textbf{90.35} & \textbf{89.62} & \textbf{10.06} & \textbf{5.882} & \textbf{19.69} & 0.2041 & 0.3994 & \textbf{0.4840} & 0.3011 \\
        
        {} & SBT & 73.06 & 78.68 & 75.77 & 9.021 & 4.540 & 16.45 & 0.1046 & 0.2325 & 0.4069 & 0.1967 \\
        
        {} & SBT w/o Token & 40.91 & 42.71 & 41.79 & 7.043 & 3.638 & 14.67 & 0.000 & 0.0146 & 0.0408 & 0.0139 \\
        
        {} & Token + SBT w/o Token & 86.06 & 90.05 & 88.01 & 9.840 & 5.541 & 18.93 & \textbf{0.2187} & \textbf{0.4082} & \textbf{0.4840} & \textbf{0.3134} \\
        
        \midrule

        \multirow{4}{*}{Transformer}& Token & \textbf{93.80} & 93.15 & \textbf{93.74} & \textbf{15.13} & \textbf{8.974} & \textbf{30.74} & \textbf{0.3615} & 0.6122 & 0.6909 & 0.4774 \\
        
        {} & SBT & 92.57 & 91.06 & 91.81 & 15.00 & 8.555 & 29.84 & 0.2216 & 0.4898 & 0.6122 & 0.3490 \\
        
        {} & SBT w/o Token & 68.47 & 58.40 & 63.03 & 11.55 & 5.204 & 22.22 & 0.0421 & 0.0146 & 0.0671 & 0.0962 \\
        
        {} & Token + SBT w/o Token & 91.39 & \textbf{93.75} & 92.55 & 14.71 & 8.656 & 29.86 & 0.3586 & \textbf{0.6327} & \textbf{0.7114} & \textbf{0.4842} \\

        \bottomrule
    \end{tabular}
\end{table*}

Table~\ref{tab:RQ2-code-clone-summarization-search} presents the overall performance of BiLSTM-based and Transformer-based models trained with four types of inputs on the three code-related tasks. From rows 3--4 and 7--8, it is observed that the models trained with Token consistently outperform those with SBT on all three tasks. For example, in terms of $F_1$, BLEU, and MRR, the BiLSTM-based models trained with Token achieve 18\%, 12\%, and 53\% improvement over the models trained with SBT, respectively. It indicates that the Token-based code representation is better than the AST-based code representation in promoting these three tasks.

From rows 4--5 and 8--9, it is observed that compared with the models trained with SBT, the performance of the models trained with SBT w/o Token consistently has a significant reduction on all three tasks. For instance, the Transformer-based models trained with SBT w/o Token reduce $F_1$, BLEU, and MRR by 31\%, 23\%, and 72\%, respectively. It means that token information plays a pivotal role in AST-based code representation.

From rows 4, 6, 8, and 10, it is observed that compared with the models trained with SBT, the performance of the models trained with Token + SBT w/o Token consistently improves on all three tasks. For example, the BiLSTM-based models trained with Token + SBT w/o Token improve $F_1$, BLEU, and MRR by 16\%, 15\%, and 59\%, respectively. It signifies that varying combinations of token and syntactic information influence the effectiveness of AST-based code representation, thereby impacting the model's performance on code-related tasks. 

From rows 3, 6, 7, and 10, it is observed that compared with the models trained with Token, the overall performance of the models trained with Token + SBT w/o Token shows improvement only on the code search tasks. However, for the tasks of code clone detection and code summarization, not only is there no enhancement, but also the performance deteriorates. Both the increase and decrease of $F_1$, BLEU, and MRR are between 1\% and 4\%. It should be noted that the values in Table~\ref{tab:RQ2-code-clone-summarization-search} are the average scores of all samples in each metric. Therefore, although the increase and decrease are not large from the average scores, it still shows that the syntactic information of AST has an impact on the code representation of certain subsets of samples, thereby affecting the model's performance on code-related tasks.

\begin{figure}[htbp]
\centering
    \subfigure[BiLSTM]
    {
        \includegraphics[width=0.25\linewidth]{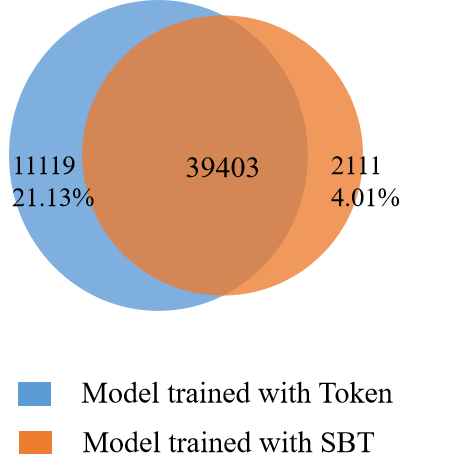}
        \label{fig:RQ2-clone-distribution-BiLSTM}
    }
    \hspace{10mm}
    \subfigure[Transformer]
    {
        \includegraphics[width=0.24\linewidth]{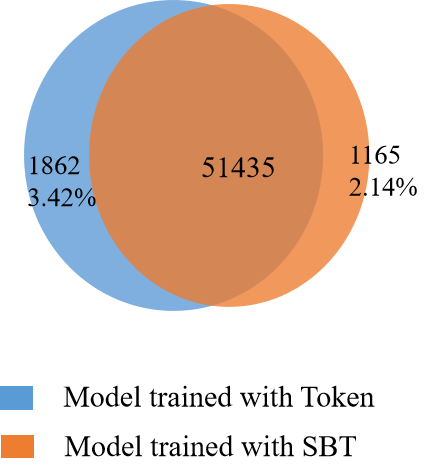}
        \label{fig:RQ2-clone-distribution-Transformer}
    }
    \caption{Venn diagram of clone pairs detected by models trained with Token and SBT}
    \label{fig:RQ2-clone-distribution}
\end{figure}

\begin{figure}[htbp]
    \centering
    \subfigure[BiLSTM]
    {
        \includegraphics[width=0.3\linewidth]{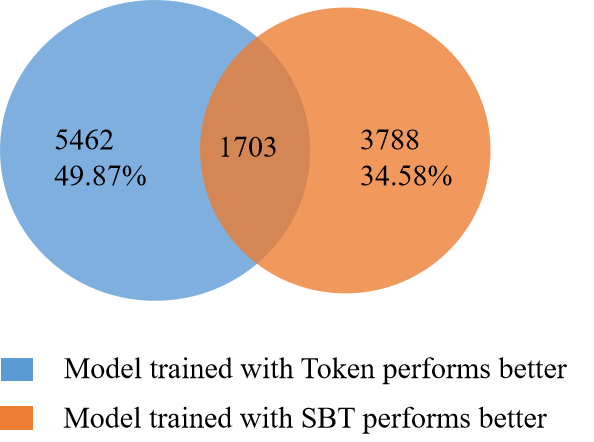}
        \label{fig:RQ2-summarization-distribution-BiLSTM}
    }
    \hspace{6mm}
    \subfigure[Transformer]
    {
        \includegraphics[width=0.3\linewidth]{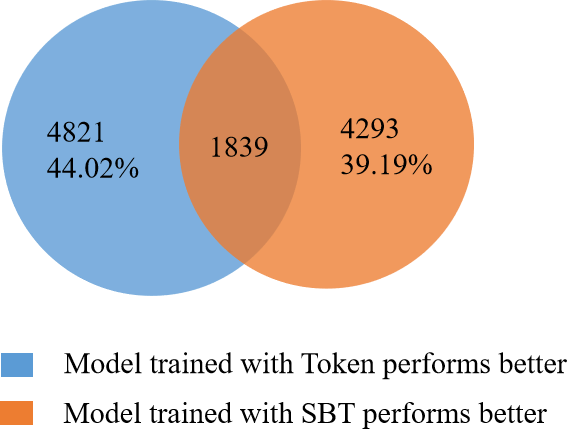}
        \label{fig:RQ2-summarization-distribution-Transformer}
    }
    \caption{Venn diagram of code summarization samples on which models trained with Token and SBT get higher BLEU scores}
    \label{fig:RQ2-summarization-distribution}
\end{figure}

\begin{figure}[htbp]
\centering
    \subfigure[BiLSTM]
    {
        \includegraphics[width=0.3\linewidth]{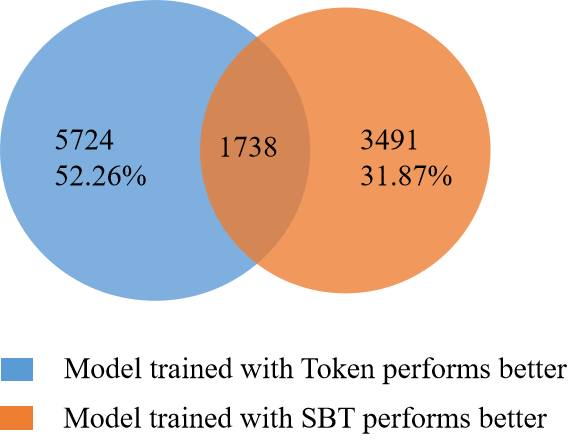}
        \label{fig:RQ2-search-distribution-BiLSTM}
    }
    \hspace{6mm}
    \subfigure[Transformer]
    {
        \includegraphics[width=0.3\linewidth]{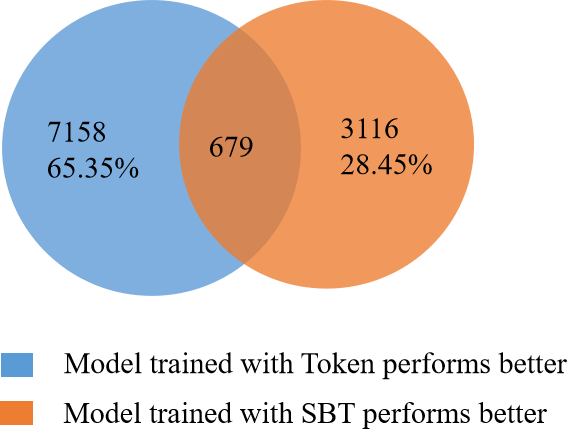}
        \label{fig:RQ2-search-distribution-Transformer}
    }
    \caption{Venn diagram of code search samples on which the ground-truth results are ranked higher by models trained with Token and SBT}
    \label{fig:RQ2-search-distribution}
\end{figure}

The work~\cite{2018-DL-Similarities-from-Code-Representations} suggests that different code features (e.g., Token, AST, CFG, and DFG) can provide an orthogonal view of the same code snippet. 
To figure out under what circumstances SBT (or syntactic information) can perform better than Token, we further drill down to analyze samples where models trained with only Token or SBT perform better. 
For the code clone detection, we analyze the distribution of clone pairs detected by the models trained with Token and SBT, respectively. Fig.~\ref{fig:RQ2-clone-distribution} presents the statistics, with the left side displaying clone pairs detected by BiLSTM-based models and the right side showcasing clone pairs detected by Transformer-based models. It is observed that whether BiLSTM or Transformer is used, there are some clone pairs that can only be detected by Token or SBT. For instance, as illustrated in Fig.~\ref{fig:RQ2-clone-distribution-Transformer}, a total of 54,462 true clone pairs have been detected. Out of them, 1,862 are exclusively identified by the model trained with Token, constituting 3.42\% of the total, while 1,165 are solely detected by the model trained with SBT, constituting 2.14\% of the total. 
The same phenomenon occurs for code summarization and code search tasks. For example, as shown in Fig.~\ref{fig:RQ2-summarization-distribution}, the BiLSTM-based and Transformer-based models trained with SBT perform better than corresponding models trained with Token in BLEU on 34.58\% and 39.19\% of code summarization samples, respectively. As shown in Fig.~\ref{fig:RQ2-search-distribution}, 31.87\% and 28.45\% of code search samples are ranked higher by the BiLSTM-based and Transformer-based models trained with SBT than corresponding models trained with Token. These phenomena all suggest that, in certain cases of code clone detection/summarization/search, the syntactic information embedded within the AST proves to be more valuable than the token information. 

To better grasp the usage of AST in code representation, thereby facilitating code-related tasks (e.g., code clone detection/summarization/search that this paper focuses on), we undertake a more detailed excavation of the characteristics of the code within samples where SBT-based models perform better.

\textbf{For the code clone detection task}, intuitively, the model trained with Token should exhibit greater accuracy in identifying true clone pairs with similar tokens. In this paper, we leverage the Jaccard index~\cite{1901-Jaccard-index} to compute the similarity between tokens of two code snippets in each clone pair. The token similarity $s$ between two code snippets can be computed as follows:
\begin{equation}
    \small
    s = JaccardIndex(code_1, code_2) = \frac{|Tok(code_1) \cap Tok(code_2)|}{|Tok(code_1) \cup Tok(code_2)|}
    \label{equ:Jaccard_Index}
\end{equation}
where $Tok(\cdot)$ is a function used to tokenize code snippets; $|\cdot|$ represents the size of the set $\cdot$. As for the tokenizer, we use the pre-trained tokenizer provided by CodeBERT~\cite{2020-CodeBERT}.

\begin{figure}[!t]
    \centering
    \subfigure[BiLSTM]
    {
        \includegraphics[width=0.44\linewidth]{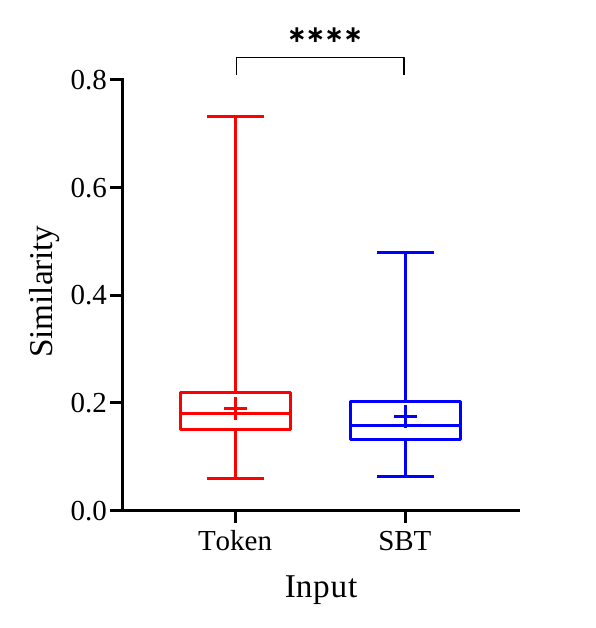}
        \label{fig:RQ2-clone-jaccard-BiLSTM}
    }
    \subfigure[Transformer]
    {
        \includegraphics[width=0.44\linewidth]{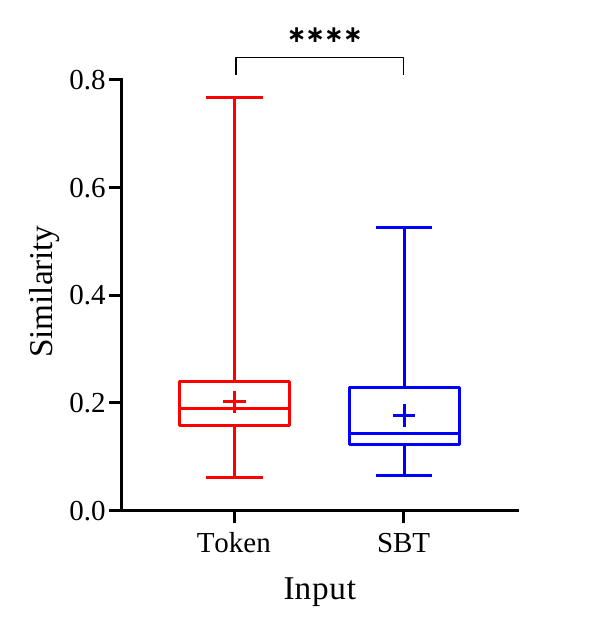}
        \label{fig:RQ2-clone-jaccard-Transformer}
    }
    \caption{Distribution of the Jaccard similarity scores between tokens of clone pairs only detected by models trained with Token or SBT. In (a) and (b), the `+' denotes the means; `****' ($p < 0.0001$) represents the differences between the two groups are extremely significant.}
    \label{fig:RQ2-clone-jaccard}
\end{figure}

Fig.~\ref{fig:RQ2-clone-jaccard} presents the box-and-whiskers plots of the Jaccard index of cloned code pairs detected by BiLSTM-based and Transformer-based models trained with Token and SBT. From the figure, it is observed that for both BiLSTM-based and Transformer-based models, the median, first quartile, and third quartile values of models trained with Token are greater than those of the models trained with SBT. 
We further perform a statistical test to check whether statistically significant differences exist. Statistical tests provide a probability value, i.e., $p$-value between 0 and 1. The lower the $p$-value, the more likely that the null hypothesis is false. It is accepted by the research community that a $p$-value under 0.05 is statistically significant enough for the hypothesis to be considered false~\cite{2014-Hitchhiker-Guide-Statistical-Test-in-SE}. Since the data of similarity scores do not follow a normal distribution in general, our analysis requires the use of non-parametric test methods. Specifically, we perform the Mann-Whitney U test~\cite{2008-Mann-Whitney-U-Test} on two groups of similarity scores at a significance level of 5\% by default. All $p$-values reported by the test for similarity scores of cloned code pairs detected by BiLSTM-based and Transformer-based models are smaller than the significant threshold value of 0.05. It means that there is a significant difference between models trained with Token and SBT in the token similarity of detected cloned code pairs.

\begin{table*}[!t]
    \caption{Results statistics by different $s$ intervals. }
    \label{tab:RQ2-clone_further2} 
    \resizebox{\linewidth}{!}{
        \begin{tabular}{ccccccccc}
            \toprule
            Model & $s$ & $\leq$ 0.1 & (0.10, 0.15] & (0.15, 0.20] & (0.20, 0.25] & (0.25, 0.30] & (0.30, 0.35] & \textgreater0.35 \\ 
            
            \midrule
            
            \multirow{4}{*}{BiLSTM} & Token & 264 & 2746 & 4109 & 2384 & 1105 & 341 & 170 \\
            
            {} & SBT & 118 & 813 & 599 & 285 & 172 & 87 & 37 \\

            \cmidrule(lr){2-9}

            {} & Average $s$ & 0.0897 & 0.1299  & 0.1742 & 0.2221 & 0.2723 & 0.3202 & 0.3967 \\


            \midrule
            
            \multirow{4}{*}{Transformer} & Token & 33 & 363 & 627 & 429 & 282 & 94 & 34 \\
            
            {} & SBT & 84 & 553 & 154 & 145 & 136 & 59 & 34 \\

            \cmidrule(lr){2-9}

            {} & Average $s$ & 0.0899 & 0.1282 & 0.1730 & 0.2244 & 0.2722 & 0.3195 & 0.3996 \\

            \bottomrule
        \end{tabular}
    }
\end{table*}

To further explore the influence of the Jaccard index, we count the number of code clones that are only detected by models trained with Token/SBT in different Jaccard index intervals. Table~\ref{tab:RQ2-clone_further2}
shows the statistical results, where the first row shows seven Jaccard index intervals. Rows 2 and 5/Rows 3 and 6 showcase the number of code clones that are only detected by models trained with Token/SBT, respectively. Rows 4 and 7 present the average Jaccard index value of all clone pairs counted in the corresponding Jaccard index interval. 
According to this table, it is observed that (1) no matter in which Jaccard index interval, there are some clone pairs that are correctly detected by the model trained with only on Token/SBT; (2) when the Jaccard index of cloned code pairs less than 0.15, there are a large number of cloned code pairs that are only detected by the model trained with SBT. Based on these observations, we can conclude that syntactic information embedded in AST can facilitate code representation and subsequent clone detection tasks, especially when the similarity between the tokens of two code snippets in a clone pair is low.

\begin{figure*}[htbp]
    \centering
    \subfigbottomskip=2pt 
    \subfigcapskip=-5pt
    \subfigure[Code Snippet $s_3$]{
        \includegraphics[width=0.4\linewidth]{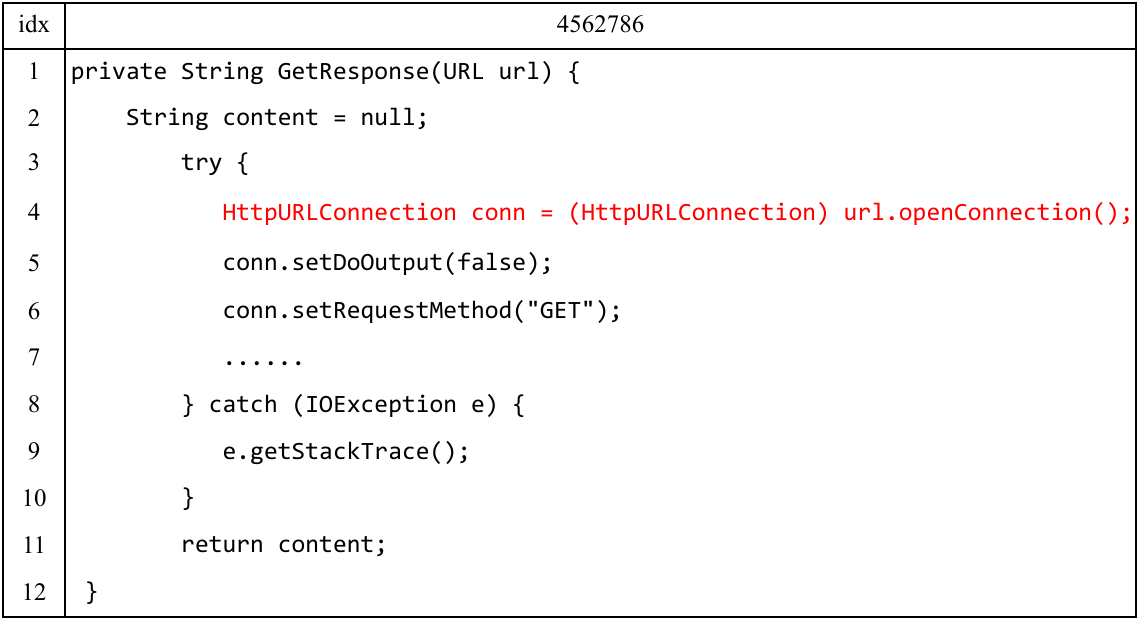}
        \label{fig:RQ2-clone-ex-func-sbt-code-1}
    }
    \subfigure[AST of the 4-$th$ statement in $s_3$]{
        \includegraphics[width=0.56\linewidth]{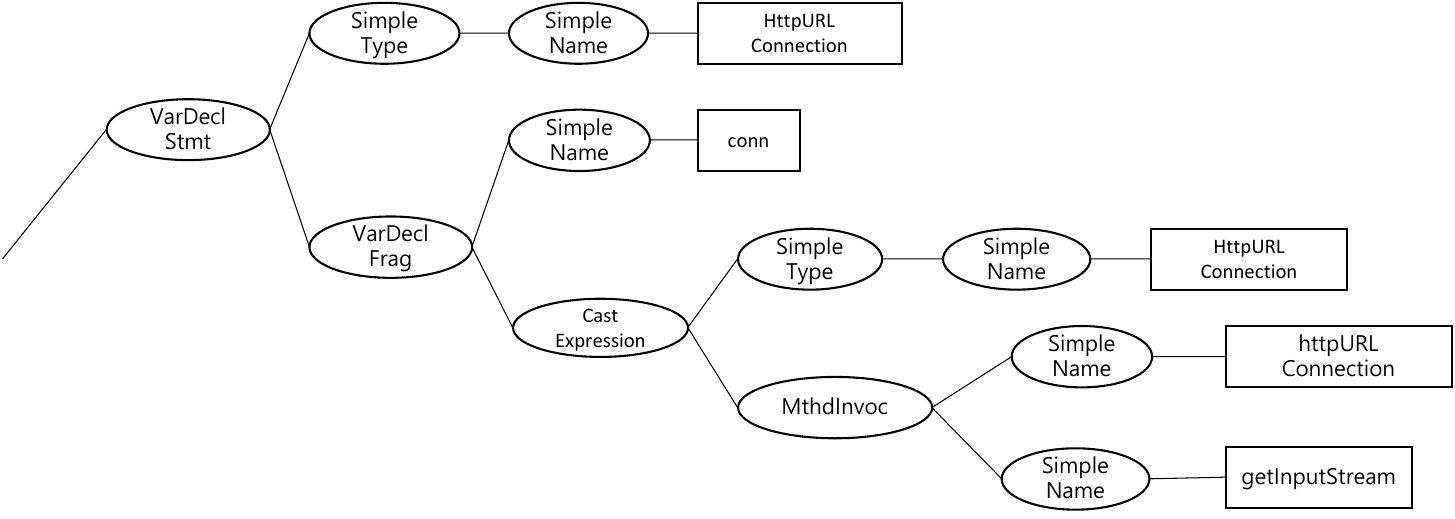}
        \label{fig:RQ2-clone-ex-func-sbt-ast-1}
    }
    \subfigure[Code Snippet $s_4$]{
        \includegraphics[width=0.4\linewidth]{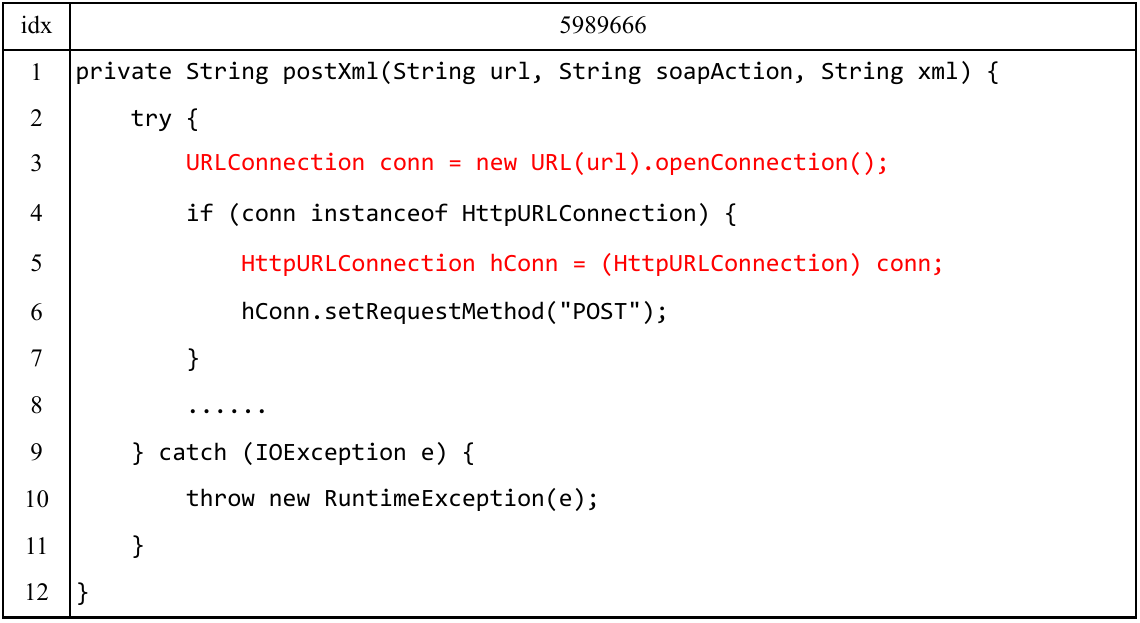}
        \label{fig:RQ2-clone-ex-func-sbt-code-2}
    }
    \subfigure[AST of the 3-$th$ and 5-$th$ statements in $s_4$]{
        \includegraphics[width=0.56\linewidth]{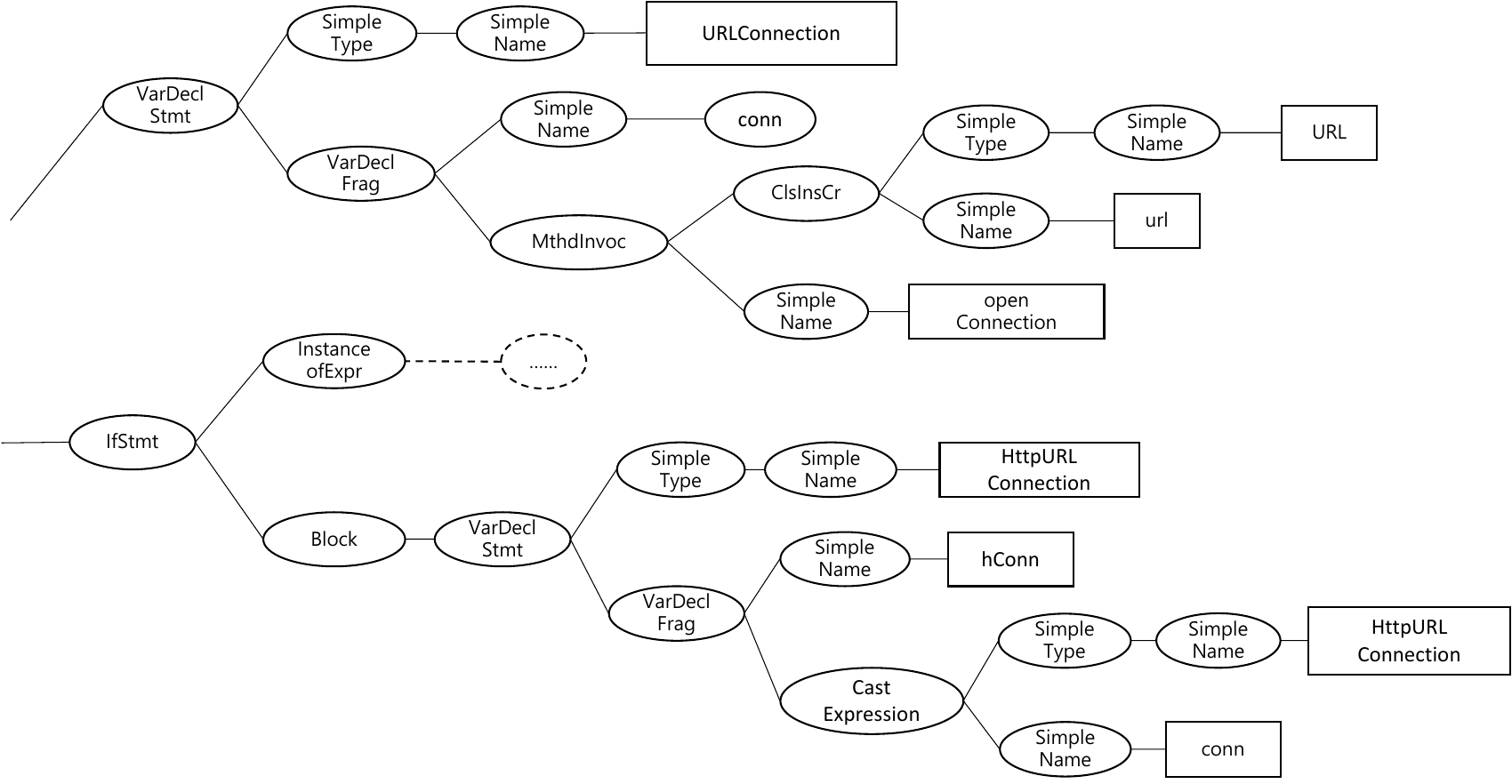}
        \label{fig:RQ2-clone-ex-func-sbt-ast-2}
    }
    \caption{An example of a clone pair detected by the Transformer-based model trained with Token}
    \label{fig:RQ2-clone-ex-func-sbt}
\end{figure*}

\begin{figure*}[htbp]
    \centering
    \subfigbottomskip=2pt
    \subfigcapskip=-5pt
    \subfigure[Code Snippet $s_5$]{
        \includegraphics[width=0.48\linewidth]{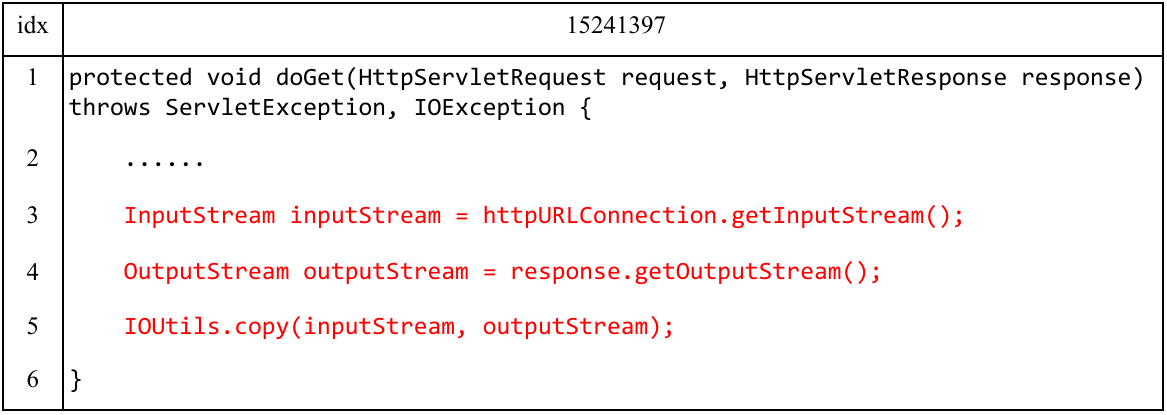}
        \label{fig:RQ2-clone-ex-sbt-func-code-1}}
    \subfigure[Code Snippet $s_6$]{
        \includegraphics[width=0.48\linewidth]{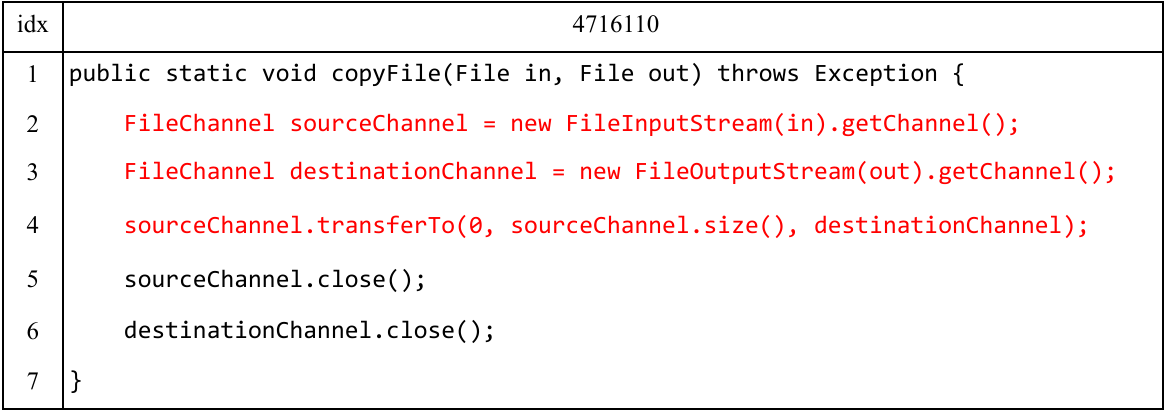}
        \label{fig:RQ2-clone-ex-sbt-func-code-2}}
    \subfigure[AST of statements in lines 3--5 of $s_5$]{
        \includegraphics[width=0.48\linewidth]{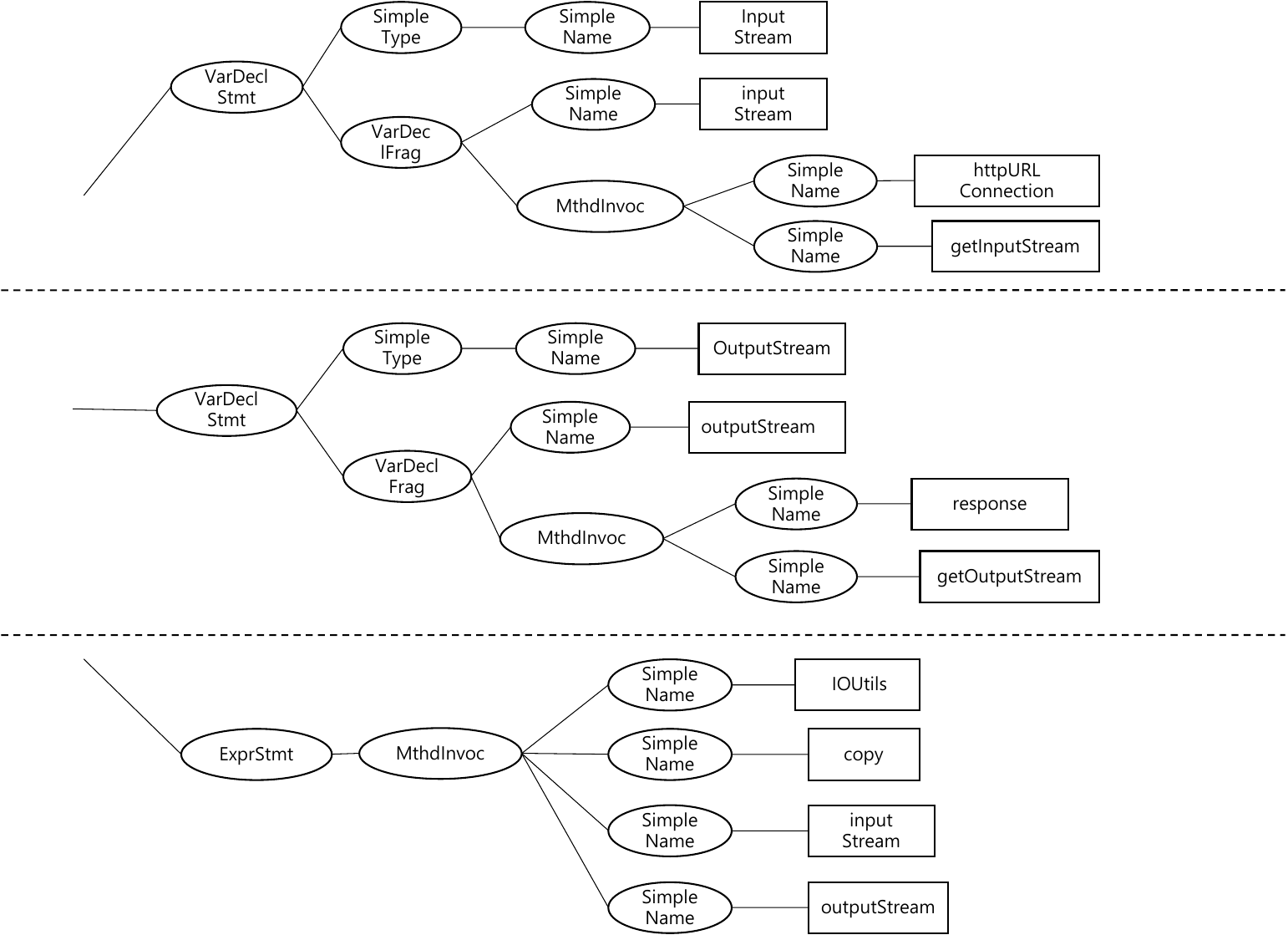}
        \label{fig:RQ2-clone-ex-sbt-func-ast-1}
    }
    \subfigure[AST of statements in lines 2--4 of $s_6$]{
        \includegraphics[width=0.48\linewidth]{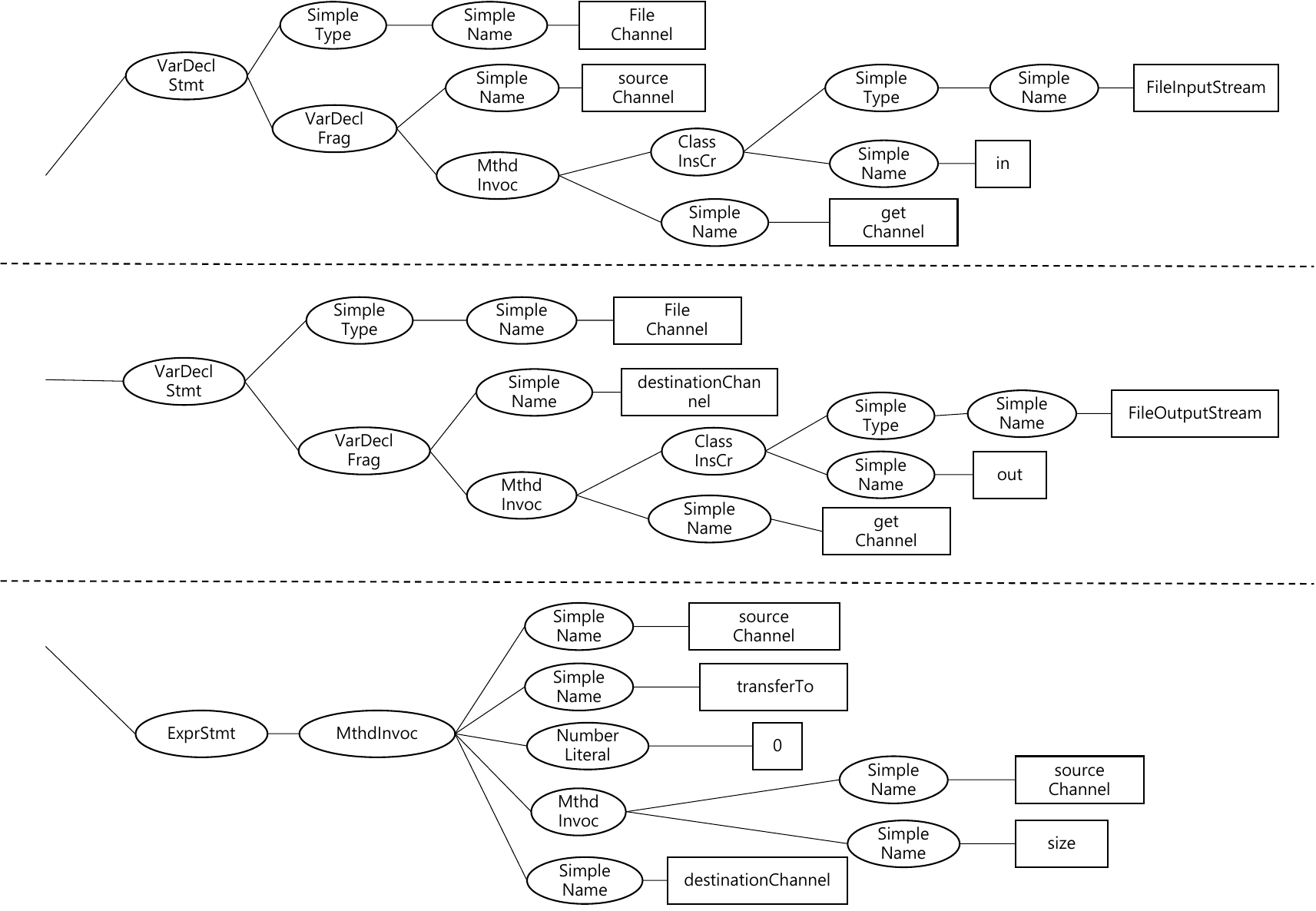}
        \label{fig:RQ2-clone-ex-sbt-func-ast-2}
    }
    \caption{An example of a clone pair detected by the Transformer-based model trained with SBT}
    \label{fig:RQ2-clone-ex-sbt-func}
\end{figure*}

Fig.~\ref{fig:RQ2-clone-ex-func-sbt} and~\ref{fig:RQ2-clone-ex-sbt-func} visually show examples of true cloned code pairs $\langle s_3, s_4\rangle$ and $\langle s_5, s_6\rangle$ detected by Transformer-based models trained with only Token and SBT, respectively. The code snippets $s_3$ and $s_4$ are from the BigCloneBench dataset. Due to the page limit, we only display the ASTs of partial code statements highlighted in red font. For example, Fig.~\ref{fig:RQ2-clone-ex-func-sbt-ast-1} showcases the AST of the 4-$th$ statement in $s_3$. From Fig.~\ref{fig:RQ2-clone-ex-func-sbt}, it is observed that although the code statements highlighted in red font in $s_3$ and $s_4$ implement the same semantics, that is, using the \textit{url} to create an HTTP connection, their ASTs are (i.e., syntactic information) are significantly different. The token similarity between $s_3$ and $s_4$ calculated by Equation~(\ref{equ:Jaccard_Index}) is 0.45, and the semantic similarity between embeddings of $s_3$ and $s_4$ produced by Transformer-based models trained with Token and SBT are 0.8562 and 0.0281, respectively. 
From Fig.~\ref{fig:RQ2-clone-ex-sbt-func}, it is observed that although the code statements highlighted in red font in $s_5$ and $s_6$ implement the different semantics, their ASTs are (i.e., syntactic information) are very similar. The token similarity between $s_5$ and $s_6$ calculated by Equation~(\ref{equ:Jaccard_Index}) is 0.2895, and the semantic similarity between embeddings of $s_5$ and $s_6$ produced by Transformer-based models trained with Token and SBT are 0.0000 and 0.9982, respectively.

\begin{table}[htbp]
    \footnotesize
    \caption{Comparison of models trained with Token and SBT on BLEU. Column Number presents the number of samples on which the models trained with Token or SBT achieve a higher BLEU score. In Column BLEU, the values in brackets refer to the BLEU scores of the model trained with another input (SBT/Token) for the same batch of samples.}
    \label{tab:RQ2-summarization_further1}
    \centering  
    \begin{tabular}{ccccc}
        \toprule
        
        Model & Input & Number & BLEU & Average $e$ \\ 

        \midrule
        
        \multirow{2}{*}{BiLSTM} & Token & 5884 & 0.1196 (0.06759) & \textbf{0.2836} \\
        
        & SBT & 4254 & 0.1198 (0.0750) & 0.2799 \\
        
        \midrule

        \multirow{2}{*}{Transformer} & Token & 4965 & 0.1695 (0.1131) & \textbf{0.2898} \\
        
        & SBT & 4476 & 0.1858 (0.1266) & 0.2775 \\

        \bottomrule
    \end{tabular}
\end{table}

\textbf{For the code summarization task}, similar to the Jaccard index used in the code clone detection task, we also seek to find an indicator that can be used to depict the characteristics of the code summarization sample, thereby guiding model developers when it is more appropriate to use Token/SBT. Intuitively, code summarization is easier if most of the words in the summary can be found in the code. In this case, it may be sufficient to train the model directly with Token. 
Therefore, we measure the ease of code summarization by the proportion of words in the summary that appear in the code snippet. 
It can be regarded as a conditional probability distribution in set theory, which is the proportion of the intersection of two sets $A$ and $B$ relative to the size of either set $A$ or $B$. In our setting, $A$ and $B$ are summaries and the corresponding code snippets. Therefore, the ease of code summarization denoted $e$ can be computed as follows:
\begin{equation}
    \small
    e = P(code|summary) = \frac{|Tok(summary) \cap Tok(code)|}{|Tok(summary)|}
    \label{equ:ease_of_code_summarization}
\end{equation}
where $Tok(\cdot)$ is a function used to tokenize summaries and code snippets; $|\cdot|$ represents the size of the set. As for the tokenizer, we use the pre-trained tokenizer provided by CodeBERT\cite{2020-CodeBERT}.

If the user expects the summaries generated by the code summarization model to contain as much token information as possible in the code snippet, that is, expects a larger $e$, then the performance of the model trained based on Token will be better than the model trained based on AST. The last column of Table~\ref{tab:RQ2-summarization_further1} shows the average $e$ of all samples in the corresponding row. It is observed that those samples on which the BiLSTM-based and Transformer-based models trained with Token perform better have higher $e$, which are 0.2836 and 0.2898, respectively.

\begin{table}[htbp]
    \caption{Results statistics by different $e$ intervals}
    \footnotesize
    \label{tab:RQ2-summarization_further2} 
        \begin{tabular}{cccccccc}
            \toprule
            Model & $e$ & [0, 0.15] & (0.15, 0.30] & (0.30, 0.45] & (0.45, 0.60] & (0.6, 0.75] \\ 
            
            \midrule
            
            \multirow{3}{*}{BiLSTM} & Token & 1548 & 1968 & 1289 & 729 & 271\\
            
            {} & SBT & 1157 & 1414 & 922 & 514 & 184 \\

            \cmidrule(lr){2-7}

            {} & Average $e$ & 0.0705 & 0.2285  & 0.3721 & 0.5221 & 0.6772 \\
            
            \midrule
            
            \multirow{3}{*}{Transformer} & Token & 1257 & 1655 & 1089 & 650 & 243\\
            
            {} & SBT & 1239 & 1481 & 977 & 520 & 193\\

            \cmidrule(lr){2-7}

            {} & Average $e$ & 0.0714 & 0.2281  & 0.3720 & 0.5224 & 0.6761 \\

            \bottomrule
        \end{tabular}
\end{table}

To further explore the influence of $e$, we count the number of samples on which models trained with Token/SBT obtain better BLEU scores in different $e$ intervals. Table~\ref{tab:RQ2-summarization_further2} shows the statistical results, where the first row shows five $e$ intervals. The larger the value of the $e$ interval from left to right, the easier the code summarization is. 
From this table, it is observed that 1) no matter in which $e$ interval, there are some code samples where the summaries generated by the model trained with only on Token/SBT have a higher BLEU score; 2) when $e$ is smaller, the model trained with SBT performs better. Based on these observations, we can conclude that syntactic information embedded in AST can facilitate code representation and subsequent code summarization tasks, especially when the developer expects the summaries generated by the model to not contain too many tokens appearing in code snippets.  
Lower $e$ values can also be understood as scenarios where code snippets require more high-level abstract summaries (i.e., without/independent of many code tokens).

\begin{figure}[t]
  \centering
  \includegraphics[width=\linewidth]{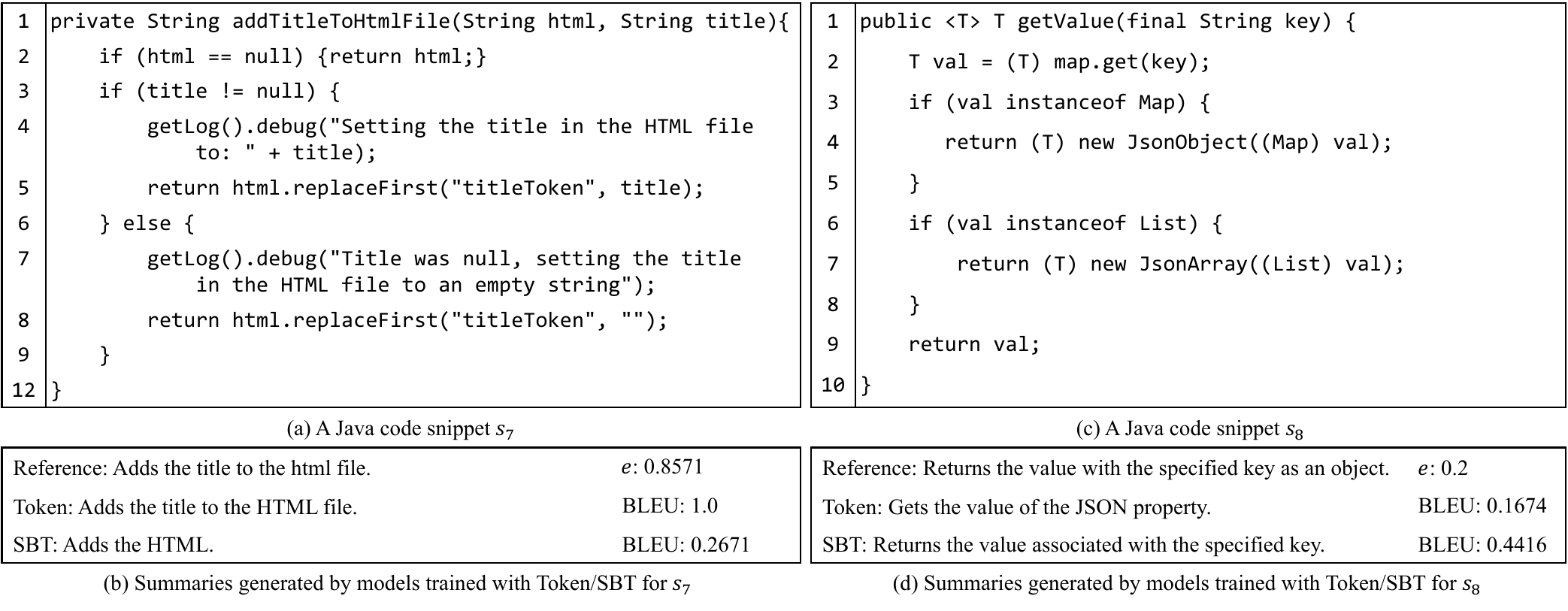}
  \caption{Case of SBT having a negative/positive effect on code summarization task}
  \label{fig:RQ2-code_summarization_Token_SBT_better_case}
\end{figure}

Fig.~\ref{fig:RQ2-code_summarization_Token_SBT_better_case} illustrates two code summarization cases, where two code snippets $s_7$ and $s_8$ in (a) and (d) are from the CodeSearchNet dataset. 
The first lines (i.e., Reference line) of Fig.~\ref{fig:RQ2-code_summarization_Token_SBT_better_case}(b) and (d) present the ground-truth summaries of $s_7$ and $s_8$. The second and third lines (i.e., Token and SBT lines) show the summaries generated by the models trained with Token and SBT for $s_7$ and  $s_8$. 
From Fig.~\ref{fig:RQ2-code_summarization_Token_SBT_better_case}(a) and (b), it is observed that compared with the summary generated by the model trained with SBT, the summary generated by the model trained with Token has a higher BLEU value of 1.0 and is closer to the reference summary. The $e$ between $s_7$ and its reference summary we compute is 0.8571. 
From Fig.~\ref{fig:RQ2-code_summarization_Token_SBT_better_case}(c) and (d), it can be seen that the summary generated by the model trained with SBT has a BLEU value of 0.4416 and is better than that generated by the model trained with SBT (0.1674). The $e$ between $s_8$ and its reference summary we compute is only 0.2. These two examples intuitively illustrate that SBT is suitable for situations where the code snippet itself lacks tokens that can be used to constitute a summary, such as when the tokens (especially the developer-defined identifiers) in the code snippet are not informative. 
The developer's expectation that when generating summaries the model does not rely on the identifiers used by the code snippet itself can be considered one of these cases. 

\begin{table}[htbp]
    \footnotesize
    \caption{Comparison of models trained with Token and SBT on MRR. Column Number presents the number of samples on which the models trained with Token or SBT achieve a higher MRR score. In Column MRR, the values in brackets refer to the MRR scores of the model trained with another input (SBT/Token) for the same batch of samples.}
    \label{tab:RQ2-search_further1}
    \centering  
    \begin{tabular}{ccccc}
        \toprule
        
        Model & Input & Number & MRR & Average $r$ \\ 

        \midrule

        \multirow{2}{*}{BiLSTM} & Token & 5724 & 0.4309 (0.1058) & {\textbf{0.1325}} \\
        
        & SBT & 3491 & 0.3196 (0.0922) & 0.1207 \\ 
        
        \midrule

        \multirow{2}{*}{Transformer} & Token & 7158 & 0.4829 (0.0979) & \textbf{0.1397} \\
        
        & SBT & 3116 & 0.3490 (0.0723) & 0.1208 \\

        \bottomrule
    \end{tabular}
\end{table}

\textbf{For the code search task}, similarly, to find out under what circumstances SBT and Token can perform better in code search, we further analyze the distribution of MRR scores obtained by the code search models trained with Token and SBT, respectively. Table~\ref{tab:RQ2-search_further1} presents the MRR scores of the BiLSTM-based and Transformer-based models on the test set of the CSN-Java dataset. Column Number presents the number of samples on which the models trained with Token or SBT achieve a higher MRR score. For instance, as shown in the Number column of rows 4 and 5, compared to the Transformer-based model trained with SBT, the Transformer-based model trained with Token achieves a higher MMR score on 7,158 samples. Conversely, the opposite can be observed in another set of 3,116 samples. 
Column MRR presents the MRR score obtained by these models on all corresponding samples, where the values in parentheses refer to the MRR scores of the model trained with another input (SBT/Token) for the same samples. For example, 0.4309 (0.1058) in the MRR column of row 2 represents that for the same samples, the MRR score obtained by the model trained with SBT is 0.1058.

Similar to the $e$ used in the code summarization task, we introduce an indicator that can be used to characterize the characteristics of the code search sample, thereby guiding model developers when it is more appropriate to use Token/SBT.  
Intuitively, the results of the code search are more relevant if most of the words in the user queries can be found in the code. In this case, it may be sufficient to train the code search model directly with Token. 
Therefore, we measure the relevance of the code search result by the proportion of words in the query that appear in the code snippet. 
Similar to the calculation of $e$, the relevance of code search represented $r$ can be computed as follows:
\begin{equation}
    \small
    r = P(code|query) = \frac{|Tok(query) \cap Tok(code)|}{|Tok(query)|}
    \label{equ:relevence_of_code_search}
\end{equation}
where $Tok(\cdot)$ is a function used to tokenize queries and code snippets; $|\cdot|$ represents the size of the token set. As for the tokenizer, we also use the pre-trained tokenizer provided by CodeBERT\cite{2020-CodeBERT}.

\begin{table}[htbp]
    \caption{Results statistics by different $r$ intervals}
    \footnotesize
    \label{tab:RQ2-search_further2} 
        \begin{tabular}{ccccccc}
        \toprule
        
        Model & $r$ & [0.025, 0.05] & (0.05, 0.075] & (0.1, 0.125] & (0.125, 0.15] & (0.15,0.175] \\ 
        
        \midrule
        
        \multirow{4}{*}{BiLSTM} & Token & 1139 & 1141 & 938   & 748 & 502 \\
        
        {} & SBT & 801  & 705  & 566  & 418  & 271 \\

        \cmidrule(lr){2-7}
        {} & average $r$ & 0.0243  & 0.0667  & 0.1111  & 0.1333 & 0.1667 \\

         \midrule
        
        \multirow{4}{*}{Transformer} & Token & 1417 & 1459 & 1187 & 977 & 603 \\
        
        {} & SBT & 750  & 642  & 516  & 330  & 197 \\

        \cmidrule(lr){2-7}
        {} & average $r$ & 0.0232  & 0.0667  & 0.1000  & 0.1333 & 0.1667 \\

        \bottomrule
        \end{tabular}
\end{table}

If the user expects the code snippets retrieved by the code search model to contain as many words as possible in the queries (the retrieved code snippets are more relevant and easier to understand), that is, expects a larger $r$, then the performance of the model trained based on Token may be better than the model trained based on AST. 
The last column of Table~\ref{tab:RQ2-summarization_further1} shows the average $r$ of all samples in the corresponding row. It is observed that those samples on which the BiLSTM-based and Transformer-based models trained with Token perform better have higher $r$, which are 0.1325 and 0.1397, respectively. 

To further explore the influence of $r$, we count the number of samples on which the ground-truth results (i.e., code snippets) are ranked higher by models trained with Token/SBT in different $r$ intervals. Table~\ref{tab:RQ2-search_further2} shows the results, where the first row shows five $r$ intervals, and the larger the value of the $r$ interval from left to right, the more relevant the code snippet is to the query. 
From this table, it is observed that 1) no matter in which $r$ interval, there are some code search samples where the ground-truth results are ranked higher by the model trained with only on Token/SBT; 2) when $r$ is smaller, the model trained with SBT performs better. Based on these observations, we can conclude that syntactic information embedded in AST can facilitate code representation and subsequent code search tasks, especially when the developer expects the matching code snippet to rank higher even if it does not contain too many tokens appearing in the query.   
Lower $r$ values can also be understood as code search scenarios where code snippets semantically match queries but contain fewer query words.

\begin{figure}[!t]
    \centering
    \includegraphics[width=\linewidth]{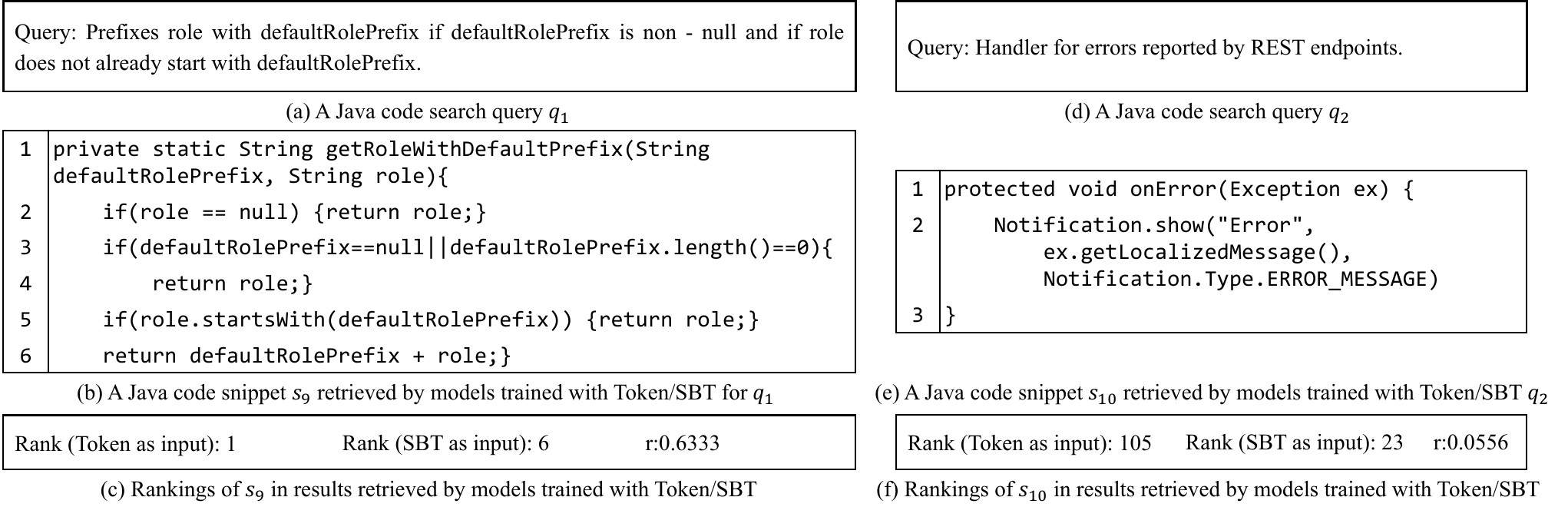}
    \caption{Case of SBT having a negative/positive effect on code search task}
    \label{fig:RQ2-code_search_Token_SBT_better_case}
\end{figure}

Fig.~\ref{fig:RQ2-code_search_Token_SBT_better_case} illustrates two code search cases, where queries $q_1$ and $q_2$ in (a) and (d) are comments from the CodeSearchNet dataset. 
Fig.~\ref{fig:RQ2-code_search_Token_SBT_better_case}(b) and (e) shows two code snippets $s_9$ and $s_{10}$ that are ground-truth results of $q_1$ and $q_2$, respectively. 
Fig.~\ref{fig:RQ2-code_search_Token_SBT_better_case}(c) and (f) show the rankings of $s_9$ and $s_{10}$ in corresponding results retrieved for $q_1$ and $q_2$ respectively by the models trained with Token/SBT. 
For example, ``Rank (Token as input): 1'' in Fig.~\ref{fig:RQ2-code_search_Token_SBT_better_case}(c) indicates that $s_9$ ranks first in the results retrieved by the model trained with Token for query $q_1$, while it ranks 6th in the results retrieved by the model trained with SBT (denoted ``Rank (SBT as input): 6'' in Fig.~\ref{fig:RQ2-code_search_Token_SBT_better_case}(c)). 
Clearly, in this case, the model trained with Token outperforms the one trained with SBT. 
The $r$ we calculate between $s_9$ and $q_1$ has a large value of 0.6333. 
From Fig.~\ref{fig:RQ2-code_search_Token_SBT_better_case}(e) and (f), it can be seen that the rankings of $s_{10}$ in the results retrieved for $q_2$ by the models trained with Token and SBT are 105 and 23, respectively. 
It indicates that in this case, the model trained with SBT outperforms the one trained with Token. 
The $r$ between $s_{10}$ and $q_2$ we compute is only 0.0556. 
These two examples intuitively illustrate that SBT is suitable for situations where the code snippet itself may not contain many tokens that may appear in the user queries, such as when the identifiers in the code snippet are not informative. 
The developer's expectation that when retrieving code snippets the model does not rely on the identifiers used by the code snippet itself can be considered one of these cases. 

\summary{Currently, Token is more helpful than SBT for code understanding and representation and facilitates subsequent code-related tasks. Token information plays a pivotal role in SBT-based code representation. The syntactic information contained in the AST is very useful on some samples, such as two code snippets with low textual similarity in code clone detection, code snippets that require more high-level abstract summaries in code summarization, and code snippets that semantically match but contain fewer query words in code search.}

\subsection{Answer to RQ3: How do AST parsing methods affect the performance of AST-based code representation on subsequent code-related tasks?}
\label{subsec:Answer_to_RQ3}
In Section~\ref{subsec:Answer_to_RQ1}, we find that the ASTs generated by different AST parsing methods are significantly different. Therefore, we further discuss the influence of these differences on AST-based code representation and the performance of models trained with AST on subsequent code-related tasks. Specifically, in order to obtain more general findings and conclusions, the experiments in this section involve all four AST parsers (i.e., JDT, srcML, ANTLR, and Tree-sitter), two AST preprocessing methods (i.e., SBT and Raw AST), two AST encoding methods (i.e., BiLSTM for encoding SBT and Child-Sum TreeLSTM for encoding Raw AST), and three tasks (i.e., code clone detection, code summarization, and code search).

\begin{table*}[!t]
    \caption{Effect of AST parsing methods on the performance of models trained with AST}
    \label{tab:RQ3-code-clone-summarization-search}
    \centering  
    \resizebox{\linewidth}{!}{
    \begin{tabular}{ccccccccccccc}
        \toprule
        
        \multirow{2}{*}{Preprocessing} &  \multirow{2}{*}{Encoding} & \multirow{2}{*}{Parsing} & \multicolumn{3}{c}{Code Clone Detection} & \multicolumn{3}{c}{Code Summarization} & \multicolumn{4}{c}{Code Search} \\

        \cmidrule(lr){4-6} \cmidrule(lr){7-9} \cmidrule(lr){10-13}
        
        & & & Recall & Precision & $F_1$ & BLEU & METEOR & ROUGE-L & SR@1 & SR@5 & SR@10 & MRR \\
        
        \midrule
        
        \multirow{4}{*}{SBT} & \multirow{4}{*}{BiLSTM} & JDT & \textbf{73.06} & 78.68 & \textbf{75.77} & \textbf{9.021} & \textbf{4.540} & 16.45 & 0.1720 & \textbf{0.4052} & \textbf{0.5276} & \textbf{0.2827} \\
        
        {} & {} & srcML & 69.77 & \textbf{80.39} & 74.70 & 8.672 & 4.021 & \textbf{17.03} & 0.1395 & 0.3023 & 0.3604 & 0.2254 \\
        
        {} & {} & ANTLR & 71.59 & 80.11 & 75.61 & 7.689 & 3.353 & 13.48 & 0.1279 & 0.2093 & 0.2674 & 0.1791 \\
        
        {} & {} & Tree-sitter & 72.56 & 79.00 & 75.64 & 8.663 & 4.333 & 15.99 & \textbf{0.1744} & 0.3605 & 0.4419 & 0.2732 \\
        
        \midrule

        \multirow{4}{*}{AST} & \multirow{4}{*}{TreeLSTM} & JDT & 84.25 & \textbf{92.18} & \textbf{88.04} & \textbf{9.781} & \textbf{5.317} & \textbf{18.54} & \textbf{0.0906} & \textbf{0.2076} & \textbf{0.2660} & \textbf{0.1524} \\
        
        {} & {} & srcML & 78.13 & 82.87 & 80.43 & 9.578 & 3.516 & 18.28 & 0.0556 & 0.1784 & 0.2485 & 0.1181 \\
        
        {} & {} & ANTLR & \textbf{87.06} & 84.04 & 85.52 & 9.372 & 3.425 & 17.79 & 0.0760 & 0.1520 & 0.2251 & 0.1226 \\
        
        {} & {} & Tree-sitter & 84.76 & 88.36 & 86.52 & 9.279 & 3.521 & 17.97 & 0.0760 & 0.1784 & 0.2193 & 0.1268 \\

        \bottomrule
    \end{tabular}
    }
\end{table*}

Table~\ref{tab:RQ3-code-clone-summarization-search} presents the overall performance of BiLSTM-based and TreeLSTM-based models trained with ASTs generated by four AST parsing methods on three subsequent code-related tasks. 
It is observed that on the code clone task, in terms of precision, the model trained with the AST generated by srcML performs the best under the SBT setting (with a value of 80.39), while the model trained with the AST generated by ANTLR performs the best (with a value of 87.06) under the AST setting; in terms of recall and $F_1$, the models trained with the AST generated by JDT achieve the largest $F_1$ values under both preprocessing settings (with values of 75.77 and 88.04 respectively), outperforming those trained with the AST generated by all other three parsing methods.  
On the code summarization task, except that the model trained with AST generated by srcML performs best in ROUGE-L under the SBT setting (with a value of 17.03), the model trained with the AST generated by JDT performs best in other cases (e.g., in all three metrics and two preprocessing settings). 
On the code search task, except that the model trained with AST generated by Tree-sitter performs best in SR@1 under the AST setting (with a value of 0.1744), the model trained with the AST generated by JDT performs best in other cases (e.g., in all four metrics and two preprocessing settings).
Overall, JDT performs best on all three tasks, achieving the highest scores on most metrics. 
Note that, in Section~\ref{subsec:Answer_to_RQ1}, we find that the tree size, tree depth, unique types, and unique tokens of the ASTs generated by JDT are the smallest. Therefore, although the ASTs generated by srcML, ANTLR, and Tree-sitter are generally richer than that of JDT, this richness could potentially impose a higher learning burden on the model at the same time, which may not be conducive to improving the model's performance on subsequent code-related tasks.

\summary{JDT, which generates ASTs with the smallest tree size, shallowest tree depth, and lowest abstraction level, yields the most favorable outcomes on all three tasks. This suggests that too many nodes in the AST or too deep trees can impede the DL model’s ability to learn code semantics from the AST.}

\subsection{Answer to RQ4: How do AST preprocessing methods affect the performance of AST-based code representation on subsequent code-related tasks?}
\label{subsec:Answer_to_RQ4}
In this section, we discuss the impact of AST preprocessing methods on AST-based code representation as well as subsequent code-related tasks. 
AST preprocessing methods we investigate can be classified into two categories: one processes ASTs into sequential data, while the other processes ASTs into structural data. 
For sequential data including BFS, SBT, and AST Path, we uniformly leverage BiLSTM, a representative sequence model, to encode them.
For structural data encompassing Raw AST and Split AST, we uniformly utilize Child-Sum TreeLSTM to encode them. For Binary Tree, we use N-ary TreeLSTM as the encoder. 

For BFS, SBT, and AST Path, we uniformly use JDT to generate ASTs for the BigCloneBench and CodeSearchNet datasets, since JDT achieves the best performance, detailed in Section~\ref{subsec:Answer_to_RQ3}. For Raw AST, Binary Tree, and Split AST, we uniformly utilize Tree-sitter to generate ASTs. 
This choice is primarily due to the necessity of splitting the source code during the conversion into Split AST. 
It is worth noting that a significant portion of the split code snippets may contain grammatical errors. In such cases, JDT and srcML are incapable of generating ASTs for code snippets with these errors. In addition, we opt for Tree-sitter over ANTLR, primarily because it demonstrated superior performance (see Section~\ref{subsec:Answer_to_RQ3}).

Table~\ref{tab:RQ4-code-clone-summarization-search} presents the overall performance of BiLSTM-based and TreeLSTM-based models on three code-related tasks. These models are trained using ASTs preprocessed by the above six AST preprocessing methods. 
Combining the results in Table~\ref{tab:RQ4-code-clone-summarization-search} with the observations in Section ~\ref{subsec:AST_preprocessing_method}, we can analyze the impact of AST node and structure information on AST-based code representation and different code-related tasks, providing some guidelines for choosing appropriate AST preprocessing methods for each code-related task.

\textbf{For the code clone detection task}, from columns 4--6 of Table~\ref{tab:RQ4-code-clone-summarization-search}, it is observed that for three sequential AST data, the BiLSTM-based model trained with SBT achieves the highest$F_1$-score of 75.77, outperforming those trained with AST Path (75.08) and BFS (72.88). For three structural AST data, the TreeLSTM-based model trained with Raw AST achieves the highest $F_1$-score of 86.52, better than those trained with Binary Tree (85.49) and Split AST (69.02). 

Recall that as detailed in Section~\ref{subsec:AST_preprocessing_method}, SBT and Raw AST retain complete node and structure information while other preprocessing methods do not.
This demonstrates that both node and structure information are crucial to identifying code clones. Therefore, the preprocessing method that preserves complete AST nodes and structure information is optimal. Moreover, while neglecting some redundant nodes (e.g., AST Path and Binary Tree) may not considerably deteriorate the model's performance, the loss of structure information (e.g., BFS and Split AST) can lead to a more significant decline in model performance, suggesting that structure information contained in AST plays a vital role in promoting AST-based code representation and the code clone task.

\textbf{For the code summarization task}, from columns 7--9 of Table~\ref{tab:RQ4-code-clone-summarization-search}, it is observed that for three sequential AST data, the BiLSTM-based trained with SBT has achieved the best performance on BLEU, METEOR, and ROUGE-L. 
Notably, the BLEU score of the model trained with SBT has exhibited an improvement of 8.09\% and 19.47\% over those of the BiLSTM-based models trained with BFS and AST Path, respectively. It indicates that complete node and structure information (SBT) is the optimal choice when generating code summaries using sequence models. Besides, BFS (complete nodes, little structure information) has exhibited superior performance compared to AST Path (partial nodes, partial structure information), suggesting that nodes play a crucial role in sequence models to generate code summaries. 

Among the three structural AST data, the TreeLSTM-based model trained with Binary Tree attains the highest scores on BLEU and ROUGE-L (i.e., 9.581 and 18.34, respectively), while the TreeLSTM-based model trained with Split AST performs the best on METEOR with a score of 4.387. 
Therefore, we can find that although Raw AST contains complete node and structure information, the TreeLSTM-based model trained with it has the worst code summarization performance. 
Compared with Raw AST, both Binary Tree and Split AST remove some redundant node/structure information or retain only essential nodes and structures, resulting in improved performance in code summarization. 
Of course, it is undeniable that both 
node and structure information are crucial to facilitate AST-based code representation and the code summarization task. 

\textbf{For the code search task}, from columns 10--13 of Table~\ref{tab:RQ4-code-clone-summarization-search}, it is observed among the three sequential AST data, the BiLSTM-based model trained with SBT attains the highest values on SR@1 and SR@10 (i.e., 0.1720 and 0.5276, respectively), whereas the BiLSTM-based model trained with AST Path has the highest values on SR@5 and MRR (i.e., 0.4383 and 0.2904, respectively). In comparison to the BiLSTM-based model trained with AST Path, the BiLSTM-based model trained with SBT has exhibited improvements of 11.47\% and 1.76\% on SR@1 and SR@10, respectively. 
Conversely, the BiLSTM-based model trained with AST has improved SR@5 and MRR by 8.17\% and 2.72\% over the BiLSTM-based model trained with SBT, respectively. 

Among the three structural AST data, the TreeLSTM-based model trained with AST yields the best results on SR@1 and SR@5 (0.0760 and 0.1784, respectively), the TreeLSTM-based model trained with Binary Tree exhibits the best performance on MRR (0.1279), and while the TreeLSTM-based model trained with Split AST exhibits the best performance on SR@10 (0.2588). 
In terms of overall performance, the TreeLSTM-based model trained with Raw AST emerges as the top performer, followed by the models trained with Binary Tree and Split AST.

Based on the performance of BiLSTM-based and TreeLSTM-based models trained with different preprocessed data, we can come to the conclusion that similar to the code clone detection task, capturing complete node and structure information is most advantageous on code search tasks. Additionally, structure information is critical in promoting AST-based code representation and the code search task.

\begin{table*}[!t]
    \caption{Overall performance of models trained with six types of AST preprocessing methods}
    \label{tab:RQ4-code-clone-summarization-search}
    \centering  
    \resizebox{\linewidth}{!}{
    \begin{tabular}{ccccccccccccc}
        \toprule
        
        \multirow{2}{*}{Parsing} &  \multirow{2}{*}{Encoding} & \multirow{2}{*}{Preprocessing} & \multicolumn{3}{c}{Code Clone Detection} & \multicolumn{3}{c}{Code Summarization} & \multicolumn{4}{c}{Code Search} \\

        \cmidrule(lr){4-6} \cmidrule(lr){7-9} \cmidrule(lr){10-13}
        
        & & & Recall & Precision & $F_1$ & BLEU & METEOR & ROUGE-L & SR@1 & SR@5 & SR@10 & MRR \\
        
        \midrule
        
        \multirow{3}{*}{JDT} & \multirow{3}{*}{BiLSTM} & BFS & 68.24 & 78.52 & 72.88 & 8.346 & 3.849 & 16.01 & 0.1453 & 0.3372 & 0.4244 & 0.2440 \\
        
        {} & {} & SBT & \textbf{73.06} & \textbf{78.68} & \textbf{75.77} & \textbf{9.021} & \textbf{4.540} & \textbf{16.45} & \textbf{0.1720} & 0.4052 & \textbf{0.5276} & 0.2827 \\
        
        {} & {} & AST Path & 72.84 & 77.45 & 75.08 & 7.551 & 4.118 & 14.07 & 0.1543 & \textbf{0.4383} & 0.5185 & \textbf{0.2904} \\
        
        \midrule

        \multirow{3}{*}{Tree-sitter} & \multirow{3}{*}{TreeLSTM} & Raw AST & \textbf{84.76} & 88.36 & \textbf{86.52} & 9.279 & 3.521 & 17.97 & \textbf{0.0760} & \textbf{0.1784} & 0.2193 & 0.1268 \\
        
        {} & {} & Binary Tree & 80.64 & \textbf{90.96} & 85.49 & \textbf{9.581} & 3.767 & \textbf{18.34} & 0.0706 & 0.1765 & 0.2353 & \textbf{0.1279}\\
        
        {} & {} & Split AST & 66.64 & 71.57 & 69.02 & 9.432 & \textbf{4.387} & 17.80 & 0.0588 & 0.1647 & \textbf{0.2588} & 0.1179 \\
        
        \bottomrule
    \end{tabular}
    }
\end{table*}

\summary{The AST data generated by different AST preprocessing methods exhibit significant differences in terms of included AST node and structure information. 
Whether it is sequential AST data or structural AST data, the AST node and structure information contained in them facilitate AST-based code representation and subsequent code-related tasks. 
In the code clone detection and code search tasks, the AST preprocessing methods that preserve complete AST node and structure information are optimal, e.g., SBT and Raw AST. The loss of structure information can lead to a more significant decline in model performance, e.g., BFS and Split AST.
In the code summarization task, when using sequence models as AST encoding methods (e.g., BiLSTM), the complete node and structure information is the best choice, i.e., Raw AST. 
Contrarily, when using tree-structured models as AST encoding methods (e.g., TreeLSTM), removing the redundant node (token) and structure information (e.g., Binary Tree) yields the most significant improvement.}

\subsection{Answer to RQ5: How do AST encoding methods affect the performance of AST-based code representation on subsequent code-related tasks?}
\label{subsec:Answer_to_RQ5}
In this section, we conduct experiments to investigate the impact of AST encoding methods on AST-based code representation and follow-up code-related tasks. 
Specifically, we investigate two types of models, i.e., sequence models and tree-structured models. 
Both of them are widely used to encode AST in existing code-related studies. 
For sequence models, we compare two classic models, BiLSTM and Transformer. Such models are not inherently designed to process tree-structured data. Therefore, AST must first be transformed into a sequential format to be fed into these models. 
For tree-structured models, we compare two representative models, i.e., TreeLSTM and AST-Trans. Such models are specifically designed to handle tree-structured data and can more effectively capture the structure information of AST. From the perspective of model architecture, BiLSTM and TreeLSTM are built on the LSTM architecture, while Transformer and AST-Trans are built on the Transformer architecture. 
Therefore, we analyze the experimental results from two perspectives: (1) between sequence models and tree-structured models, which one is more effective in modeling AST; and (2) between LSTM and Transformer, which architecture is more suitable for building AST-based code representation model. 
For a fair comparison, we uniformly employ JDT to generate ASTs for the BigCloneBench and CodeSearchNet datasets, since JDT achieves the best performance, detailed in Section~\ref{subsec:Answer_to_RQ3}. 
For BiLSTM and Transformer, we convert AST into SBT to adapt to the models. For TreeLSTM and AST-Trans, we feed the Raw AST into the models.

Table ~\ref{tab:RQ5-code-clone-summarization-search} shows the overall performance of models built on the four AST encoding methods on three code-related tasks. 
From rows 3 and 4, it is observed that compared with the models built on BiLSTM, the models built on Transformer overall perform better on all three tasks. From rows 5 and 6, it can be seen that the model built on TreeLSTM outperforms that built on AST-Trans on the code clone detection tasks, but the reverse is true on the code summarization and code search tasks. 
Combining the above observations, we can come to the conclusion that Transformer architecture is more advantageous than LSTM architecture in building the AST-based code representation model.

Comparing the performance of BiLSTM and TreeLSTM, it is observed from Row 3 and 5 of Table~\ref{tab:RQ5-code-clone-summarization-search} that the models built on TreeLSTM achieve better performance in the code clone detection and code summarization tasks, while the model built on BiLSTM achieves better performance in the code search task. 
It indicates that tree-structured models are better suited for the code clone detection and code summarization tasks, while sequence models are better suited for the code search task. However, a similar phenomenon does not appear in the comparison between Transformer and AST-Trans, as shown in rows 4 and 6 of Table~\ref{tab:RQ5-code-clone-summarization-search}. 
Different from Transformer, AST-Trans takes the relationship matrix as input in addition to SBT, and uses tree-structured attention to model the structure information of AST. Nevertheless, we believe that there is still room for improvement in enhancing the Transformer architecture to better model AST structures.

\begin{table*}[!t]
    \caption{Overall performance of models built on four different AST encoding methods}
    \label{tab:RQ5-code-clone-summarization-search}
    \centering  
    \resizebox{\linewidth}{!}{
    \begin{tabular}{ccccccccccccc}
        \toprule
        
        \multirow{2}{*}{Parsing} &  \multirow{2}{*}{Preprocessing} & \multirow{2}{*}{Encoding} & \multicolumn{3}{c}{Code Clone Detection} & \multicolumn{3}{c}{Code Summarization} & \multicolumn{4}{c}{Code Search} \\

        \cmidrule(lr){4-6} \cmidrule(lr){7-9} \cmidrule(lr){10-13}
        
        & & & Recall & Precision & $F_1$ & BLEU & METEOR & ROUGE-L & SR@1 & SR@5 & SR@10 & MRR \\
        
        \midrule
        
        \multirow{4}{*}{JDT} & \multirow{2}{*}{SBT} & BiLSTM & 73.06 & 78.68 & 75.77 & 9.021 & 4.540 & 16.45 & 0.1720 & 0.4052 & 0.5276 & 0.2827 \\
        
        {} & {} & Transformer & \textbf{92.57} & 91.06 & \textbf{91.81} & \textbf{15.00} & \textbf{8.555} & \textbf{29.84} & \textbf{0.2216} & \textbf{0.4898} & \textbf{0.6122} & \textbf{0.3490} \\
        
        \cmidrule(lr){2-13}

        {} & \multirow{2}{*}{Raw AST} & TreeLSTM & 84.25 & \textbf{92.18} & \textbf{88.04} & 9.781 & \textbf{5.317} & 18.54 & 0.0906 & 0.2076 & 0.2660 & 0.1524 \\
        
        {} & {} & AST-Trans & \textbf{87.66} & 87.96 & 87.81 & \textbf{10.30} & 4.710 & \textbf{20.11} & \textbf{0.1241} & \textbf{0.2701} & \textbf{0.3430} & \textbf{0.2015} \\
        
        \bottomrule
    \end{tabular}
    }
\end{table*}

\summary{Among the four AST encoding methods investigated in this paper, Tranformer performs best overall. Tree-structured models (e.g., TreeLSTM) and sequence models (e.g., Transformer) each have their own advantages in modeling AST. This paper finds that tree-structured models are better suited for the code clone detection and code summarization task, while sequence models are better suited for the code search task. However, there is still room for improvement in designing a model built on Transformer architecture that is suitable for modeling structure information of AST.}

\section{Threats to Validity}
\label{sec:threats_to_validity}

\subsection{Threats to Internal Validity}
\label{subsec:threats_to_internal_validity}
The threats to internal validity lie in the implementation of AST preprocessing methods and AST encoding methods. We cannot guarantee that our implementation of these methods is completely correct and shares the same performance as they claimed. We have done our best to mitigate these threats. 

Specifically, in terms of AST preprocessing methods, we implement the algorithms to convert source code into SBT~\cite{2018-Deep-Code-Comment-Generation}, AST Path~\cite{2018-Path-based-Representation-Predicting-Program-Properties}, and Binary Tree~\cite{2017-Supervised-Deep-Features-for-Clone-Detection} according to the algorithms described in the corresponding paper. This is necessary because either the original paper does not provide code or the code provided in the original paper is limited to a specific AST parsing tool. As BFS is a widely used and simple algorithm, we implement the algorithm to convert source code into BFS on ourselves as well. As for split AST, there are several preprocessing methods to split the complete AST into smaller components, such as ASTNN~\cite{2019-ASTNN}, CAST~\cite{2021-CAST} and BASTS~\cite{2021-BASTS}. 
The split AST of ASTNN consists of AST nodes of one statement, which is considered to ignore the long-term dependencies of multiple statements and has been proven to be less effective than the split AST of CAST~\cite{2021-CAST} and BASTS~\cite{2021-BASTS}. Between CAST and BASTS, considering the need to not be limited to a specific AST parser, we opt for the split AST preprocessing method of BASTS, which is proposed by Lin et al.~\cite{2021-BASTS}. We use the implementation code provided by Lin et al. to process source code into split code, and then generate ASTs for the split code using an AST parsing tool.

In terms of AST encoding models, we call the interface provided by PyTorch to implement BiLSTM and Transformer. TreeLSTM is modified based on the code provided by Lin et al.~\cite{2021-BASTS}. The original code is implemented using TensorFlow, and we convert it into a PyTorch-based implementation. As to AST-Trans~\cite{2022-AST-trans}, we reuse the model part and data processing part of the code provided in the original paper, and slightly modify the code to adapt to the code framework of CodeXGLUE. We have published the specific implementation code for other researchers to check and use~\cite{2023-AST4PLU}.

\subsection{Threats to External Validity}
\label{subsec:threats_to_external_validity}
The threats to external validity lie in the generalizability of our findings. In order to reduce these threats, we considered three popular types of code-related tasks, including a code-to-code matching task, i.e., code clone detection, a text-to-code matching task, i.e., code search, and a code-to-text generating task, i.e., code summarization.

Another potential threat to external validity is the variety of AST parsing/preprocessing/encoding methods we investigated, which may affect the generalization of our findings to other methods. To reduce this threat, we consider multiple representative AST parsing/preprocessing/encoding methods in our experiments, which are commonly used for various code-related tasks.
In the future, we will conduct experiments on more AST parsing/preprocessing/encoding methods to generalize our conclusions.

The selection of the experimental dataset also poses a threat to external validity. To mitigate this threat, we select two classic and representative datasets, i.e., BigCloneBench and CodeSearchNet, as our experimental datasets. Both of them are widely used to evaluate the effectiveness of code models in the field of software engineering. In addition, we mainly focus on the Java programming language in this paper for the following reasons: (1) Java is one of the most popular programming languages~\footnote{\url{https://www.tiobe.com/tiobe-index/}}; (2) some existing AST parsing tools have limitations on programming languages, but almost every AST parsing tool can generate ASTs for Java code. In summary, investigating the influence of ASTs on Java code representation learning has a potential impact on real-world software development. Therefore, there may be slight differences when adapting our findings to other programming languages and datasets. We leave exploration in other programming languages and datasets to future work.

\subsection{Threats to Construct Validity}
\label{subsec:threats_to_construct_validity}
The threats to construct validity mainly lie in the evaluation metrics we use in our work. 
In order to ensure the effectiveness of the evaluation metrics, we consider the metrics that are widely used on three code-related tasks. Additionally, to avoid implementation errors, we use public libraries or reuse the implementation code provided by previous researchers to calculate these metrics.
For the code clone detection task, we use the interface provided by Scikit-learn to calculate the precision, recall, and $F_1$-score. For the code summarization task, we calculate BLEU using the code provided in the CodeXGLUE~\cite{2021-CodeXGLUE}. We calculate METEOR and ROUGE-L using the code provided by Chen et al.~\cite{2015-Microsoft-coco-captions}. 
For the code search task, we calculate MRR using the code provided in the CodeXGLUE code framework. We calculate SR@k ($k = 1, 5, 10$) using the code provided by Sun et al.~\cite{2022-TranCS}.

\section{Related Work}
\label{sec:related_work}

\subsection{Code Features}
\label{subsec:code_features}
Code features refer to a standardized data format that is derived from the source code. Code features usually describe certain aspects of the source code, e.g., the lexical aspect, syntactic aspect, semantic aspect, etc. 
They are designed to facilitate the processing and analysis of the source code, and can be used for a variety of code-related tasks, such as code clone detection~\cite{2016-DL-Code-Clone-Detection, 2020-Functional-Code-Clone-Detection}, code search~\cite{2018-Neural-Code-Search, 2021-Multimodal-Representation-NCS, 2022-TranCS}, code summarization~\cite{2018-Deep-Code-Comment-Generation, 2021-Project-Level-Encoding-Code-Summarization, 2023-EACS}, etc.
The choice of code features is crucial and non-trivial as it requires taking into account the available DL model and the characteristics of the code-related tasks that need to be solved~\cite{2022-code-representation-survey}. 
Due to differences in model architectures, different DL models accept different formats of inputs, and their learning capabilities vary with different code features. Furthermore, different code-related tasks focus on different aspects of the source code. 

Code features that are widely used in code-related tasks can be divided into three categories: token-based features, tree-based features, and graph-based features, as detailed below. 

\textbf{Token-based features.} Token-based features extract partial or complete tokens from the source code as code features. Tokens are bags of words that are parsed from the source code. 
Commonly used partial tokens are method names~\cite{2018-DeepCodeSearch, 2018-Neural-Code-Search, 2021-TabCS}, API sequences~\cite{2018-TL-CodeSum, 2018-DeepCodeSearch, 2021-MuCoS, 2021-Search-for-Compatible-Code}. 
Compared with the complete token, partial tokens include user-defined identifiers, are simpler, and are easier for the model to learn. 
But it should be noted that partial tokens lose contextual syntax and semantics, while complete tokens contain complete code semantics. 
Therefore, many researchers are prone to use the complete code token sequence~\cite{2018-Neural-Framework-Code-Summarization, 2021-FCCA, 2021-TabCS, 2021-Multimodal-Representation-NCS}. 
They usually refer to the complete code token sequence directly as ``Token''. In this paper, we also follow this habit. 
Since Token contains rich lexical information of the source code, they are frequently used by researchers to solve code-related tasks~\cite{2021-TabCS, 2021-FCCA, 2021-Multimodal-Representation-NCS, 2018-Neural-Framework-Code-Summarization, fang2021self, 2018-DeepCodeSearch, 2020-Multi-Perspective-Architecture-for-CS}.

\textbf{Tree-based features.} Tree-based features mainly refer to parsing the syntax tree of the source code and using it as code features. 
The syntax tree can be further divided into two categories: concrete syntax tree (CST) and abstract syntax tree (AST). 
A CST is an ordered, rooted tree that represents the syntactic structure of the source code according to some context-free grammar~\cite{1997-Abstract-Syntax-from-Concrete-Syntax}. Some studies use CST as code features to learn code representation and assist in solving code-related tasks~\cite{2020-MISIM, 2021-TPTrans, 2023-Tree-Transformer}. For example, Peng et al.~\cite{2021-TPTrans} propose to learn code representation from CST and apply learned code representation to the code summarization task. 
Ye et al.~\cite{2020-MISIM} propose a context-aware semantic structure called CASS, which is built on CST. They use CASS to learn code representation and solve the code similarity analysis task. 

An AST is a simplified representation of a CST and focuses on the essential elements of the program's structure~\cite{1997-Abstract-Syntax-from-Concrete-Syntax, 2007-Compilers-principles-techniques}. Compared to CST, AST provides a more concise and structured representation of the program's semantics, making it easier to analyze and manipulate. 
Currently, AST is one of the most commonly used code features in code representation learning~\cite{2022-code-representation-survey, 2022-Learning-Program-Semantics-Empirical-Study}. 
AST has been widely used to solve many code-related tasks, such as code classification~\cite{2019-ASTNN, 2022-UAST}, code clone detection~\cite{2019-ASTNN, 2019-Recursive-Aggregation-of-AST-for-Clone-Detection, 1998-Clone-Detection-Using-ASTs}, code search~\cite{2019-Multi-modal-Attention-for-Code-Rerieval, 2021-CRaDLe, 2020-Multi-Perspective-Architecture-for-CS, 2021-TabCS}, code summarization~\cite{2021-BASTS, 2021-CAST}, code property (e.g., method and variable names) prediction~\cite{2018-Path-based-Representation-Predicting-Program-Properties}, etc. 
In many cases, ASTs are preprocessed before being fed to the DL model, to adapt to the model input format or further emphasize certain features in the AST. For example, some existing works traverse an AST using BFS or DFS to get the list of all the AST nodes~\cite{2021-TabCS, 2023-Fold2Vec, 2021-API2Com}. Many techniques that chose to convert AST into SBT~\cite{2018-Deep-Code-Comment-Generation, 2018-Deep-Code-Comment-Generation, 2019-Ast-attendgru, 2020-Hybrid-DeepCom, 2021-Code-Summarization-for-Smart-Contracts, 2020-Retrieve-and-Refine-Comment-Generation}. Some techniques~\cite{2019-Code2Vec, 2019-Code2seq, 2021-Authorship-Attribution-Language-agnostic-Approach} use a collection of paths on the tree (i.e., AST Path) to represent an AST. The purpose is to extract key information in the AST and reduce the length of the model input. There are also a lot of techniques that convert AST into simple tree structures, such as Binary Trees~\cite{2017-Supervised-Deep-Features-for-Clone-Detection, 2018-Improving-Code-Summarization-via-DRL, 2019-Multi-modal-Attention-for-Code-Retrieval, 2021-FCCA} and Split AST~\cite{2019-ASTNN, 2021-CAST, 2021-BASTS}, to facilitate model learning.

\textbf{Graph-based features.} 
Graph-based features refer to extracting the graph information from the source code as code features. 
Existing code-related works have mined multiple types of graph information to enhance code representation and solve code-related tasks. 
For example, the control flow graph (CFG) can represent all possible execution paths for the program and is commonly used to capture control dependencies between code elements~\cite{2018-DeepSim, 2019-Multi-modal-Attention-for-Code-Retrieval, 2021-CRaDLe}. 
The data flow graph (DFG) describes the process of how data flows and is processed. Many works~\cite{2019-Multi-modal-Attention-for-Code-Retrieval, 2021-CRaDLe, 2018-DeepSim} use DFG to capture the data dependencies between code elements. 
The program dependence graph (PDG) depicts the data dependencies and control dependencies of each operation in a program. Gu et al.~\cite{2021-CRaDLe} utilize PDG to extract code structures. 
In addition to the well-known graph information mentioned above, some researchers have also mined some new types of graph information. For example, Zhou et al.~\cite{2019-Devign} represent various sub-graphs, including AST, CFG, DFG, and Natural Code Sequence (NCS), into one joint graph to get a code property graph (CPG). 
Allamanis et al.~\cite{2018-Learning-Represent-Programs-with-Graphs} enhance ASTs with different data flow information by constructing diverse types of edges among the nodes, such as connecting variable nodes where the variable is last written.

In this paper, we focus on a tree-based feature, i.e., AST. We conduct a quantitative evaluation and qualitative analysis of the extent to which AST facilitates code representations and subsequent code-related tasks.

\subsection{Code Representation Learning}
\label{subsec:code_representation_learning_related_work}
As mentioned in Section~\ref{subsec:code_representation_learning}, code representation learning aims to convert source code features into distributed, real-valued vector representation, i.e., code representation. 
Such numerical representation condenses the semantics of the code, and facilitates the solution of subsequent code-related tasks. 
In the previous section, we have discussed in detail the first process of code representation learning (i.e., code feature extraction). Next, we mainly introduce the second process, i.e., code feature representation. 

\textbf{Code representation for token-based features.}
Token-based features are sequential data. 
Most works transfer sequence models from the NLP field to encode token-based features. 
For instance, Gu et al.~\cite{2018-DeepCodeSearch} introduce the Recurrent Neural Networks (RNN)~\cite{2010-RNN-based-Language-Model} to encode method names and API sequences. LeClair et al.~\cite{2019-Ast-attendgru} utilize a GRU layer to encode Token (i.e., the complete code token sequence). 
Wei et al.~\cite{2020-Retrieve-and-Refine-Comment-Generation} first map the one-hot embedding token sequence into a word embedding sequence and then use a BiLSTM to process word embedding sequences. Cai et al.~\cite{2021-Search-for-Compatible-Code} and Shuai et al.~\cite{2020-CARLCS} employ BiLSTM to encode method names and API sequences, respectively. 
 
There are also some works that first transform token-based features into other forms before selecting a suitable encoding model. For example, Cheng et al.~\cite{2022-CSRS} and Deng et al.~\cite{2022-FcarCS} convert method name, API sequence, and Token into vector matrices, and then use the Convolutional Neural Network (CNN)~\cite{1989-CNN} to encode them.

\textbf{Code representation for tree-based features.} As mentioned in Section~\ref{sec:introduction}, many works preprocess the raw AST before feeding it to the model for learning, with the purpose of simplifying the complexity of the AST. When the raw AST is converted into sequential AST data (e.g., SBT and AST Path) by AST preprocessing methods, most works encode it using a sequence model. For example, Hu et al.~\cite{2018-Deep-Code-Comment-Generation} use LSTM to encode the SBT of the corresponding AST. LeClair et al.~\cite{2019-Ast-attendgru} utilize a GRU layer to encode SBT-AO. SBT-AO is a modified version of SBT in which all the code structure remain intact, but all words (except official Java API class names) in the code are replaced with a special <OTHER> token. 
Alon et al.~\cite{2019-Code2seq} adopt BiLSTM to encode AST Path. 
Bertolotti et al.~\cite{2023-Fold2Vec} split the AST into sub-trees and then use the pre-order visit to get lists of non-terminal and terminal nodes in the AST. Then they use a multi-head global attention layer to encode the terminal node list and use BiLSTM to encode the non-terminal node list. 
Shahbazi et al.~\cite{2021-API2Com} flatten ASTs into sequences by tree traversal and encode the flattened ASTs with Transformer. 

When the raw AST is split into simple structural data while retaining the tree structure, most works encode it using a tree-structured model. For instance, Mou et al.~\cite{2016-TBCNN} propose a novel tree-based convolutional neural network (TBCNN), in which a set of subtree feature detectors, called the tree-based convolution kernel, slides over the entire AST to extract structural information of a program. Dynamic pooling is applied to gather information over different parts of the AST. Finally, a hidden layer and an output layer are added. Many works~\cite{2017-Supervised-Deep-Features-for-Clone-Detection, 2018-Improving-Code-Summarization-via-DRL, 2019-Multi-modal-Attention-for-Code-Retrieval, 2021-FCCA} adopt TreeLSTM (including N-ary TreeLSTM) to encode binary trees of the corresponding AST. Zhang et al.~\cite{2019-ASTNN} split each AST into a sequence of statement trees (ST-trees) and encode ST-trees with a Recursive Neural Network (RvNN)~\cite{2011-Parsing-Natural-Scenes-with-RvNN} to learn vector representations of ST-trees. Then the statement representations are fed into a Bidirectional Gated Recurrent Unit (BiGRU) to learn the representation of the code fragment. 
Similar to~\cite{2019-ASTNN}, Shi et al.~\cite{2021-CAST} hierarchically split a large AST into a set of subtrees according to a pre-defined AST splitting rule and utilize an RvNN to encode the subtrees. Then, they aggregate the embeddings of subtrees by reconstructing the split ASTs to get the representation of the complete AST. 
Shido et al.~\cite{2019-Code-Summarization-with-Extended-Tree-LSTM} propose Multi-way TreeLSTM to handle ASTs with an arbitrary number of ordered children. Specifically, they add an ordinary chain-like LSTM to each gate before linear transformation to flexibly adapt to a node that has an arbitrary number of ordered children.
Lin et al.~\cite{2021-BASTS} split the code of a method based on the blocks in the dominator tree of the control flow graph, and generate a split AST for each code split. Each split AST is then encoded by a Child-Sum TreeLSTM. 
Some researchers have tried to modify the architecture of the Transformer to facilitate the learning of structural information.
For example, LeClair et al.~\cite{2021-SiT} add additional edges to the AST to further represent control flow and data dependency, and then adopt an adjacency matrix to represent the AST. They propose a structure-induced Transformer (SiT), which contains three structure-induced self-attention network (Si-SAN) layers. Code sequences and corresponding adjacency matrices are passed into the SiT encoder to get code embeddings.
Gong et al.~\cite{2022-SCRIPT} propose a StruCtural RelatIve Position guided Transformer, named SCRIPT. SCRIPT leverages ASTs to obtain the structural relative positions between tokens, which are then fed into two types of transformer encoders. One transformer encoder directly adjusts the input based on the structural relative distance, while the other transformer encodes the structural relative positions while computing the self-attention scores. Finally, these two types of transformer encoders are stacked together to learn source code representations.
Tang et al.~\cite{2022-AST-trans} denote the ancestor-descendent and sibling relationship matrices among AST nodes through two relationship matrices. They apply tree-structured attention instead of the standard self-attention to dynamically allocate weights for relevant nodes and exclude irrelevant nodes based on these two relationships. 

There are also works that treat ASTs as graphs, thus encoding ASTs with graph neural networks (GNNs). For instance, LeClair et al.~\cite{2020-Improved-Code-Summarization-Via-GNN} encode the AST nodes and edges with ConvGNN and allow the nodes of the AST to learn representations based on their neighboring nodes.
Zhou et al.~\cite{2022-Automatic-Source-Code-Summarization-With-GNN} treat the AST as an undirected graph and utilize Graph Attention Networks (GAN) as the encoder.
Yang et al.~\cite{2021-Code-Summarization-for-Smart-Contracts} transform the source code into ASTs and SBT sequences, and use graph convolutional neural network (GCN) and Transformer as the encoder, respectively. 
These works essentially treat a tree as a special type of graph.

\textbf{Code representation for graph-based features.} The Graph Neural Network (GNN) and its variants, including Graph Convolutional Network (GCN), Graph Attention Network (GAT), and Gated Graph Neural Network (GGNN), are the most commonly used neural network architecture in conjunction with graph-based code features in code-related tasks. 
For example, Allamanis et al.~\cite{2018-Learning-Represent-Programs-with-Graphs} represent program source code as graphs and use different edge types to model syntactic and semantic relationships between different tokens. Gated Graph Neural Network (GGNN) is applied to encode the constructed program graph. 
Zhao et al.~\cite{2018-DeepSim} encode the code semantics represented by control flow and data flow into a single semantic matrix, and train a specially designed feed-forward neural network to learn code representations. 
Zhou et al.~\cite{2019-Devign} use GNN to learn representations for the CPG. 
Hua et al.~\cite{2021-FCCA} adopt the Graph Convolutional
Network (GCN)~\cite{2017-GCN} to encode CFGs. 
Liu et al.~\cite{2023-Graphsearchnet} construct directed graphs for programs based on ASTs. They propose a multi-head attention module to further improve the expression of bidirectional GGNN (BiGGNN) and feed the graphs into two BiGGNN encoders to learn the vector representations. 

In this paper, we focus on a tree-based code representation, i.e., AST-based code representation. We conduct comprehensive experiments to explore the impact of the choice of popular AST preprocessing and encoding methods on AST-based code representation as well as subsequent code-related tasks.

\section{Conclusion}
\label{sec:conclusion}
In this paper, we first quantitatively evaluate the contribution of AST-based code representation on three popular code-related tasks, i.e., code clone detection, code search, and code summarization. The results show that the models trained with AST-based code representation consistently perform worse across all three tasks compared with the models trained with Token-based code representation. 
Then, we conduct a qualitative analysis of scenarios/cases in which AST-based code representation performs better than Token-based code representation in the three code-related tasks. We find that the models trained with AST-based code representation outperform models trained with Token-based code representation in certain subsets of samples across all three tasks. For example, clone pairs of code snippets have low textual similarity in code clone detection.
Finally, we conduct comprehensive experiments to evaluate and reveal the impact of the choice of various AST parsing, preprocessing, and encoding methods on AST-based code representation and subsequent code-related tasks. The results demonstrate that the impact of different methods at different stages varies for different code-related tasks. We believe that the results and findings of this paper can provide useful guidance for subsequent researchers to use AST to solve code-related tasks.

\section*{Acknowledgment}
This work is supported by National Natural Science Foundation of China (61932012, 62141215, 62372228).

\bibliographystyle{ACM-Reference-Format}
\bibliography{reference}

\end{document}